%% file: dice_lattice.tex
\documentclass[notitlepage,english,aps,prx,tightenlines,floatfix,longbibliography,10pt,citeautoscript,twocolumn,colorlinks=true,urlcolor=blue,citecolor=blue, linkcolor=blue]{revtex4-2}

\pdfoutput=1

\usepackage{graphicx}
\usepackage{dcolumn}
\usepackage{bm}
\usepackage{color}
\usepackage[dvipsnames]{xcolor}
\usepackage{physics}
\usepackage{orcidlink}
\usepackage{hyperref}
\usepackage{amssymb,ulem,amsmath}
\newcommand{\qql}{\textquotedblleft}
\newcommand{\qqr}{\textquotedblright}
\newcommand{\dg}{\dagger}
\newcommand{\ham}{\mathcal{\hat H}}

\begin{document}

\title{Signatures of many-body localization of quasiparticles in a flat band superconductor}

\author{Koushik Swaminathan\orcidlink{https://orcid.org/0000-0003-4932-9977}}

\affiliation{Department of Applied Physics, Aalto University, FI-00076 Aalto, Finland}

\author{Poula Tadros}

\affiliation{Department of Applied Physics, Aalto University, FI-00076 Aalto, Finland}

\author{Sebastiano Peotta\orcidlink{https://orcid.org/0000-0002-9947-1261}}
\email{sebastiano.peotta@aalto.fi}	
\affiliation{Department of Applied Physics, Aalto University, FI-00076 Aalto, Finland}

\begin{abstract}
We construct a class of exact eigenstates of the Hamiltonian obtained by projecting the Hubbard interaction term onto the flat band subspace of a generic lattice model. These exact eigenstates are many-body states in which an arbitrary number of localized fermionic particles coexist with a  sea of mobile Cooper pairs with zero momentum. By considering the dice lattice as an example, we provide evidence that these exact eigenstates are, in fact, a manifestation of local integrals of motions of the projected Hamiltonian. In particular, the spin and particle densities retain memory of the initial state for a very long time if localized unpaired particles are present at the beginning of the time evolution. This shows that many-body localization of quasiparticles and superfluidity can coexist even in generic two-dimensional lattice models with flat bands, for which it is not known how to construct local conserved quantities. Our results open new perspectives on the old condensed matter problem of the interplay between superconductivity and localization. 
\end{abstract}

\maketitle

\section{Introduction}

Strongly correlated quantum many-body systems remain a foremost challenge in physics and  hold the key to understand fascinating phenomena such as high-temperature (high-$T_{\rm c}$) superconductivity.~\cite{Bednorz1986} The microscopic origin of high-temperature superconductivity remains a topic of active research because it is still very difficult to accurately simulate the model Hamiltonians that are believed to describe the relevant low-energy degrees of freedom of high-$T_{\rm c}$ superconducting materials. One of these model Hamiltonians, the Fermi-Hubbard model~\cite{HubbardJohn1963} which describes copper-based high-$T_{\rm c}$ superconductors (so-called cuprates), has become the favorite test subject in the field of strongly correlated systems.~\cite{LeBlanc2015,Arovas2022} Numerical exact diagonalization is particularly challenging in the case of the Hubbard model with repulsive interactions, due to many competing orders.~\cite{Kivelson1998,Fradkin2015,Vanhala2018a} The type of order that manifests in the ground state is sensitive to the size of the finite cluster used in exact diagonalization, making the extrapolation to the thermodynamic limit problematic. On the other hand, quantum Monte Carlo suffers from the sign problem in the case of repulsive interactions.~\cite{LeBlanc2015}

To overcome these difficulties, new approaches have been developed, such as the quantum simulation of the Hubbard model using ultracold gases in optical lattices~\cite{Esslinger2010,Mazurenko2017}, in addition to new advanced numerical methods
based on tensor network states,~\cite{Schollwock2011,Stoudenmire2012} for instance. Recent numerical results indicate that superconducting long-range order does not seem to occur in the repulsive Hubbard model on a simple square lattice with only nearest-neighbor hoppings.~\cite{Qin2020} Thus, to properly account for the superconductive properties of copper-based superconductors it becomes important to consider more realistic model Hamiltonians, for instance, by augmenting the pure Hubbard model with next-nearest-neighbor hoppings,~\cite{Ponsioen2019,Jiang2019} or by considering a lattice model with more than one band. In the case  of cuprates, a three-band model describing both the copper $d$-orbitals  and the oxygen $p$-orbitals in the copper-oxide planes of cuprates seems the most appropriate.~\cite{Varma1987,Mattheiss1987,Emery1987,Kung2016,Valkov2016,Adolphs2016}

\begin{figure}
	\includegraphics[scale=0.9]{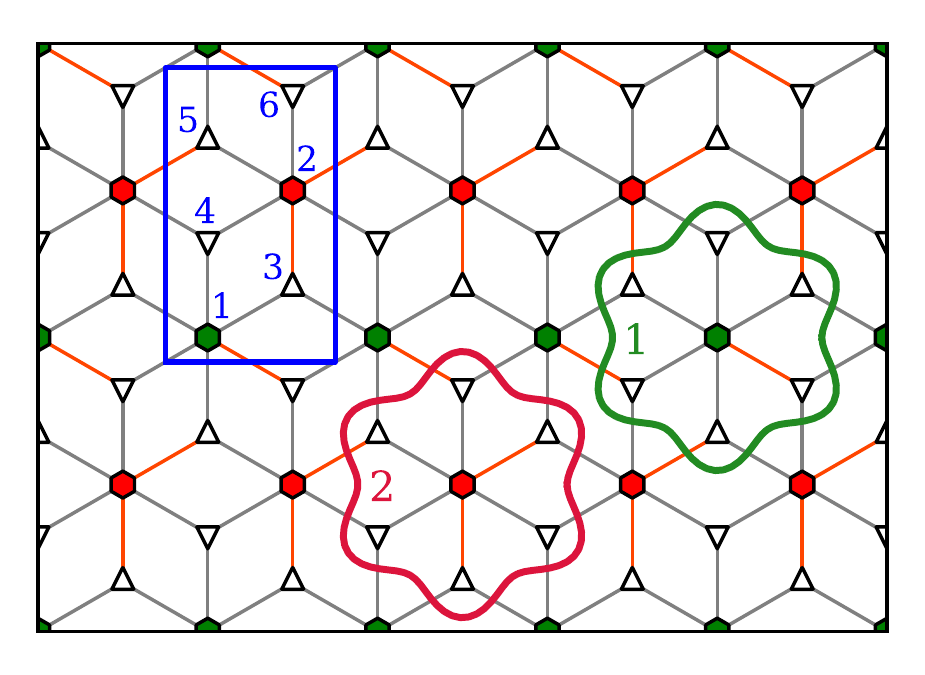}
	\caption{\label{fig:dice} Schematic of the dice lattice. Six-fold coordinated sites (\textit{hub} sites, orbital index $\alpha = 1,2$) are denoted by hexagons, while three-fold coordinated sites (\textit{rim} sites) are denoted by triangles. The rim sites can be further divided into two triangular sublattices denoted by up- ($\alpha = 3,5$) and down-pointing triangles ($\alpha = 4,6$), respectively. The bonds denote nearest-neighbour hoppings which are all real and have the same absolute value $t=1$. The color of the bond denotes the sign of the hopping amplitude, grey for $+t$ and red for $-t$. The rectangular box is the magnetic unit cell, given by the fundamental vectors $\vb{a}_{j = 1,2}$~\eqref{eq:fund_vec_rect}. Two Wannier functions of the two lowest flat bands of the dice lattice Hamiltonian are also shown. The Wannier function $w_{n\vb{l}}$, with $n = 1,2$, is centered on the hub site $\alpha =n$ in unit cell $\vb{l}$ and is nonzero only on the same hub site and the adjacent rim sites, therefore it is compactly localized. These \qql flower states\qqr form the orthonormal basis used in the expansion of the projected Hamiltonian~\eqref{eq:Ham}.}
\end{figure}

Multiband lattice models contain more than one orbital per unit cell, and are therefore more difficult to simulate numerically. On the other hand, they can harbor new qualitative effects. For instance, in a multiband lattice model, the strongly correlated regime can be achieved by reducing the bandwidth of a partially filled band to zero, obtaining a so-called flat band.~\cite{Leykam2018} Crucially, due to interfering hopping paths that lead to particle localization, the vanishing of the bandwidth is not necessarily accompanied by the vanishing of the hopping  amplitudes in the lattice model, as it would in the case of a single band/orbital lattice model. This means that the Wannier functions that span the flat band subspace can have a large overlap and by projecting the interaction term on this subspace one obtains a nontrivial purely \textit{quartic} Hamiltonian, that is a linear combination of products of four fermionic field operators.~\cite{Huber2010,Tovmasyan2016} The range of the terms in this Hamiltonian is determined by how fast the Wannier functions decay with distance.

The Hubbard interaction term projected on the flat band subspace of a lattice model, called the projected Hamiltonian from here on, is the subject of the present work. This class of quartic Hamiltonians essentially poses a many-body problem and are in general, difficult to study. Nevertheless, it is possible to derive several interesting exact results. One example is provided in this work, in which we derive exact eigenstates of generic projected Hamiltonians in the spin imbalanced case, that is for different number of spin up and down particles $N_\uparrow \neq N_\downarrow$. This new result builds on previous related work.~\cite{Tovmasyan2016,Tovmasyan2018,Herzog-Arbeitman2022}

Recently, there has been a lot of interest on the topic of superconductivity and superfluidity occurring in flat or quasi-flat bands. The original motivation is that the critical temperature of the superconducting transition is enhanced by the high density of states of a quasi-flat band~\cite{Heikkila2011,Kopnin2011} (the density diverges in the strict flat band limit, nevertheless, the critical temperature remains finite). Moreover, lattice models with flat bands have been realized with ultracold gases in optical lattices, which are a flexible platform for exploring superfluid phenomena. Two notable examples are the Lieb lattice,~\cite{Taie2015,Ozawa2017} which also describes the atomic structure of the copper-oxide planes of cuprates, and the kagome lattice.~\cite{Jo2012,Leung2020} Proposals to realize other lattice models with flat bands, such the dice lattice, have been put forward.~\cite{Moller2018} Finally, the discovery of superconductivity in twisted bilayer graphene~\cite{Cao2018} has stimulated a lot of theoretical and experimental efforts in the new field of moir\'e materials,~\cite{Torma2022} in which flat bands and band structures with exotic properties are engineered through the precise twisting of layers of two-dimensional materials, such as graphene, with respect to each other. The same idea has also been implemented in ultracold gas experiments.~\cite{Meng2021}

Flat bands are rather peculiar, since, in the absence of interactions they are necessarily insulating regardless of the filling. Thus, one way to characterize flat bands is as translationally invariant or disorder-free Anderson insulators. On the other hand, a partially filled flat band generally becomes a superconductor with a high critical temperature as soon as an attractive Hubbard interaction is switched on.~\cite{Julku2016,Liang2017a,Hofmann2020,Peri2021} The reason being that two-body bound states become delocalized and their inverse effective mass is proportional to the interaction strength, and to the overlap between distinct Wannier functions of the flat band.~~\cite{Vidal2000,Doucot2002,Tovmasyan2016,Torma2018}   

Recent works~\cite{Tovmasyan2018,Herzog-Arbeitman2022} have pointed out that, even if the ground state of a partially filled flat band in the presence of interactions is superconducting/superfluid, the quasiparticle excitations can be localized, namely they have infinite effective mass. Quasiparticle excitations carry a nonzero spin and have fermionic statistics. These localized quasiparticles coexist with mobile two-body bound states, which are bosons. 
In other words, the dispersion of quasiparticles is flat as in the noninteracting case. Evidence of localization of quasiparticles in flat band superconductors mainly comes from the analytical results obtained in Refs.~\onlinecite{Tovmasyan2018,Herzog-Arbeitman2022} and from the quasiparticle dispersion obtained from the Bogoliubov-de Gennes Hamiltonian of mean-field theory, for instance, in the case of the Lieb lattice.~\cite{Julku2016}

The phenomenon of localization of quasiparticles in 
flat band superconductors draws an interesting connection with many-body localization, a subject which has attracted considerable interest,~\cite{Gornyi2005,Basko2006,Oganesyan2007,Pal2010,Imbrie2016,Nandkishore2015,Altman2015,Alet2018} thanks to its observation in ultracold gas experiments.~\cite{Kondov2015,Schreiber2015,Choi2016} Loosely speaking, many-body localization is simply Anderson localization in the presence of interactions. More precisely, it is now established that the defining property of many-body localization is the presence of an extensive number of local integrals of motion that completely block particle transport and prevent thermalization in a quantum many-body system.~\cite{Imbrie2016,Serbyn2013,Huse2014,Rademaker2017,Imbrie2017} These local integrals of motion generally appear for large enough disorder, however, examples of disorder-free systems in which many-body localization occurs are also known.~\cite{Schiulaz2015,Smith2017,Mondaini2017,Brenes2018,Osborne2023} Methods to design lattice models possessing local integrals of motion from the bottom up have been proposed as well.~\cite{Danieli2020,Danieli2022}

Interestingly, for some specific one-dimensional lattice models with flat bands, it is possible to analytically construct local integrals of motion, also called conserved quantities, thereby establishing a rigorous connection between the localization of quasiparticles in flat band superconductors and many-body localization.~\cite{Tovmasyan2018} The main difference with standard many-body localization is that the mobile two-body bound states present ergodic behavior and thermalize, contrary to the localized quasiparticles. Analogous analytical results regarding local conserved quantities in the case of lattice models in two or more dimensions is lacking at present. This important open question is addressed here with an exact diagonalization study of a specific lattice model with flat bands, the dice lattice,~\cite{Horiguchi1974,Sutherland1986,Vidal1998,Vidal2001,Moller2012,Payrits2014,Andrijauskas2015,Wu2021} see Fig.~\ref{fig:dice}. Through the analysis of the eigenstate spectrum and the time evolution of three- and four-particle states, we find evidence for the presence of local conserved quantities in this model.

The present work is organized as follows.
In Section~\ref{sec:exact_generic}, we provide new analytic results for the projected Hamiltonian of generic lattice models with flat bands. In order the keep our work self-contained, in Section~\ref{sec:spin_balanced} we introduce the concept of projected Hamiltonian for the attractive Hubbard interaction and derive its exact ground state in the spin balanced case $N_\uparrow = N_\downarrow$ ($N_\sigma$ is the number of particles with spin $\sigma$). 
The method used to obtain this result is slightly different from the original work~\cite{Tovmasyan2016,Herzog-Arbeitman2022} and is more suited to derive the new exact eigenstates in the spin imbalanced case $|N_\uparrow - N_\downarrow|\geq 1$, which are presented in Section~\ref{sec:exact_localized}. In these exact eigenstates, localized quasiparticles are placed at arbitrary positions on top of a sea of Cooper pairs with zero momentum. The special case $|N_\uparrow - N_\downarrow| = 1$ has been already discussed in Ref.~\onlinecite{Herzog-Arbeitman2022}. The assumptions under which these results hold are presented in detail in Section~\ref{sec:spin_balanced}, particularly, the so-called uniform pairing condition introduced in Ref.~\onlinecite{Tovmasyan2016}.

In Section~\ref{sec:dice_lattice_analytic}, a specific lattice model with flat bands, the two-dimensional dice lattice with a perpendicular magnetic field,~\cite{Vidal1998,Vidal2001} is introduced. For a specific value of the magnetic flux per unit cell, the two lowest bands of the dice lattice are perfectly flat and degenerate. A special property of the dice lattice is that the flat band subspace is spanned by Wannier functions that can be taken to be compactly localized, that is, nonvanishing only on a finite number of lattice sites. This property makes the dice lattice particularly amenable to analytical and numerical studies. Indeed, in Section~\ref{sec:uniform_pairing_dice_lat}, we find that the general analytical results presented in Section~\ref{sec:exact_generic} remain valid in the case of the dice lattice even if the uniform paring condition is partially lifted. This is a straightforward consequence of the compact nature of the Wannier functions. In Section~\ref{sec:proj_Ham_dice}, we provide a convenient representation of the projected Hamiltonian of the dice lattice. The projected Hamiltonian can be written in terms of operators that belong to three different classes: i) localized spins, ii) on-site singlets and iii) bond singlets. The on-site singlets and the bond singlets are two-body bound states moving on triangular and kagome lattices, respectively. The projected Hamiltonian also contains a term that converts on-site singlets to bond singlets and vice versa. In preparation for the numerical results presented in the following sections, the two-body problem for the projected Hamiltonian of the dice lattice is solved in Appendix~\ref{app:two_body_problem}. 

Section~\ref{sec:numerical_results} presents numerical results obtained by exact diagonalization of the projected Hamiltonian of the dice lattice for particle number $N_\uparrow + N_\downarrow >2$ and $N_\uparrow \neq N_\downarrow$. 
First, in Section~\ref{sec:energy_spectrum} the spectrum of the Hamiltonian is investigated. It is found that, while the ground state is perfectly degenerate, in agreement with the general analytical results of Section~\ref{sec:exact_generic}, the excited states form multiplets that are only quasidegenerate. We interpret this as evidence of local conserved quantities, with the lifting of the degeneracy possibly due to finite size effects.
In Section~\ref{sec:dynamics},  the focus is on the nonequilibrium dynamics. The main result is that the spin and particle densities do not thermalize, but rather retain memory of the initial positions of the single unpaired particles. This is another piece of evidence for the presence of local conserved quantities enforcing quasiparticle localization.

Finally, in Section~\ref{sec:conclusions}, we summarize and discuss the main results of this work and point out interesting directions for future studies.

\section{Exact many-body eigenstates in lattice models with flat bands}
\label{sec:exact_generic}

\subsection{Flat band projected Hamiltonian} 
\label{sec:spin_balanced}

In this section, we introduce the technique of projecting the Hubbard interaction term of a lattice Hamiltonian onto the subspace corresponding to a flat band or a group of degenerate flat bands, which are obtained by diagonalizing the noninteracting term of the same Hamiltonian. This method is just the first step in a systematic expansion known as the Schrieffer-Wolff transformation.~\cite{Tovmasyan2016} The small parameter in this perturbative expansion is $U/E_{\rm gap}\ll 1$, where $U$ is the coupling constant of the interaction, in our case the Hubbard interaction, and $E_{\rm gap}$ is the energy gap, that is, the minimal energy interval separating the group of degenerate flat bands from all other bands. 
The projected Hamiltonian provides an accurate description of the low-energy degrees of freedom of the many-body system if the parameter $U/E_{\rm gap}$ is small enough. This is called the isolated band limit.

Using the algebra of projected field operators, it is shown how to construct exact eigenstates of the projected Hamiltonian with an arbitrary number of Cooper pairs with zero momentum. The derivation of this result, presented originally in Ref.~\onlinecite{Tovmasyan2016} in the case where the particle number is not fixed (grand canonical ensemble) and in  Ref.~\onlinecite{Herzog-Arbeitman2022} for fixed particle number (canonical ensemble), is repeated here with some variations since it serves as a starting point for obtaining a new class of exact eigenstates in  Section~\ref{sec:exact_localized}. The notation used here is very close to that of Refs.~\onlinecite{Tovmasyan2016,Tovmasyan2018}.

Consider a generic translationally invariant lattice model with a Hamiltonian of the form $\ham = \ham_{\rm free} + \ham_{\rm int}$, where $\ham_{\rm free}$ is the noninteracting term and $\ham_{\rm int}$ an Hubbard interaction term of the form
\begin{equation}
	\ham_{\rm int} = -\sum_{\vb{i},\alpha}U_\alpha\hat{n}_{\vb{i}\alpha\uparrow}\hat{n}_{\vb{i}\alpha\downarrow}\,, \quad \hat{n}_{\vb{i}\alpha\sigma} = \hat{c}^\dg_{\vb{i}\alpha\sigma} \hat{c}_{\vb{i}\alpha\sigma}\,,
\end{equation} 
with $\hat{c}_{\vb{i}\alpha\sigma}$ fermionic field operators. Here, we focus on the case of attractive interactions $U_\alpha > 0$. The vector of integers $\vb{i} = (i_1, i_2)^T$ labels the unit cells and $\alpha = 1,\dots,N_{\rm orb}$ labels the different orbitals inside the unit cell. For instance, the dice lattice, which is studied in detail in Sections~\ref{sec:dice_lattice_analytic} and~\ref{sec:numerical_results}, contains $N_{\rm orb} = 6$ orbitals in its unit cell, as shown in Fig.~\ref{fig:dice}. 
For spin-$1/2$ fermions, the spin index takes two possible values $\sigma = \uparrow,\,\downarrow$. Without loss of generality, we restrict our presentation to two-dimensional systems for definiteness. 

The noninteracting term $ \ham_{\rm free}$ is quadratic in the field operators and can be diagonalized by solving the corresponding single-particle problem. If we also assume that the component $\sigma$ of the spin is a conserved quantity, then the noninteracting term  can be written in a diagonalized form as 
\begin{equation}
	\ham_{\rm free} = \sum_{n,\vb{k}} \sum_{\sigma = \uparrow,\downarrow}\varepsilon_{n\vb{k}\sigma} \hat{f}^\dg_{n\vb{k}\sigma}\hat{f}_{n\vb{k}\sigma}\,,
\end{equation}
where $\hat{f}_{n\vb{k}\sigma}$ are new fermionic operators associated to the eigenstates  $\psi_{n\vb{k}\sigma}(\vb{i},\alpha)$ of the noninteracting (single-particle) Hamiltonian. Due to translational invariance, these eigenstates are Bloch plane waves labeled by the band index $n$ and the quasimomentum $\vb{k} = (k_x, k_y)^T$. We assume that the bands in a given set $\mathcal{F}$ are all degenerate and flat, that is,
\begin{equation}
	\varepsilon_{n\vb{k}\sigma} = \varepsilon_0 = \mathrm{const.}\qq{for} n \in \mathcal{F}\qq{and} \sigma = \uparrow,\,\downarrow\,.
\end{equation}
We require the Hamiltonian to be  time-reversal symmetric, therefore all the flat bands are spin degenerate. The group of degenerate flat bands is also isolated, that is, separated from all other bands by a finite band gap $E_{\rm gap}$.

Since we are interested in the low energy properties of the system when the flat bands in the set $\mathcal{F}$ are partially filled and $U_\alpha/E_{\rm gap}\ll1$, it is useful to introduce projected field operators~\cite{Tovmasyan2016}
\begin{equation}
\label{eq:proj_field_op}
\begin{split}
\bar{c}_{\vb{i}\alpha\sigma} &= \frac{1}{\sqrt{N_{\rm c}}}\sum_{n \in \mathcal{F}}\sum_{\vb{k}}\psi_{n\vb{k}\sigma}(\vb{i},\alpha)\hat{f}_{n\vb{k}\sigma} \\
&= \sum_{n \in \mathcal{F}} \sum_{\vb{l}}w_{n\sigma}(\vb{i}-\vb{l},\alpha)\hat{d}_{n\vb{l}\sigma}\,.
\end{split}
\end{equation}
The projected field operators are obtained by expanding the field operators $\hat{c}_{\vb{i}\alpha\sigma}$ in the basis of Bloch plane waves and retaining only those terms corresponding to the flat bands in the set $\mathcal{F}$. An alternative expansion is in terms of the Wannier functions $w_{n\sigma}(\vb{i}-\vb{l},\alpha)$ and the associated field operators $\hat{d}_{n\vb{l}\sigma}$. The Wannier functions are obtained  from the Bloch plane waves as
\begin{equation}
	w_{n\sigma}(\vb{i},\alpha) = \frac{1}{N_{\rm c}}\sum_{\vb{k}} \psi_{n\vb{k}\sigma}(\vb{i},\alpha)\,,\label{eq:Bloch_to_Wannier}
\end{equation}
with $N_{\rm c}$ the number of unit cells in the lattice.~\cite{Marzari2012} If, and only if, the bands are flat, will the Wannier functions be eigenstates of the noninteracting Hamiltonian. The Wannier functions provide a convenient basis of local wave functions on which the projected Hamiltonian can be expanded, as shown below in the case of the dice lattice.
An important property of Wannier functions is that a complete basis for the subspace of a given band $n$ is obtained by taking the translates of just a single Wannier function, that is $\{w_{n\sigma}(\vb{i}-\vb{l},\alpha)\,|\, \vb{l} \in \mathbb{Z}^2\}$. Moreover, this basis is orthonormal since
\begin{equation}
\sum_{\vb{i},\alpha} w^*_{n_1\sigma}(\vb{i}-\vb{l}_1,\alpha)
w_{n_2\sigma}(\vb{i}-\vb{l}_2,\alpha)	=\delta_{n_{1},n_{2}}\delta_{\vb{l}_1,\vb{l}_2}\,.
\end{equation}

A good description of the low energy properties for small $U_\alpha/E_{\rm gap}$ is given by the projected Hamiltonian, in our case is the Hubbard interaction term projected on the subspace of the degenerate flat bands.
The projected Hamiltonian expanded in the Wannier function basis reads
\begin{widetext}
\begin{equation}\label{eq:Ham}
	\begin{split}
		&\mathcal{\overline H}_{\rm int} = -\sum_{\vb{i},\alpha}U_\alpha\bar{n}_{\vb{i}\alpha\uparrow}\bar{n}_{\vb{i}\alpha\downarrow} =
		-\sum_{\vb{i},\alpha}U_\alpha \bar{c}^\dagger_{\vb{i}\alpha\uparrow}\bar{c}^\dagger_{\vb{i}\alpha\downarrow}\bar{c}_{\vb{i}\alpha\downarrow}\bar{c}_{\vb{i}\alpha\uparrow}\\
		&= -\sum_{n_1,\dots,n_4}\sum_{\vb{l}_1,\dots,\vb{l}_4}\bigg(\sum_{\vb{i},\alpha} U_\alpha
		w^*_{n_1\uparrow}(\vb{i}-\vb{l}_1,\alpha)
		w^*_{n_2\downarrow}(\vb{i}-\vb{l}_2,\alpha)
		w_{n_3\downarrow}(\vb{i}-\vb{l}_3,\alpha)
		w_{n_4\uparrow}(\vb{i}-\vb{l}_4,\alpha)
		\bigg)\hat{d}^\dagger_{n_1\vb{l}_1\uparrow}\hat{d}^\dagger_{n_2\vb{l}_2\downarrow}\hat{d}_{n_3\vb{l}_3\downarrow}\hat{d}_{n_4\vb{l}_4\uparrow}\,.
	\end{split}
\end{equation}
\end{widetext} 
Here, the sum over the band indices is restricted to the group of degenerate flat bands ($n_1,\dots,n_4 \in \mathcal{F}$).

It is important to note that the projected field operators defined in~\eqref{eq:proj_field_op} satisfy modified anticommutation relations
\begin{gather}
	\{\bar{c}_{\vb{i}\alpha\sigma},\bar{c}_{\vb{j}\beta\sigma'}\} = \{\bar{c}_{\vb{i}\alpha\sigma}^\dagger,\bar{c}_{\vb{j}\beta\sigma'}^\dagger\} = 0\,,
	\label{eq:cbar_anticomm1}\\
	\{\bar{c}_{\vb{i}\alpha\sigma},\bar{c}_{\vb{j}\beta\sigma'}^\dagger\} = \delta_{\sigma,\sigma'}P_\sigma(\vb{i}-\vb{j},\alpha,\beta)\,,
    \label{eq:c_cbar_anticomm2}\\ 
	P_\sigma(\vb{i}-\vb{j},\alpha,\beta) = \sum_{n,\vb{l}} w_{n\sigma}(\vb{i}-\vb{l},\alpha) w^*_{n\sigma}(\vb{j}-\vb{l},\beta)\,,
\end{gather}
where $P_\sigma$ is the single-particle operator projecting on the $\mathcal{F}$ subspace with spin $\sigma$. As a consequence of the above comutation relations, we have the identity
\begin{equation}
	\label{eq:n_squared}
	\bar{n}_{\vb{i}\alpha\sigma}^2 = P_\sigma(\vb{0},\alpha,\alpha)\bar{n}_{\vb{i}\alpha\sigma}\,, \qq{with} \bar{n}_{\vb{i}\alpha\sigma} = \bar{c}_{\vb{i}\alpha\sigma}^\dagger \bar{c}_{\vb{i}\alpha\sigma}\,,
\end{equation}  
which should be compared with  $\hat{n}_j^2 = \hat{n}_j$, valid for the usual fermionic operators $\hat{c}_j,\,\hat{c}_j^\dg$ and $\hat{n}=\hat{c}_j^\dagger\hat{c}_j$. 

In order to construct exact eigenstates of the projected Hamiltonian, we introduce the creation operator of a Cooper pair in a zero momentum state
\begin{equation}
	\label{eq:b_def}
\begin{split}
\hat{b}^\dagger &= \sum_{n\in \mathcal{F}}\sum_{\vb{k}}
\hat{f}^\dagger_{n\vb{k}\uparrow}\hat{f}^\dagger_{n,-\vb{k},\downarrow} = \sum_{n\in \mathcal{F}}\sum_{\vb{j}}
\hat{d}^\dagger_{n\vb{j}\uparrow}\hat{d}^\dagger_{n\vb{j}\downarrow} \\ &= \sum_{\vb{i},\alpha} \bar{c}_{\vb{i}\alpha\uparrow}^\dagger\bar{c}_{\vb{i}\alpha\downarrow}^\dagger\,.
\end{split}
\end{equation}
The equivalence of the three different expansions of the operator $\hat{b}^\dg$ shown above
is a consequence of time-reversal symmetry, which for spin-$1/2$ particles amounts to the following relations between wave functions with opposite spin
\begin{gather}
w_{n\uparrow}(\vb{i},\alpha) = w_{n\downarrow}^*(\vb{i},\alpha) \overset{\rm def}{=} w_n(\vb{i},\alpha)\,,\label{eq:TR_wannier}\\
\psi_{n\vb{k}\uparrow}(\vb{i},\alpha) = \psi_{n,-\vb{k},\downarrow}^*(\vb{i},\alpha)  \overset{\rm def}{=} \psi_{n\vb{k}}(\vb{i},\alpha)\,,\label{eq:TR_bloch} \\
P_\uparrow(\vb{i}-\vb{j},\alpha,\beta) = P^*_\downarrow(\vb{i}-\vb{j},\alpha,\beta) \overset{\rm def}{=}
P(\vb{i}-\vb{j},\alpha,\beta)\,.
\end{gather}
The following commutation relations are essential for what follows and are only valid in the case of time-reversal symmetry,
\begin{gather}
	[\bar{c}_{\vb{i}\alpha\uparrow},\hat{b}^\dagger] = \bar{c}_{\vb{i}\alpha\downarrow}^\dagger\,,\qquad
	[\bar{c}_{\vb{i}\alpha\downarrow},\hat{b}^\dagger] = -\bar{c}_{\vb{i}\alpha\uparrow}^\dagger\,, \label{eq:comm_b_1}\\
	[\bar{c}^\dagger_{\vb{i}\alpha\sigma},\hat{b}^\dagger] = 0\,,\\
	[\bar{n}_{\vb{i}\alpha\sigma},\hat{b}^\dagger] = \bar{c}_{\vb{i}\alpha\uparrow}^\dagger\bar{c}_{\vb{i}\alpha\downarrow}^\dagger\,.
\end{gather}  
To prove our results, we need an additional assumption, namely that the  following condition, introduced for the first time in Ref.~\onlinecite{Tovmasyan2016}, holds
\begin{gather}
	U_\alpha P(\vb{0},\alpha,\alpha) = U_\beta P(\vb{0},\beta,\beta) = E_{\rm p}\,,\quad \forall \alpha,\beta \in \mathcal{S} \label{eq:uniform_pairing_1}\\
	P(\vb{0},\alpha,\alpha) = 0\,,\quad {\rm if} \quad \alpha \notin \mathcal{S}\label{eq:uniform_pairing_2}\,.
\end{gather}
Here $\mathcal{S}$ is the set of orbitals on which the wave functions of the group of degenerate flat bands $\mathcal{F}$ are nonzero. Note that this condition, called the \qql uniform pairing condition\qqr,~\cite{Tovmasyan2016,Herzog-Arbeitman2022} can always be met by adjusting the orbital-dependent Hubbard couplings $U_\alpha$.

Assuming time-reversal symmetry, conservation of the spin component $\sigma$ and the uniform pairing condition, it is shown that $\hat{b}\,,\hat{b}^\dagger$ are ladder operators for the projected Hamiltonian $\mathcal{\overline H}_{\rm int}$
\begin{equation}
	\label{eq:ladder_op_b_dg}
	\begin{split}
		[\,\mathcal{\overline H}_{\rm int},\hat{b}^\dagger] &= - \sum_{\vb{i},\alpha} U_\alpha \bar{c}_{\vb{i}\alpha\uparrow}^\dagger\bar{c}_{\vb{i}\alpha\downarrow}^\dagger[  \bar{c}_{\vb{i}\alpha\downarrow}\bar{c}_{\vb{i}\alpha\uparrow},
		\hat{b}^\dagger] \\
		&= -\sum_{\vb{i},\alpha} U_\alpha \bar{c}_{\vb{i}\alpha\uparrow}^\dagger\bar{c}_{\vb{i}\alpha\downarrow}^\dagger(\bar{c}_{\vb{i}\alpha\downarrow}\bar{c}_{\vb{i}\alpha\downarrow}^\dagger - 
		\bar{c}_{\vb{i}\alpha\uparrow}^\dagger \bar{c}_{\vb{i}\alpha\uparrow}) \\
		&= -\sum_{\vb{i},\alpha} U_\alpha P(\vb{0},\alpha,\alpha) \bar{c}_{\vb{i}\alpha\uparrow}^\dagger\bar{c}_{\vb{i}\alpha\downarrow}^\dagger\\
		&= -E_{\rm p} \sum_{\vb{i},\alpha} \bar{c}_{\vb{i}\alpha\uparrow}^\dagger\bar{c}_{\vb{i}\alpha\downarrow}^\dagger = -E_{\rm p}\hat{b}^\dagger\,.
	\end{split}
\end{equation}
Here we have used, in order, the commutation relations~\eqref{eq:comm_b_1} and then~\eqref{eq:cbar_anticomm1} and~\eqref{eq:c_cbar_anticomm2} and finally the uniform pairing condition~\eqref{eq:uniform_pairing_1}. It follows that the states
\begin{equation}
		\label{eq:ket_pair}
	\ket{N_{\rm p}} = (\hat{b}^\dagger)^{N_{\rm p}} \ket{\emptyset}\,,
\end{equation}
with $N_{\rm p}$ Cooper pairs in the same zero momentum state, are exact eigenstates of the projected Hamiltonian with energy $-E_{\rm p}N_{\rm p}$. We denote by $\ket{\emptyset}$ the vacuum, the state with no particles, while the positive Cooper pair binding energy $E_{\rm p} >0$ has been introduced in~\eqref{eq:uniform_pairing_1}. A simple argument,~\cite{Tovmasyan2016,Herzog-Arbeitman2022} not repeated here, shows also that these are eigenstates with minimal energy for any fixed number of pairs $N_{\rm p}$. 

\subsection{Exact eigenstates with localized quasiparticles}
\label{sec:exact_localized}

In this section we present a new result, that is a new class of exact eigenstates of the projected Hamiltonian, characterized by the presence of  unpaired localized particles on top of a sea of Cooper pairs. The proof is very straightforward and uses the fact that the Cooper pair creation operator $\hat{b}^\dg$~\eqref{eq:b_def} is a ladder operator, as shown in~\eqref{eq:ladder_op_b_dg}. The exact eigenstates in~\eqref{eq:ket_pair} are generated by repeatedly applying the Cooper pair creation operator to the vacuum $\ket{\emptyset}$, which is an eigenstate of $\mathcal{\overline H}_{\rm int}$. Instead of the vacuum, one can start from the following eigenstates of the projected Hamiltonian
\begin{gather}
	\ket{\mathcal{I}_\sigma} = \bigg(\prod_{(n,\vb{l})\in \mathcal{I}_\sigma}\hat{d}^\dagger_{n\vb{l}\sigma}\bigg)\ket{\emptyset}\,\\
	\mathcal{\overline H}_{\rm int} \ket{\mathcal{I}_\sigma}=0\,.
\end{gather}
where $\mathcal{I}_\sigma$ is an arbitrary set of Wannier functions. The second equation above follows from the fact that the Hubbard interaction term is nonzero only if particle with opposite spins are present. The state $\ket{\mathcal{I}_\sigma}$ corresponds to an arbitrary arrangement of particles all with the same spin $\sigma$, therefore it is a trivial eigenstate of the projected Hamiltonian with eigenvalue zero. We can then apply~\eqref{eq:ladder_op_b_dg} to show that
the states of the form
\begin{equation}
	\label{eq:exact_eigen_localized}
	\ket{\mathcal{I}_\sigma,N_{\rm p}}  = (\hat{b}^\dagger)^{N_{\rm p}} \ket{\mathcal{I}_\sigma} 
\end{equation}
are also eigenstates of $\mathcal{\overline H}_{\rm int}$, that is
\begin{equation}
	\label{eq:main1}
	\mathcal{\overline H}_{\rm int}\ket{\mathcal{I}_\sigma,N_{\rm p}} =- N_{\rm p}E_{\rm p}\ket{\mathcal{I}_\sigma,N_{\rm p}}\,.
\end{equation}
Note that the energy eigenvalue does not depend on the set $\mathcal{I}_\sigma$, only on the number of pairs $N_{\rm p}$, exactly in the same way as for the eigenstates in~\eqref{eq:ket_pair}. These states are all orthogonal
and their normalization constant can be easily calculated.~\cite{Herzog-Arbeitman2022} The case in which $\mathcal{I}_\sigma$ consists of a single element, that is a single unpaired particle, has already been  derived in Ref.~\onlinecite{Herzog-Arbeitman2022} using a different method. 

Despite the simplicity of the proof, the result in~\eqref{eq:main1} is rather nontrivial since the exact eigenstates~\eqref{eq:exact_eigen_localized} describe mobile Cooper pairs coexisting with localized particles all with the same spin, which are arbitrarily arranged in space. These localized particles are a source of disorder for the Cooper pairs. Therefore, this result is a promising starting point for investigations on the interplay of superconductivity and disorder from a genuine many-body perspective. However, one has to first answer the question on whether unpaired particles remain localized if the Cooper pairs have a finite momentum, for instance if a supercurrent is present in the system. One way to address this question is by considering specific lattice models with flat bands, such as the dice lattice shown if Fig.~\ref{fig:dice}, which is the subject of the following sections.

\section{Dice lattice}
\label{sec:dice_lattice_analytic}

The goal of this section is to study the Hamiltonian obtained by projecting the Hubbard interaction term on the two lowest degenerate flat bands of the dice lattice.
The dice lattice with its hopping amplitudes shown in Fig.~\ref{fig:dice} has a band structure composed of six flat bands, which are two by two degenerate.~\cite{Tovmasyan2018}
A convenient property of the dice lattice is that the Wannier functions of the two lowest flat bands can be taken to be compactly localized and not just exponentially decaying. Compactly localized  means that a Wannier function of a lattice model is nonzero only on a finite number of sites as shown in Fig.~\ref{fig:dice}. Thus, the projected Hamiltonian~\eqref{eq:Ham} contains only a finite number of nearest-neighbor terms. Due to its special properties, in particular, the existence of compactly localized Wannier functions, the dice lattice with the hopping amplitudes shown in Fig.~\ref{fig:dice} has been studied in a number of previous works,~\cite{Horiguchi1974,Sutherland1986,Vidal1998,Vidal2001,Moller2012,Payrits2014,Andrijauskas2015,Tovmasyan2018,Wu2021} particularly within the context of superconducting wire networks and Josephson junction arrays.~\cite{Abilio1999,Korshunov2001,Cataudella2003,Korshunov2004,Korshunov2005}

If the sign of the the hopping matrix elements shown in Fig.~\ref{fig:dice} is ignored, the underlying Bravais lattice is triangular with fundamental vectors 
\begin{gather}
	\label{eq:fund_vec_v}
	\vb{v}_1 = \sqrt{3}a\pmqty{1\\0}\,,\qquad \vb{v}_2 = \sqrt{3}a\pmqty{\frac{1}{2} \\ \frac{\sqrt{3}}{2}}\,, \\
	\label{eq:fund_vec_v3}
	 \vb{v}_3 = \vb{v}_2 -\vb{v}_1 = \sqrt{3}a\pmqty{-\frac{1}{2} \\ \frac{\sqrt{3}}{2}}\,.
\end{gather}
Here, $a$ is the length of the edge of an elementary rhombus in the dice lattice, that is, the distance between two nearest-neighbor lattice sites.
Any pair of vectors chosen from the above three is sufficient to generate the triangular lattice. However, for future calculations, it is convenient to introduce a third vector $\vb{v}_3$, which is a linear combination of the other two.

The magnetic field is encoded in a lattice model by means of Peierls phases in the hopping amplitudes. In the case of the dice lattice shown in Fig.~\ref{fig:dice}, the Peierls phases are all real, that is, they are simply a sign $\pm 1$, corresponding to the bond colors in the figure. Therefore, the magnetic flux through each elementary rhombus of the lattice is a half-flux quantum. The uniform magnetic field partially breaks the translational symmetry of the dice lattice and the magnetic unit cell has an area twice as large as the unit cell of the triangular lattice~\eqref{eq:fund_vec_v}.
The underlying Bravais lattice is rectangular and a possible choice for the fundamental vectors is 
\begin{equation}
	\label{eq:fund_vec_rect}
	\vb{a}_1 = \vb{v}_1 = \sqrt{3}a\pmqty{1\\0}\,,\qquad \vb{a}_2 = 2\vb{v}_2-\vb{v}_1 = \pmqty{0 \\ 3a}\,.
\end{equation} 

For our calculations, we only need the the Bloch functions of the two lowest degenerate flat bands~\cite{Tovmasyan2018}
\begin{gather}
	\begin{split}
	&\psi_{n\vb{k}}(\vb{i},\alpha) = e^{i\vb{k}\cdot\vb{r}_{\vb{i}}} g_{n\vb{k}}(\alpha)\,, \\ 
	&\qq*{with}\vb{r}_{\vb{i}} = i_1\vb{a}_1+i_2\vb{a}_2\,,\quad \vb{i} = (i_1, i_2)^T
	\,, 
	\end{split}\\
	\ket{g_{1,\vb{k}}} = c\qty\big(\varepsilon_0, 0, 1+e^{ik_1},1,  e^{ik_2},e^{ik_2}(e^{ik_1}-1))^T\,,\label{eq:g_1_dice}\\
	\ket{g_{2,\vb{k}}} = c\qty\big(0,\varepsilon_0,-1,1+e^{-ik_1},1-e^{-ik_1},1)^T\,.\label{eq:g_2_dice}
\end{gather}
We denote with $\ket{g_{n\vb{k}}}$ a vector with components $g_{n\vb{k}}(\alpha)$, that is, $\ket{g_{n\vb{k}}} = \big(g_{n\vb{k}}(1),\,g_{n\vb{k}}(2),\,\dots,g_{n\vb{k}}(N_{\rm orb}) \big)^T$ and we define $k_i = \vb{k}\cdot \vb{a}_i$. In the above equations, the normalization factor of the Bloch functions is given by 
\begin{equation}
	\label{eq:norm_fact}
	c = \frac{1}{\sqrt{\varepsilon_{0}^2 + 6}}\,,
\end{equation}
where
\begin{equation}
	\label{eq:epsilon_minus}
	\varepsilon_0  = \frac{1}{2}\left(\varepsilon_{\rm h} - \sqrt{\varepsilon_{\rm h}^2+24}
	\right)
\end{equation}
is the energy of the two lowest flat bands of the dice lattice. The parameter $\varepsilon_{\rm h}$ is the on-site energy of the hub sites, while the rim sites have zero on-site energy (see Fig.~\ref{fig:dice} for the definition of hub and rim sites). The unit of energy is the positive hopping between the nearest-neighbors in the dice lattice. Due to the fact that the Bloch functions in \eqref{eq:g_1_dice} and \eqref{eq:g_2_dice} are polynomials in the coefficients $e^{\pm ik_i}$, the Wannier functions, obtained from~\eqref{eq:Bloch_to_Wannier}, are compactly localized. The Wannier function generated by $\ket{g_{n,\vb{k}}}$ is centered on hub site $\alpha = n = 1,2$ and has nonvanishing weight only on this hub site and the adjacent rim sites, as shown in Fig.~\ref{fig:dice}.

\subsection{Uniform pairing condition and exact eigenstates in the dice lattice}
\label{sec:uniform_pairing_dice_lat}

The uniform pairing condition~\eqref{eq:uniform_pairing_1}-\eqref{eq:uniform_pairing_2} is required to show that the Cooper pair operator $\hat{b}^\dg$ is a ladder operator for the projected Hamiltonian of a generic lattice~\eqref{eq:ladder_op_b_dg}. As a consequence of the compact localization of the dice lattice Wannier functions, it is shown here that the uniform pairing condition can be partially relaxed. Indeed, the commutation relation in~\eqref{eq:ladder_op_b_dg} is valid in the case of the dice lattice if the orbital-dependent Hubbard couplings  
take the following form
\begin{gather}
	U_{\rm h} = U_1 = U_2\,,  \label{eq:dice_upc_1} \\
	U_{\rm r} = U_3 = U_4 = U_5 = U_6\,. \label{eq:dice_upc_2}
\end{gather}
Here, $U_{\rm r}$ and $U_{\rm h}$ are the coupling constants of the Hubbard interaction on the rim and hub sites, respectively, and can take arbitrary values. For the dice lattice, the diagonal elements of the projector $P$ are given by
\begin{equation}
	P(\vb{0}, \alpha, \alpha) = 
	\begin{cases}
        \varepsilon_0^2c^2 & \qq*{for} \alpha = 1,\,2\,,\\
		3c^2 & \qq*{for} \alpha = 3,\,4,\,5,\,6\,.
	\end{cases}
\end{equation}
Therefore, under the conditions~\eqref{eq:dice_upc_1}-\eqref{eq:dice_upc_2}, the uniform pairing condition is satisfied for the sublattices formed by the hub and rim sites separately, but is not generally satisfied if $\alpha = 1,\,2$  and $\beta = 3,\,4,\,5,\,6$ in~\eqref{eq:uniform_pairing_1}. 

The key result that allows us to modify the derivation in~\eqref{eq:ladder_op_b_dg} and make it work for arbitrary values of the coupling constants $U_{\rm r}$ and $U_{\rm h}$ is the following
\begin{equation}\label{eq:dice_lat_special}
	\sum_{\vb{i}}\sum_{\alpha = 1,2} \bar{c}_{\vb{i}\alpha\uparrow}^\dagger\bar{c}_{\vb{i}\alpha\downarrow}^\dagger = \varepsilon_0^2c^2 \hat{b}^\dagger=  \frac{\varepsilon_0^2}{\varepsilon_0^2+6}\hat{b}^\dagger\,.
\end{equation}
Note how the sum over the orbitals is restricted to the hub sites only. The above equation is a consequence of the fact that distinct Wannier functions have zero overlap on the hub sites. From~\eqref{eq:dice_lat_special}, it is easy to show that the commutation relation $[\,\mathcal{\overline H}_{\rm int},\hat{b}^\dagger] = -E_{\rm p}\hat{b}^\dagger$  holds with a modified expression for the Cooper pair binding energy $E_{\rm p}$ for arbitrary values of $U_{\rm r}$ and $U_{\rm h}$.

Another consequence of the compact nature of the Wannier functions of the dice lattice is that it is possible to construct an even wider class of eigenstates similar to~\eqref{eq:exact_eigen_localized} with the difference being that the localized particles can have both up or down spins. This is the case since particles with opposite spin localized on non-overlapping Wannier functions do not interact. Thus, the following states are eigenstates with eigenvalue zero of the projected Hamiltonian
\begin{equation}
	\label{eq:updown_exact}
	\ket{\mathcal{I}_\uparrow, \mathcal{I}_\downarrow} = \bigg(\prod_{(n,\vb{l})\in \mathcal{I}_\uparrow}\hat{d}^\dagger_{n\vb{l}\uparrow}\bigg)\bigg(\prod_{(n',\vb{l}')\in \mathcal{I}_\downarrow}\hat{d}^\dagger_{n'\vb{l}'\downarrow}\bigg)\ket{\emptyset}\,,
\end{equation}
under the conditions that all the Wannier functions in the set $\mathcal{I}_\uparrow$ have zero overlap with the Wannier functions in the set $\mathcal{I}_\downarrow$. However, this is not true in general for arbitrary lattice models, since the Wannier functions have usually exponential tails and particle with opposite spins do interact slightly even if far apart. We can then apply the ladder operator $\hat{b}^\dg$ to the above states and obtain the following exact eigenstates of the dice lattice projected Hamiltonian
\begin{equation}
	\ket{\mathcal{I}_\uparrow, \mathcal{I}_\downarrow,N_{\rm p}} = (\hat{b}^\dagger)^{N_{\rm p}}\ket{\mathcal{I}_\uparrow, \mathcal{I}_\downarrow}\,.
\end{equation}
The energy eigenvalue of this state is given by $-E_{\rm p}N_{\rm p}$ as in~\eqref{eq:main1} and is independent from the number, position and spin of the localized particles.

\subsection{Projected Hamiltonian of the dice lattice}
\label{sec:proj_Ham_dice}

\begin{figure}
	\includegraphics[scale=0.8]{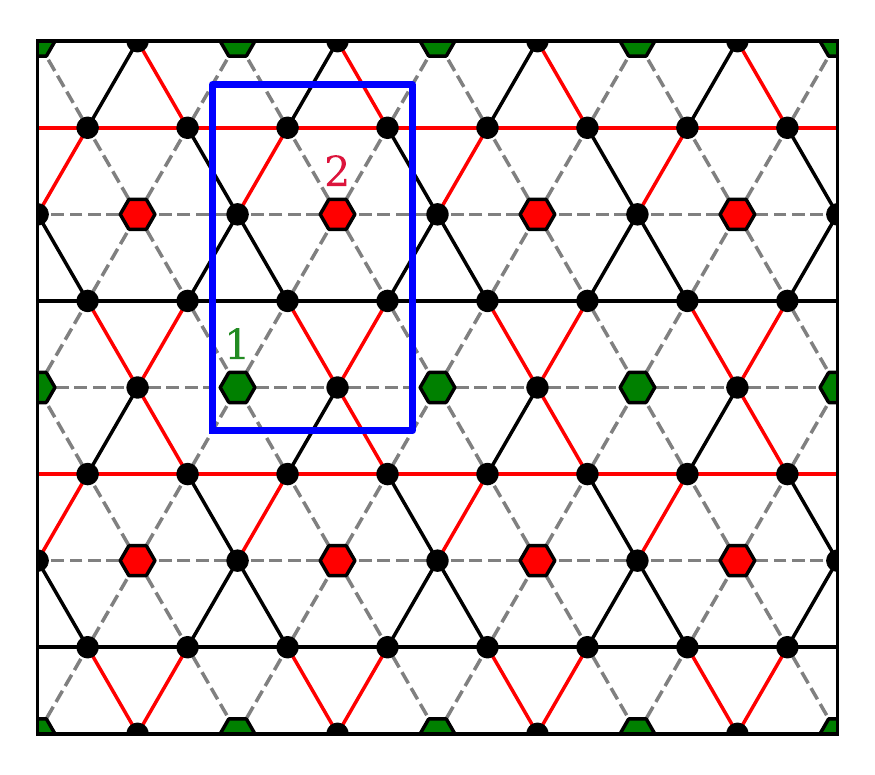}
	\caption{\label{fig:full_lattice} Graphical representation of the term $\mathcal{\hat{H}}_{\rm tri.}+\mathcal{\hat{H}}_{\rm kag.}$ in the projected Hamiltonian $\mathcal{\overline H}_{\rm int}$~\eqref{eq:dice_proj_Ham}. The Wannier functions $w_{n = 1,\vb{l}}$ and $w_{n = 2,\vb{l}}$ are represented by the green and red hexagons, respectively. One unit cell (shown as the blue rectangle) contains one green site and one red site, corresponding to the two nonequivalent Wannier functions.  The dashed lines connecting green and red sites with each other represent the nearest-neighbor interaction and hopping terms of on-site pairs and on-site spins in $\mathcal{\hat{H}}_{\rm tri.}$~\eqref{eq:H_tri}. In particular the on-site pairs hop between nearest-neighbors in the triangular lattice composed by green and red sites. The black sites sit on the middle of the bonds connecting the green and red sites and form a kagome lattice. The operator $\hat{B}^+_{\langle n_1\vb{l}_1, n_2\vb{l}_2\rangle}$ creates a singlet on the bond $\langle n_1\vb{l}_1, n_2 \vb{l}_2\rangle$, thus the bond singlets live on the kagome lattice formed by the black dots. The black and red bonds represent terms of the form $\hat{B}^+_{\langle n_1\vb{l}_1,n_2\vb{l}_2\rangle}\hat{B}_{\langle n_1\vb{l}_1,n_3\vb{l}_3\rangle}^-$ in the bond singlet hopping Hamiltonian $\mathcal{\hat{H}}_{\rm kag.}$~\eqref{eq:H_kag}. The sign of the hopping amplitude of the bonds connecting the black sites is given by $-s(n_1\vb{l}_1|n_2\vb{l}_2,n_3\vb{l}_3)$ (see Eq.~\eqref{eq:s_coeff_def}) and is equal to $-1$ for the black bonds and to $+1$ for the red bonds. The terms of the form $\hat{B}_{n_1\vb{l}_1}^+\hat{B}_{\langle n_2\vb{l}_2,n_3\vb{l}_3\rangle}^-$ in $\hat{\mathcal{H}}_{\rm tri.-kag.}$~\eqref{eq:H_tri-kag} are not represented in this figure for clarity. See Fig.~\ref{fig:transformation} in Appendix~\ref{app:two_body_problem} instead.}
\end{figure}

Here, we provide the projected Hamiltonian of the dice lattice in a form which is rather compact and convenient for subsequent considerations and computations. To this end, we introduce three different sets of operators that are linear combinations of products of two Wannier function operators $\hat{d}_{n\vb{l}\sigma}$ and $\hat{d}_{n\vb{l}\sigma}^\dg$. The first is the set of on-site spin operators, that is, the spin operators relative to a single Wannier function. The on-site spin operators of the Wannier function labeled by $n\vb{l}$ are defined as
\begin{gather}
		\hat{S}_{n\vb{l}}^z = \frac{1}{2}\qty(\hat{d}_{n\vb{l}\uparrow}^\dg\hat{d}_{n\vb{l}\uparrow} - \hat{d}_{n\vb{l}\downarrow}^\dg\hat{d}_{n\vb{l}\downarrow}) = \frac{1}{2}\qty(\hat{\rho}_{n\vb{l}\uparrow} - \hat{\rho}_{n\vb{l}\downarrow})\,,\\
	\hat{S}^+_{n\vb{l}} = \hat{d}_{n\vb{l}\uparrow}^\dg \hat{d}_{n\vb{l}\downarrow} = (\hat{S}^-_{n\vb{l}})^\dg\,,\\
	\hat{S}^x_{n\vb{l}} = \frac{1}{2}\qty\big(\hat{S}^+_{n\vb{l}}+ \hat{S}^-_{n\vb{l}})\,,\qquad \hat{S}^y_{n\vb{l}} = \frac{1}{2i}\qty\big(\hat{S}^+_{n\vb{l}}- \hat{S}^-_{n\vb{l}})\,.
\end{gather}
In the first equation above, we have introduced the spin-resolved occupation number operator $\hat{\rho}_{n\vb{l}\sigma} = \hat{d}_{n\vb{l}\sigma}^\dg\hat{d}_{n\vb{l}\sigma}$ of a Wannier function. For later use, it is convenient to also introduce the occupation number operator summed over the spin components as $\hat{\rho}_{n\vb{l}} = \hat{\rho}_{n\vb{l}\uparrow} + \hat{\rho}_{n\vb{l}\downarrow}$.

The second set is composed of operators that create and annihilate a pair of particles with opposite spins on the same Wannier function. These operators have been introduced, for instance, in Ref.~\onlinecite{Tovmasyan2018} and are defined as
\begin{gather}
	\hat{B}^z_{n\vb{l}} = \frac{1}{2}(\hat{\rho}_{n\vb{l}} -1)\,,\\
 	\hat{B}^+_{n\vb{l}} = \hat{d}^\dg_{n\vb{l}\uparrow}\hat{d}^\dg_{n\vb{l}\downarrow} = (\hat{B}^-_{n\vb{l}})^\dg \,, \\
	\hat{B}^x_{n\vb{l}} = \frac{1}{2}\qty\big(\hat{B}^+_{n\vb{l}}+ \hat{B}^-_{n\vb{l}})\,,\qquad \hat{B}^y_{n\vb{l}} = \frac{1}{2i}\qty\big(\hat{B}^+_{n\vb{l}}- \hat{B}^-_{n\vb{l}})\,.
\end{gather}
We call these operators the on-site pair operators (or on-site singlets) and it is easy to check that they obey the same $\mathrm{SU}(2)$ algebra as the on-site spin operators. Moreover, the on-site spin operators and the on-site pair operators commute with each other: $[\hat{S}^\alpha_{n\vb{l}}, \hat{B}_{n\vb{l}}^\beta] = 0$  for $\alpha, \beta = x,y,z$.

The third and final set is composed of operators of the form
\begin{equation}
\label{eq:bond_singlet_operator}
\begin{split}
&\hat{B}^+_{\langle n_1\vb{l}_1,n_2\vb{l}_2\rangle} = \hat{d}^ \dg_{n_1\vb{l}_1\uparrow} \hat{d}^\dg_{n_2\vb{l}_2\downarrow} - \hat{d}^\dg_{n_1\vb{l}_1\downarrow}\hat{d}^\dg_{n_2\vb{l}_2\uparrow} \\
&= 	
\hat{d}^ \dg_{n_1\vb{l}_1\uparrow} \hat{d}^\dg_{n_2\vb{l}_2\downarrow} + \hat{d}^\dg_{n_2\vb{l}_2\uparrow}\hat{d}^\dg_{n_1\vb{l}_1\downarrow} = \qty\big(\hat{B}^-_{\langle n_1\vb{l}_1,n_2\vb{l}_2\rangle})^\dg\,.
\end{split}
\end{equation}
The Wannier functions are centered on the hub sites of the dice lattice and thus form a triangular lattice. This triangular lattice is shown in Fig.~\ref{fig:full_lattice} as green and red sites, corresponding to the $w_{n = 1,\vb{l}}$ and $w_{n=2,\vb{l}}$ Wannier functions, respectively. In~\eqref{eq:bond_singlet_operator} we denote with $\langle n_1\vb{l}_1,n_2\vb{l}_2\rangle$ a pair of nearest-neighbor Wannier functions, that is, two Wannier functions that overlap on exactly two rim sites. The operator $\hat{B}^+_{\langle n_1\vb{l}_1,n_2\vb{l}_2\rangle}$ creates a pair of particles that are delocalized on these two Wannier functions and whose total spin (the eigenvalue of $(\hat{\vb{S}}_{n_1\vb{l}_1}+\hat{\vb{S}}_{n_2\vb{l}_2})^2$ with $\hat{\vb{S}}_{n\vb{l}} = (\hat{S}_{n\vb{l}}^x, \hat{S}_{n\vb{l}}^y, \hat{S}_{n\vb{l}}^z)^T$) is zero, i.e., they form a singlet state. Indeed, this is evident from~\eqref{eq:bond_singlet_operator}, since the state $\hat{B}^+_{\langle n_1\vb{l}_1,n_2\vb{l}_2\rangle} \ket{\emptyset}$ is symmetric under the exchange of the orbital degree of freedom $1\leftrightarrow 2$ and antisymmetric under the exchange of the spin $\uparrow\leftrightarrow \downarrow$.

Using the operators introduced above, we can now provide the projected Hamiltonian of the dice lattice. For convenience we break it down into three parts
\begin{equation}
	\label{eq:dice_proj_Ham}
	\mathcal{\overline H}_{\rm int} = \hat{\mathcal{H}}_{\rm tri.} + \hat{\mathcal{H}}_{\rm kag.} + \hat{\mathcal{H}}_{\rm tri.-kag.}\,. 
\end{equation}
The Hamiltonian $\hat{\mathcal{H}}_{\rm tri.}$ describes the hopping of the on-site pairs (singlets) on the triangular lattice formed by the Wannier function centers, namely the sublattice formed by the hub sites of the dice lattice, and the spin exchange interaction between localized particles on neighboring Wannier functions. It is given by
\begin{equation}
	\label{eq:H_tri}
	\begin{split}
	&\mathcal{\hat{H}}_{\rm tri.} = - A\sum_{n,\vb{l}} \hat{B}^+_{n\vb{l}}
	\hat{B}^-_{n\vb{l}} \\
	&-4\sum_{\langle n\vb{l}, n'\vb{l}' \rangle}\bigg[\qty(\hat{B}_{n\vb{l}}^z+\frac{1}{2})\qty(\hat{B}_{n'\vb{l}'}^z
	+\frac{1}{2}) \\
	&\qquad\qquad\qquad  + \frac{1}{2} \qty(\hat{B}_{n\vb{l}}^+ \hat{B}_{n'\vb{l}'}^- +\hat{B}_{n\vb{l}}^- \hat{B}_{n'\vb{l}'}^+)\bigg]
	\\
	&+4\sum_{\langle n\vb{l}, n'\vb{l}' \rangle}\qty[\hat{S}_{n\vb{l}}^z \hat{S}_{n'\vb{l}'}^z +\frac{1}{2} \qty(\hat{S}_{n\vb{l}}^+ \hat{S}_{n'\vb{l}'}^- +\hat{S}_{n\vb{l}}^- \hat{S}_{n'\vb{l}'}^+)]\,.
	\end{split}
\end{equation}
The first term, proportional to the parameter $A = 6 + U_{\rm h}\varepsilon^4_-/U_{\rm r}$, gives the binding energy of an on-site pair. The energy scale used for the projected Hamiltonian is $U_{\rm r}c^4 = 1$ and $A$ is the only free parameter. The terms in the second and third lines describe the nearest-neighbor interaction and the hopping of on-site pairs on the triangular lattice. Taken together, they can be written as an isotropic Heisenberg exchange term $\hat{\vb{B}}_{n\vb{l}} \cdot \hat{\vb{B}}_{n'\vb{l}'}$ for the pseudospin associated to the on-site pair operators $\hat{\vb{B}}_{n\vb{l}} = (\hat{B}^x_{n\vb{l}},\hat{B}^y_{n\vb{l}},\hat{B}^z_{n\vb{l}})^T$  with additional terms. Finally, the last line is the antiferromagnetic Heisenberg exchange interaction of on-site spins, which can be written as $\hat{\vb{S}}_{n\vb{l}} \cdot \hat{\vb{S}}_{n'\vb{l}'}$.

The Hamiltonian  $\mathcal{\hat{H}}_{\rm tri.}$ takes the same form as the projected Hamiltonian of the Creutz ladder~\cite{Tovmasyan2018} with the only difference being that the Wannier functions in the latter case  form a one-dimensional chain instead of a triangular lattice. As in the case of the Creutz ladder, $\mathcal{\hat{H}}_{\rm tri.}$ possesses an extensive number of local integrals of motions given by the operators $\hat{\vb{S}}_{n\vb{l}}^2$, which have eigenvalue $3/4$ if exactly one particle is present in the Wannier function labeled by $n\vb{l}$ and zero otherwise. This means that particles can move in the lattice only if they form on-site pairs and are localized otherwise. However, these integrals of motion do not commute with the full projected Hamiltonian due to the additional terms in $\hat{\mathcal{H}}_{\rm kag.} + \hat{\mathcal{H}}_{\rm tri.-kag.}$, as we will see below. Note that, due to the identity $\hat{\vb{S}}_{n\vb{l}}^2 + \hat{\vb{B}}_{n\vb{l}}^2 = \frac{3}{4}\hat{1}$, the operators $\hat{\vb{B}}_{n\vb{l}}^2$ do not form a set of independently conserved quantities.

The second term $\mathcal{\hat{H}}_{\rm kag.}$ in the projected Hamiltonian describes the hopping of bond singlets and takes the form
\begin{equation}
\label{eq:H_kag}
\begin{split}
&\hat{\mathcal{H}}_{\rm kag.} = -\sum_{\langle 1,2,3\rangle}\bigg[s(1|2,3)\qty(\hat{B}_{\langle 1,2\rangle}^+\hat{B}_{\langle 1,3\rangle}^-+\mathrm{H.c.}) \\ &\hspace{1cm} + \text{cyclic permutations of } (1, 2, 3)\bigg]\,.	
\end{split}
\end{equation}
Here, we have used the abbreviations $1 \equiv n_1\vb{l}_1$, $2 \equiv n_2\vb{l}_2$ and $3 \equiv n_3\vb{l}_3$ and we denote by $\langle n_1\vb{l}_1, n_2\vb{l}_2,n_3\vb{l}_3 \rangle$
a triplet of Wannier functions that are two by two nearest-neighbors. Therefore their centers form an equilateral triangle whose sides are given by the three vectors $\vb{v}_{i = 1,2,3}$ in~\eqref{eq:fund_vec_v}-\eqref{eq:fund_vec_v3}. The sum $\sum_{\langle 1,2,3\rangle}$ runs over all such triangles in the triangular lattice. The quantity $s( n_1\vb{l}_1| n_2\vb{l}_2, n_3\vb{l}_3) = \pm 1$ depends on the overlap of the Wannier functions forming a triangle of nearest-neighbors and is defined as
\begin{equation}
\label{eq:s_coeff_def}
	\begin{split}
	&U_{\rm r}c^4s(n_1\vb{l}_1|n_2\vb{l}_2,n_3\vb{l}_3)\\
	&=\sum_{\vb{i},\alpha}U_\alpha
	w^2_{n_1}(\vb{i}-\vb{l}_1,\alpha)
	w_{n_2}(\vb{i}-\vb{l}_2,\alpha)
	w_{n_3}(\vb{i}-\vb{l}_3,\alpha)\,.
	\end{split}
\end{equation}
Note that we have used the fact that the Wannier functions of the dice lattice are real.
The quantity $s( n_1\vb{l}_1| n_2\vb{l}_2, n_3\vb{l}_3)$ is visualized in Fig.~\ref{fig:full_lattice} as the color of the bonds connecting the sites in a kagome lattice, which is the lattice on which the bond singlets live. Due to the signs of the bond singlet hoppings, the Hamiltonian $\hat{\mathcal{H}}_{\rm kag.}$ does not have the symmetry of the triangular lattice as in the case of $\hat{\mathcal{H}}_{\rm tri.}$, but instead has the same translational symmetry of the original dice lattice given by the fundamental vectors $\vb{a}_{i = 1,2}$~\eqref{eq:fund_vec_rect}.

The last term in~\eqref{eq:dice_proj_Ham} describes the processes by which on-site pairs are converted into bond singlets and vice-versa,
\begin{equation}
	\label{eq:H_tri-kag}
	\begin{split}
		&\hat{\mathcal{H}}_{\rm tri.-kag.} = -\sum_{\langle 1,2,3\rangle}\bigg[s(1|2,3)\qty(\hat{B}_{1}^+\hat{B}_{\langle 2,3\rangle}^-+\mathrm{H.c.}) \\ &\hspace{3cm} + \text{cyc. perm. of } (1, 2, 3)\bigg]\,.	
	\end{split}
\end{equation}
Due to the sign factor $s(1|2,3)$, the translational symmetry of this term is the same as $\hat{\mathcal{H}}_{\rm kag.}$. By allowing the motions of bond singlets, in which two neighboring Wannier functions are occupied each by a single particle, and the conversion of on-site pairs to bond singlets, the projected Hamiltonian of the dice lattice does not possess obvious local integrals of motion as in the case of $\mathcal{\hat{H}}_{\rm tri.}$ discussed above. Our main goal is to investigate whether some sort of conserved quantities associated to localized particles are nevertheless present in the projected Hamiltonian. 

An important first step is the solution of the two-particle problem, which is worked out in Appendix~\ref{app:two_body_problem}. For three or more particles, one has to resort to numerical methods. In the next section, we analyse the energy spectrum and the nonequilibrium dynamics of the projected Hamiltonian in the case of three and four particles.

\section{Exact diagonalization results for the dice lattice}
\label{sec:numerical_results}

In some previous works,~\cite{Tovmasyan2018} it has been established that unpaired particles are always localized due to the presence of local conserved quantities in specific one-dimensional models. These are the same conserved quantities of the Hamiltonian term $\mathcal{\hat{H}}_{\rm tri.}$~\eqref{eq:H_tri}, as discussed in Section~\ref{sec:proj_Ham_dice}.  However, they are spoiled by the additional terms $\mathcal{\hat{H}}_{\rm kag.}+\mathcal{\hat{H}}_{\rm tri.-kag.}$ associated with the bond singlets, which are not present in the one-dimensional case. Therefore, in the case of the dice lattice, it is not clear if local conserved quantities exist.
In this section, we address this important question using exact diagonalization. 

In Section~\ref{sec:energy_spectrum}, we provide numerical results for the energy spectrum of the projected Hamiltonian of the dice lattice in case of an imbalance between spin up and down particles ($N_\uparrow \neq N_\downarrow$). It is shown that the excited states form well-separated and almost degenerate groups of states. This is similar to, but at the same time different from, the perfect degeneracy of the exact eigenstates derived analytically in Section~\ref{sec:exact_localized}, which are found to form the ground state manifold in the case of spin imbalance. It is argued that the lack of degeneracy in the excited states is a finite-size effect.

In Section~\ref{sec:dynamics}, we focus on the time evolution from an initial state that includes both on-site pairs and single unpaired particles. It is shown that, in the long time limit, memory of the initial positions of the unpaired particles persists in both the particle density and the spin density. This is a clear signature of nonergodic behavior and many-body localization, which is a consequence of the presence of the local conserved quantities in many-body quantum systems.

All of the numerical results shown here have been obtained using the exact diagonalization package QuSpin.~\cite{Weinberg2017,Weinberg2019} Exact diagonalization is performed on a rectangular cluster with periodic boundary conditions composed of  $N_{\rm c} = N_1N_2$ unit cells, 
where $N_1$ and $N_2$ are the numbers of unit cells in the horizontal and vertical directions, respectively. For instance, a finite cluster with $N_1 = 4$, $N_2=2$ is shown in Fig.~\ref{fig:full_lattice}.  

\begin{figure*}
    \centering\includegraphics[width = \textwidth]{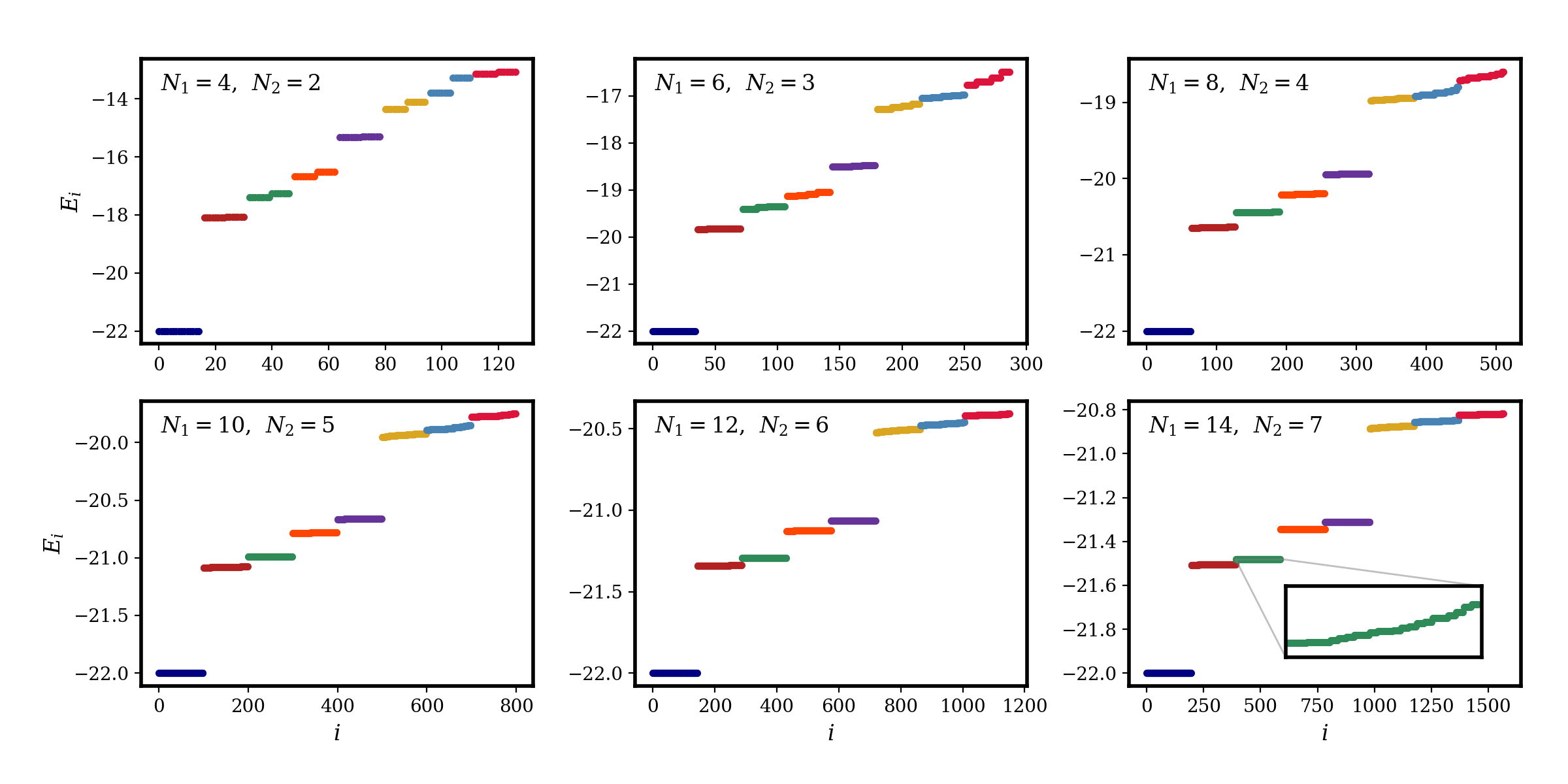}
    \caption{ \label{fig:spectrum_2up_1down} Lowest energy eigenvalues of the projected dice lattice Hamiltonian $\mathcal{\overline H}_{\rm int}$~\eqref{eq:dice_proj_Ham} obtained from exact diagonalization. Different systems sizes are shown, while the number of particles is fixed to two particles with spin up ($N_\uparrow=2$) and one particle with spin down ($N_\downarrow=1$). The system size is determined by a pair of integers, the number of unit cells in the horizontal $N_1$ and vertical $N_2$ directions, respectively. Only the lowest $16N_{\rm c}$  eigenvalues, with $N_{\rm c} = N_1N_2$ the number of unit cells, are shown in each panel. The lowest $2N_{\rm c}$ eigenstates are perfectly degenerate and have energy $E_0 = -E_{\rm p} = -22$ (the only free parameter in the projected Hamiltonian has been fixed to $A=10$). These are the exact eigenstates with localized quasiparticles presented in Section~\ref{sec:exact_localized}, see~\eqref{eq:exact_eigen_localized}. In this specific case, they take the form $\ket{N_{\rm p}=1,\mathcal{I}_\uparrow=\{(n,\vb{l})\}} = \hat{b}^\dagger\hat{d}^\dagger_{n\vb{l}\uparrow}\ket{\emptyset}$. On the other hand, the excited states can be grouped in sets of states with approximately the same energy, which have been denoted with different colors. The number of states in these sets of quasidegenerate states is the same as the degeneracy of the ground state. By increasing the system size, the deviation from perfect degeneracy in each set seems to decrease. However, even for the largest system sizes there is no perfect degeneracy in each set of excited states, as shown in the inset in the lower right panel.}
\end{figure*}

\begin{figure*}
    \centering \includegraphics[width = \textwidth]{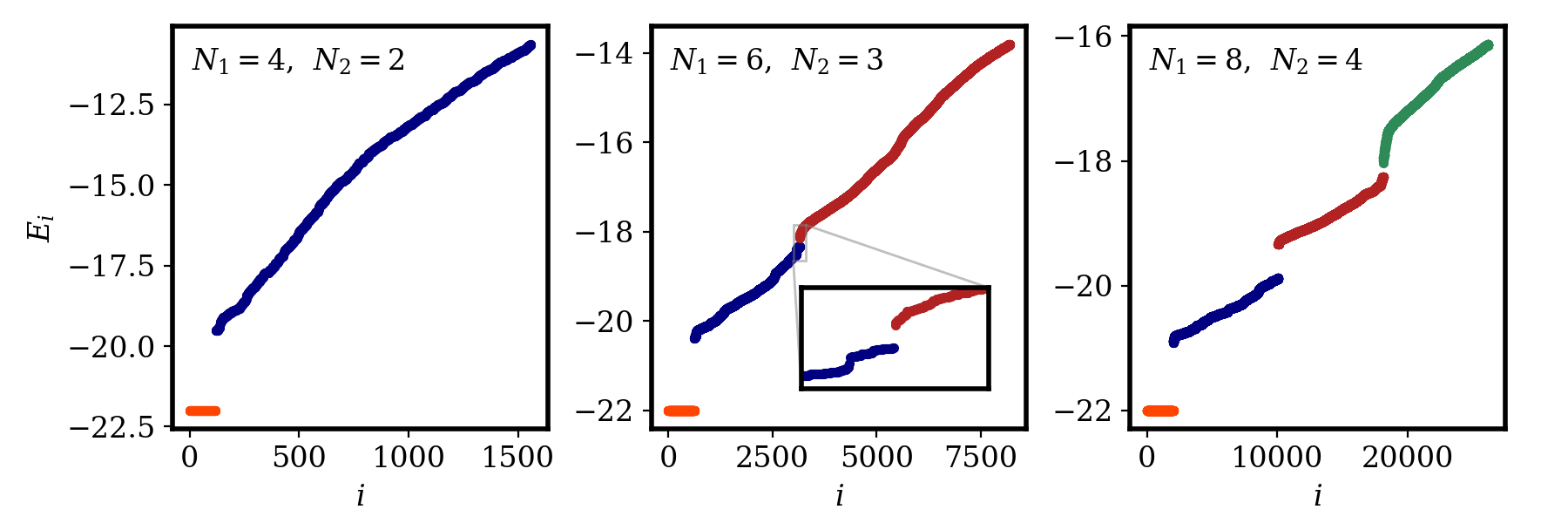}
    \caption{\label{fig:spectrum_3up_1down} Same as in Fig.~\ref{fig:spectrum_2up_1down} in the case of three particles with spin up $N_\uparrow = 3$ and one particle with spin down $N_\downarrow = 1$. Three different system sizes are shown: $N_c = N_1 N_2 = 8, 18, 32$. The lowest $13 \binom{2 N_c}{2}$ eigenvalues are shown in each panel. The lowest $\binom{2 N_c}{2}$ eigenstates are perfectly degenerate and have an energy $E_0 = -E_{\rm p} = -22$  for $A = 10$. These are the exact eigenstates in~\eqref{eq:exact_eigen_localized}, which for the given number of particles take the form $\ket{N_{\rm p}=1,\mathcal{I}_\uparrow=\{(n_1, \vb{l}_1) ,(n_2, \vb{l}_2) \} } = \hat{b}^\dagger\hat{d}^\dagger_{n_1\vb{l}_1 \uparrow} \hat{d}^\dagger_{n_2\vb{l}_2 \uparrow} \ket{\emptyset}$. On the other hand, the excited states can be grouped in sets of states separated by energy gaps, which have been denoted with different colors. Both the number of well-separated sets of excited states and the energy gaps increase with the system size. One such set is visible for $N_1 = 6, N_2 =3$, which is colored in blue. Two such sets are visible for $N_1 = 8, N_2 =4$.  The number of states in these sets is exactly four times larger than the degeneracy of the ground state.}
\end{figure*}

\subsection{Energy spectrum}
\label{sec:energy_spectrum}

In Fig.~\ref{fig:spectrum_2up_1down}, the lowest eigenvalues of the projected dice lattice Hamiltonian with three particles ($N_\uparrow = 2$, $N_\downarrow = 1$) are shown for various system sizes. The lowest $\binom{2N_{\rm c}}{N_\uparrow -N_\downarrow} = 2N_{\rm c}$ eigenstates are exactly degenerate and take the form
\begin{equation}
\label{eq:exact_localized_1}
    \ket{N_{\rm p}=1,\mathcal{I}_\uparrow=\{(n,\vb{l})\}} = \hat{b}^\dagger\hat{d}^\dagger_{n\vb{l}\uparrow}\ket{\emptyset}\,.
\end{equation}
These are precisely the exact eigenstates with localized quasiparticles presented in Sec.~\ref{sec:exact_localized}, see~\eqref{eq:exact_eigen_localized}. In the balanced case ($N_\uparrow = N_\downarrow$),  it is known~\cite{Tovmasyan2016,Herzog-Arbeitman2022} that the states of the form~\eqref{eq:ket_pair} have minimal energy and thus are ground states of the projected Hamiltonian for given number of particles. No analogous statement exists for the imbalanced case, nevertheless one can reasonably expect states of the form~\eqref{eq:exact_eigen_localized} and~\eqref{eq:exact_localized_1} to have minimal energy. Our numerical simulations show that this is indeed the case, at least for some values of the parameter $A$ in the projected Hamiltonian~\eqref{eq:dice_proj_Ham}.

In Fig.~\ref{fig:spectrum_2up_1down}, the most interesting information is contained in the excited state spectrum for which there are no available analytical results. The excited states can be grouped into sets of states with approximately the same energy. Distinct sets are indicated with different colors in Fig.~\ref{fig:spectrum_2up_1down}. The number of states in these quasidegenerate sets is exactly equal to the degeneracy of the ground state. On increasing the system size, the deviation from perfect degeneracy in each set of states seems to decrease. However, it is noted that even for the largest computed system size, there is no perfect degeneracy in each set of excited states. For example, for system size $N_1 =14, N_2 = 7$ the energy width of one of these set of states is roughly $2 \times 10^{-3}$ (see inset in Fig.~\ref{fig:spectrum_2up_1down}). 

This peculiar structure in the excited state spectrum is not entirely surprising and can be understood by comparing with what one would obtain in the case of the one-dimensional models in which local integrals of motion are known to exist.~\cite{Tovmasyan2018} In these models, the eigenvalues $s_{n\vb{l}}(s_{n\vb{l}}+1)$ of the operators $\hat{\vb{S}}_{n\vb{l}}^2$, that is the total spin on the Wannier function labeled by $n\vb{l}$, are good quantum numbers. The lowest lying states have $s_{n\vb{l}} = 1/2$ only for a single Wannier function and $s_{n\vb{l}} = 0$ otherwise since there is only a single unpaired particle. For any given state of this form, one can obtain a distinct eigenstate with the same energy by applying a translation, due to the translational invariance of the projected Hamiltonian. The new eigenstate is distinct since the localized unpaired particle present in the original state has been moved to a different Wannier function ($s_{n\vb{l}} =1/2\to s_{n,\vb{l}+\vb{j}} =1/2$ for a translation by $\vb{j}$ unit cells). Thus the lowest lying states have degeneracy which is at least equal to the number of unit cells $N_{\rm c}$.

The numerical results shown in Fig.~\ref{fig:spectrum_2up_1down} suggest that local integrals of motion are present in some approximate sense for the projected Hamiltonian of the dice lattice. The mechanism behind the lifting of the degeneracy is at the present not well understood, except for the fact that it is caused by the motion of the bond singlets and their interaction with localized single particles.
As mentioned above, perfect degeneracy is approached in each set of excited states on increasing the system size. Thus, it might by the case that the projected Hamiltonian of the dice lattice possesses exact local  integrals of motion only in the limit of infinite system size. We conjecture that these encode the localization of quasiparticles in wave functions that extend over rather large distances. In fact their range, or localization length, should be comparable to the largest size shown in Fig.~\ref{fig:spectrum_2up_1down} in order to explain the finite size effects. On the other hand, in the one-dimensional models of Ref.~\onlinecite{Tovmasyan2018}, the quasiparticle wave functions are compactly localized since they coincide with the Wannier functions.

It is interesting to explore what happens with increasing imbalance. In Fig.~\ref{fig:spectrum_3up_1down}, for instance,  we show the energy spectrum for $N_\uparrow -N_\downarrow = 2$. The ground state manifold is composed of states of the form 
\begin{equation}
    \ket{N_{\rm p}=1,\mathcal{I}_\uparrow=\{(n_1, \vb{l}_1) ,(n_2, \vb{l}_2) \} } = \hat{b}^\dagger\hat{d}^\dagger_{n_1\vb{l}_1 \uparrow} \hat{d}^\dagger_{n_2\vb{l}_2 \uparrow} \ket{\emptyset}\,,
\end{equation}
and has degeneracy $\binom{2N_{\rm c}}{N_\uparrow-N_\downarrow} = N_{\rm c}(2N_{\rm c}-1)$.
In the spectra shown in Fig.~\ref{fig:spectrum_3up_1down}, the excited states are separated from the ground state manifold by a large energy gap. 
With increasing system size, sets of excited states well-separated from each other by energy gaps start to appear. Both their number and the energy gaps that separates them increase with the system size. In the figure, one such set is visible for $N_1 = 6, N_2 = 3$, but two such sets are evident for $N_1 = 8, N_2 = 4$. In each case, the number of eigenstates in these sets is exactly four times larger than the degeneracy of the ground state manifold. As in the three particle case, the energy width of the well-separated sets of states reduces with increasing  system size and it is possible that they approach perfect degeneracy for an infinitely large system. Alternatively, they could break up in several degenerate subsets.

In summary, we observe for $N_\uparrow -N_\downarrow \neq  0$ that the excited states form almost degenerate multiplets with a degeneracy that is an integer multiple of the ground state degeneracy.  We attribute the deviation from perfect degeneracy to finite-size effects that lead to the breaking of the local integrals of motion present in an infinite lattice. Some rather strong finite-size effects also appear in the long time dynamics, as shown in the next section.

\begin{figure*}
    \centering
     \includegraphics[scale =0.85]{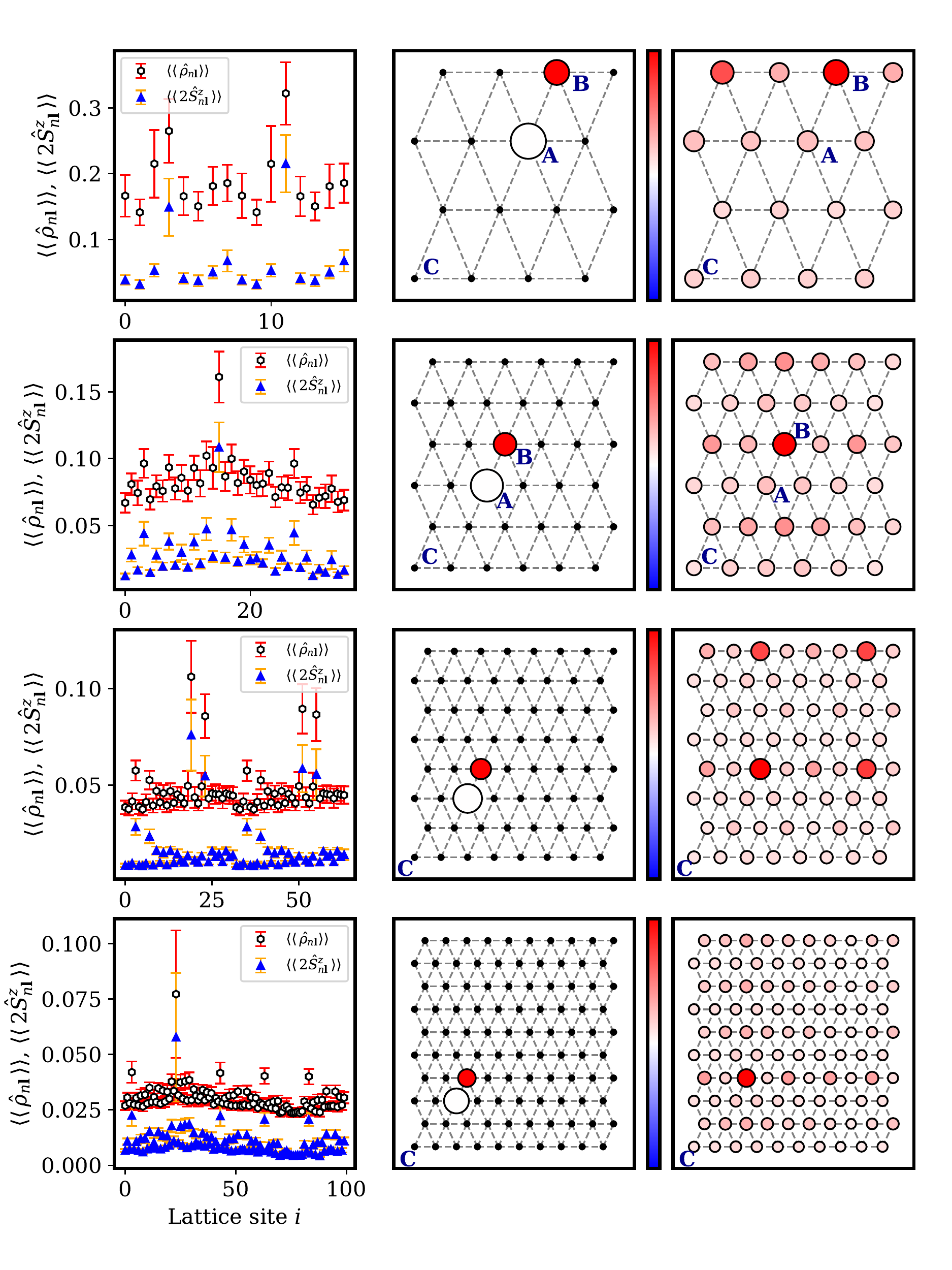}
    \caption{Time evolution of the projected dice lattice Hamiltonian for three particles. Different systems sizes are shown in each row (from top to bottom:  $N_1 \times N_2= 4 \times 2$, $6 \times 3$, $8 \times 4$ and $10 \times 5$), while the number of particles is fixed to $N_\uparrow=2$ $N_\downarrow=1$. The first panel shows the time averaged particle and spin densities at each site averaged from time $t_0$ to $t_1 = 10000$ according to \eqref{eq:time-average_observable}. $t_0 = 1000$ is selected for the first three system sizes while $t_0 = 3000$ is chosen for $N_1 \times N_2 = 10 \times 5$. The error bars indicate the standard deviation of each quantity over the same time interval, as given by \eqref{eq:stddev}. The panels in the second column show the initial state of the system $\ket{\Psi(0)} = \hat{d}^\dagger_{n_A\vb{l}_A\uparrow} \hat{d}^\dagger_{n_A\vb{l}_A\downarrow} \hat{d}^\dagger_{n_B\vb{l}_B\uparrow} \ket{\emptyset}$. The diameter of the circle represents the relative particle density at a site and the spin density is represented by a color scale, with red being up spin and blue being down spin. The panels in the third column depict the time averaged particle and spin densities on the lattice. The time averaged particle and spin densities are normalized to one at site $B$. The spin density is nonzero across all sites of the lattice, but remains largest at site $B$, where the particle with spin up was initially located.} 
    \label{fig:densities_2up_1down}
\end{figure*}

\begin{figure}
    \centering
    \includegraphics[width = \columnwidth]{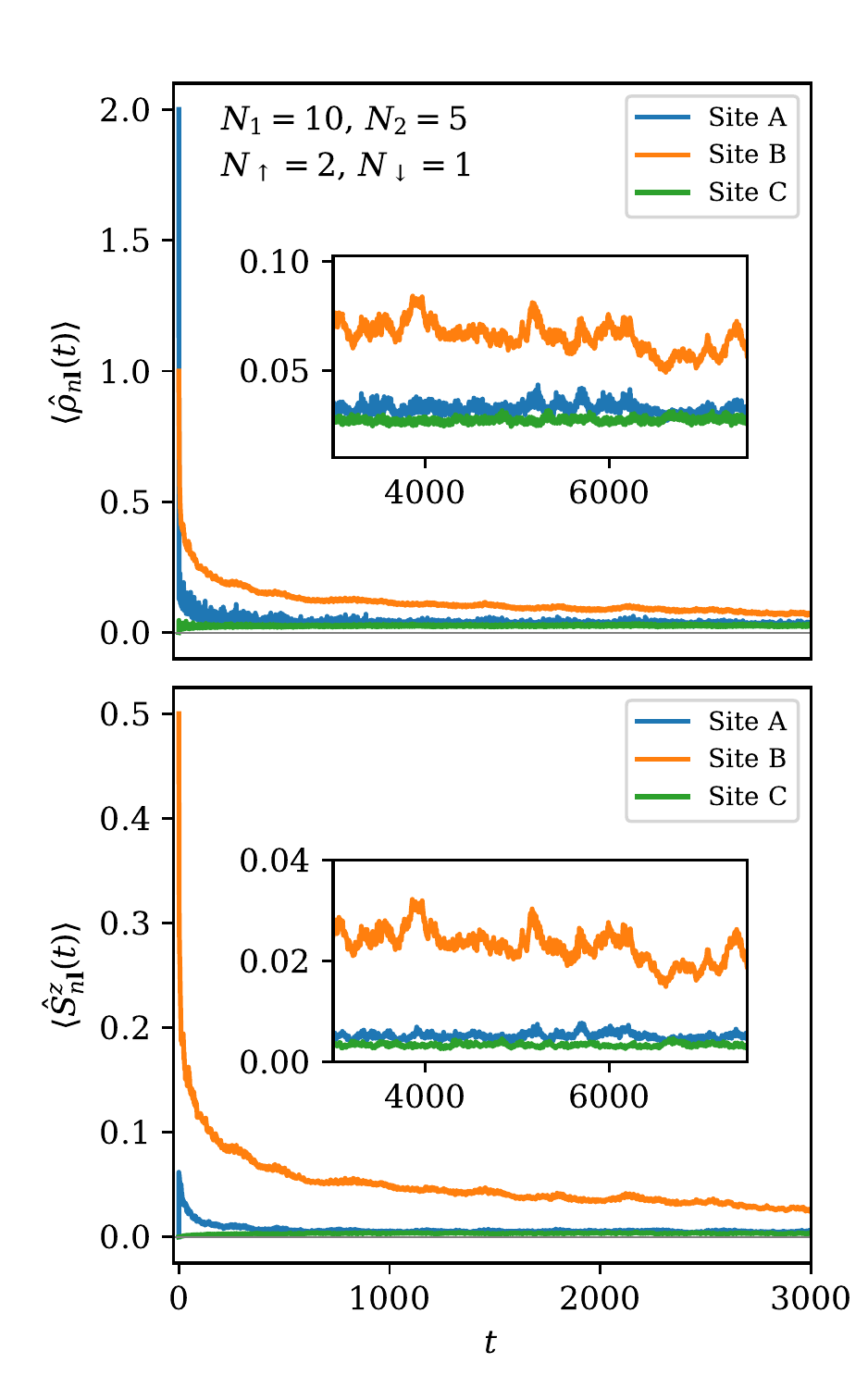}
    \caption{Time evolution  with three particles for system size $N_1=10, \,N_2=5$. The initial state is shown in the lowest row of Fig.~\ref{fig:densities_2up_1down}. The upper panel shows the particle density from $t=0$ until $t_0 = 3000$ for three sites denoted as $A, B$ and $C$ in Fig.~\ref{fig:densities_2up_1down}. The inset shows the rolling average of the particle density in the same sites from $t_0 = 3000$ until $t = 7500$. The lower panel depicts the spin densities of sites $A, B$ and $C$ from  $t=0$ until $t_0 = 3000$, with the inset depicting the rolling average of the spin densities from $t_0 = 3000$ until $t = 7500$. The rolling average in the insets are taken with a time window of $\Delta t = 3.0$, to smooth out fluctuations. The rolling averages are used for the sake of clarity in presentation.
    }
    \label{fig:time-evol_2up_1down}
\end{figure}

\subsection{Nonequilibrium dynamics}
\label{sec:dynamics}

The time evolution of the particle and spin densities obtained by evolving a three-particle initial state of the form
\begin{equation}\label{eq:initial_state}
   \ket{\Psi(0)} = \hat{d}^\dagger_{n_A\vb{l}_A\uparrow} \hat{d}^\dagger_{n_A\vb{l}_A\downarrow} \hat{d}^\dagger_{n_B\vb{l}_B\uparrow} \ket{\emptyset},
\end{equation} 
with the projected Hamiltonian $\mathcal{\overline H}_{\rm int}$ is shown in Figs.~\ref{fig:densities_2up_1down},~\ref{fig:time-evol_2up_1down}.
This initial state can be equivalently described as either an on-site singlet present at site $A$ with an additional spin-up particle at site $B$, or as a bond singlet between sites $A$ and $B$ plus a spin-up particle at site $A$, namely
\begin{equation} 
\label{eq:initial_state_eqv}
\begin{split}
\ket{\Psi(0)} &= \hat{B}^+_{n_A\vb{l}_A} \hat{d}^\dagger_{n_B\vb{l}_B\uparrow}\ket{\emptyset} \\ &= - \hat {B}^+_{\langle n_A\vb{l}_A, n_B\vb{l}_B\rangle}\hat{d}^\dagger_{n_A\vb{l}_A\uparrow}\ket{\emptyset}\,.
\end{split}
\end{equation}
These representations give us insight into the expected behavior of the time-evolved system. During the time evolution, the bond singlet can move away from sites $A$ and $B$, leaving behind the particle with spin up at site $A$. Thus, the spin density at site $A$ becomes nonzero (in the initial state the spin density is nonzero only on site $B$). This differs from when only two particles are present in the system, in which case the spin density remains at zero at all sites where it is initialized to zero. Therefore, no spin transport occurs with two particles, whereas some form of spin transport takes place with three particles, as seen in Fig.~\ref{fig:densities_2up_1down}.

In Fig.~\ref{fig:densities_2up_1down}, the time averaged particle and spin densities are shown for different system sizes. Here, the time average of an observable (computed numerically) is defined as
\begin{equation} \label{eq:time-average_observable}
    \langle \langle \hat{O} \rangle \rangle = \frac{1}{t_1 -t_0} \int_{t_0}^{t_1} \langle \hat{O}(t) \rangle dt \, ,
\end{equation}
where $\langle \hat{O}(t) \rangle $ is the expectation value of an observable $\hat{O}$ at time $t$, $t_0$ is an initial cutoff time, which is sufficiently large to ensure that the observable is close to its long-time asymptotic value and $t_1$ is the end time. In the first column of Fig.~\ref{fig:densities_2up_1down}, the standard deviation of the time average of an observable is also shown, and is defined as
\begin{equation} \label{eq:stddev}
    \sigma_{\hat{O}} = \sqrt{\langle {\langle \hat{O}(t) \rangle}^2 \rangle - {\langle \langle \hat{O} \rangle \rangle}^2}.
\end{equation}
In Fig.~\ref{fig:time-evol_2up_1down}, it can be seen that, for a cluster of size $N_1=10$ $N_2=5$, the particle and spin densities approach
their respective  equilibrium value at around $t=3000$, which is chosen as initial cutoff time $t_0$. The time scale to reach equilibrium is extremely long and increases with the system size. 
For instance, we take $t_0 = 1000$ for the other system sizes shown in Fig.~\ref{fig:densities_2up_1down}. 

From Fig.~\ref{fig:densities_2up_1down}, it can be observed that the particle and spin densities diffuse throughout the system and becomes nonzero on all lattice sites. However, at site $B$ where the particle with spin up is initially placed, the particle and spin densities are significantly larger than those at the other sites in the lattice, including site $A$, the initial position of the on-site pair. This behaviour persists as the system size is increased, as seen in Fig.~\ref{fig:densities_2up_1down}. As expected, with an increase in system size, the fluctuations of the particle and spin densities in all lattice sites also decrease, with the exception of site $B$. A surprising result for system size $N_1 = 8, \, N_2 = 4$ is the appearance of  three other sites with particle and spin densities comparable to the ones of site $B$. These three sites are obtained by translating site $B$  by linear combinations of the lattice vectors $\vb{R}_1/2$ and $\vb{R}_2/2$, with $\vb{R}_1 = N_1\vb{a}_1$ and $\vb{R}_2 = N_2\vb{a}_2$ the vectors defining the size and shape of the finite cluster with periodic boundary conditions. A similar behavior is observed also for system size $N_1 = 4$, $N_2 = 2$, but not in the case of the other system sizes shown in Fig.~\ref{fig:densities_2up_1down}. Thus, the long-time dynamics appears to be very sensitive to finite size effects. The origin of this phenomenon is unclear at present and deserves further investigations.

\begin{figure*} \includegraphics[width = \textwidth]{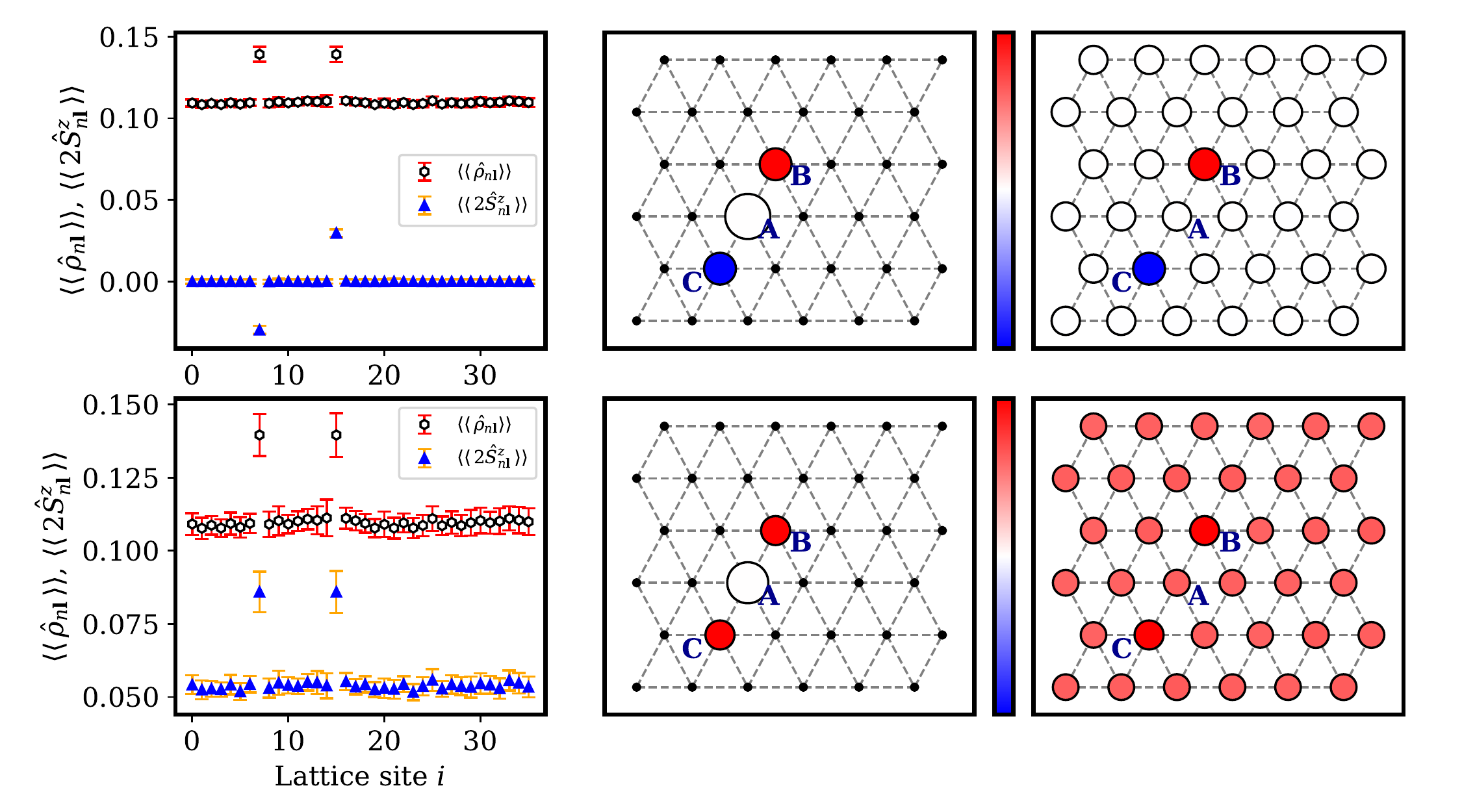}
    \caption{\label{fig:four_particle_densities}Same as in Fig.~\ref{fig:densities_2up_1down} but for four particles $N_\uparrow + N_\downarrow = 4$ and system size $N_1 =6, N_2 =3$. The time interval for the time averaged particle and spin densities is taken to be between $t_0 = 100$ and $t_1 = 5000$. The initial state for the top row is $\ket{\Psi(0)} = \hat{d}^\dagger_{n_A\vb{l}_A\uparrow} \hat{d}^\dagger_{n_A\vb{l}_A\downarrow} \hat{d}^\dagger_{n_B\vb{l}_B\uparrow} \hat{d}^\dagger_{n_C\vb{l}_C\downarrow} \ket{\emptyset}$ and includes two unpaired particles with opposite spin in sites $B$ and $C$. The initial state for the bottom row is $\ket{\Psi(0)} = \hat{d}^\dagger_{n_A\vb{l}_A\uparrow} \hat{d}^\dagger_{n_A\vb{l}_A\downarrow} \hat{d}^\dagger_{n_B\vb{l}_B\uparrow} \hat{d}^\dagger_{n_C\vb{l}_C\uparrow} \ket{\emptyset}$ in which the unpaired particles have both spin up. As in Fig.~\ref{fig:densities_2up_1down}, the particle and spin densities retain memory of the initial positions of the unpaired particles (site $B$ and site $C$) in the long time limit. On the other hand, the particle and spin densities become almost uniform throughout the remaining sites. A difference compared to the three particle case is that the time scale required to reach the long time limit is about ten times smaller and is given approximately by $t_0$.}
\end{figure*}

Analogous results are obtained in the case of four particles $N_\uparrow+N_\downarrow = 4$. The two unpaired particles in the initial state can have either the opposite or same spin. Numerical results for both initial states are shown in Fig.~\ref{fig:four_particle_densities}. Again, memory of the initial positions of the two unpaired particles persists in the particle and spin densities, which are otherwise approximately uniform over all remaining sites in the long time limit.

The results of Figs.~\ref{fig:densities_2up_1down}-\ref{fig:four_particle_densities} indicate a peculiar coexistence of ergodic and nonergodic behavior in the same system. On one hand, we have that in the long time limit the spin density distribution is highly nonuniform, in particular it peaks at site $B$. Thus, memory of the initial position of the particle with spin up is retained during the time evolution. If the time evolution were to be completely ergodic, we would expect the long-time average of the system to coincide with the average value given by the thermal equilibrium state that is the usual Gibbs ensemble. This result is known as the ergodic theorem and is the fundamental postulate of statistical mechanics~\cite{moore2015ergodic}. The equation for the long-time average of an observable stated in~\eqref{eq:time-average_observable} is its usual definition in the context of the ergodic theorem. The results of Figs.~\ref{fig:densities_2up_1down}-\ref{fig:four_particle_densities} do not coincide with the expected thermal equilibrium average, which for a translationally invariant system as studied here, would see the particle density and spin density be uniformly distributed. We can conclude from this that the states do not thermalize and the time evolution displays non-ergodic behaviour, which we interpret as a signature of many-body localization~\cite{abanin2019colloquium, Alet2018, altman2018many}. This characteristic signature of many-body localization has been observed in ultracold gas experiments~\cite{Kondov2015,Schreiber2015,Choi2016}.
On the other hand, any information regarding the initial position (site $A$) of the on-site pair is completely lost after a sufficiently long time. Therefore, the nonergodic behavior is limited to fermionic quasiparticles, while two-body bound states behave ergodically, as one would expect.

Our results also suggest a violation of the eigenstate thermalization hypothesis (ETH), which asserts that the thermalization of all initial states of a system upon time evolution implies that all many-body eigenstates are thermal as well~\cite{nandkishore2015many}. The initial state~\eqref{eq:initial_state} used to study the nonequilibrium dynamics is a linear combination of highly excited eigenstates of the system, and the absence of thermalization of the initial state indicates that at least some of the many-body eigenstates are non-thermal as well. The violation of ETH is commonly associated with many-body localization~\cite{abanin2019colloquium, Alet2018}.


Finally, it is interesting to note that the time scale for the equilibration of the particle and spin densities, which is very long as shown in Fig.~\ref{fig:time-evol_2up_1down}, seems to be related to the energy scale associated to the lifting of the degeneracy in the spectra shown in Fig.~\ref{fig:spectrum_2up_1down} and discussed in Section~\ref{sec:energy_spectrum}. The time scale is approximately given by the initial cutoff $t_0$ used for the time average~\eqref{eq:time-average_observable}. Indeed, we take $t_0 = 1000$ for 
$N_1 =8$, $N_2 = 4$ and $t_0 = 3000$ for 
$N_1 =10$, $N_2 = 5$.
Correspondingly, the 
energy width of the quasidegenerate set of states is roughly three times larger for the smaller system size. For instance, in the case of the sets colored in green in Fig.~\ref{fig:spectrum_2up_1down}, the energy widths are $\Delta E = 4.8 \times 10^{-3}$ and $\Delta E=1.9 \times 10^{-3}$, respectively. The orders or magnitude of $t_0$ and $1/\Delta E$ are also in good agreement.
Thus, a possible explanation is that the long equilibration times are a manifestation of the local integrals of motion, which are slightly broken by finite size effects.

\subsection{Finite size scaling}
\label{sec:finite_size}

In Sec~\ref{sec:dynamics}, we calculate the time evolution of the particle density and spin density over long times and observe signatures of many-body localization. However, the phenomena of many-body localization has been debated to be simply a finite size effect~\cite{Sels2023} and that at infinite system size, many-body localization would cease to exist. To understand the impact of system size, and as a consequence, finite-size effects on the signatures observed in our model, we perform a scaling analysis. 

\begin{figure}[h]
    \centering 
    \includegraphics[scale = 0.55]{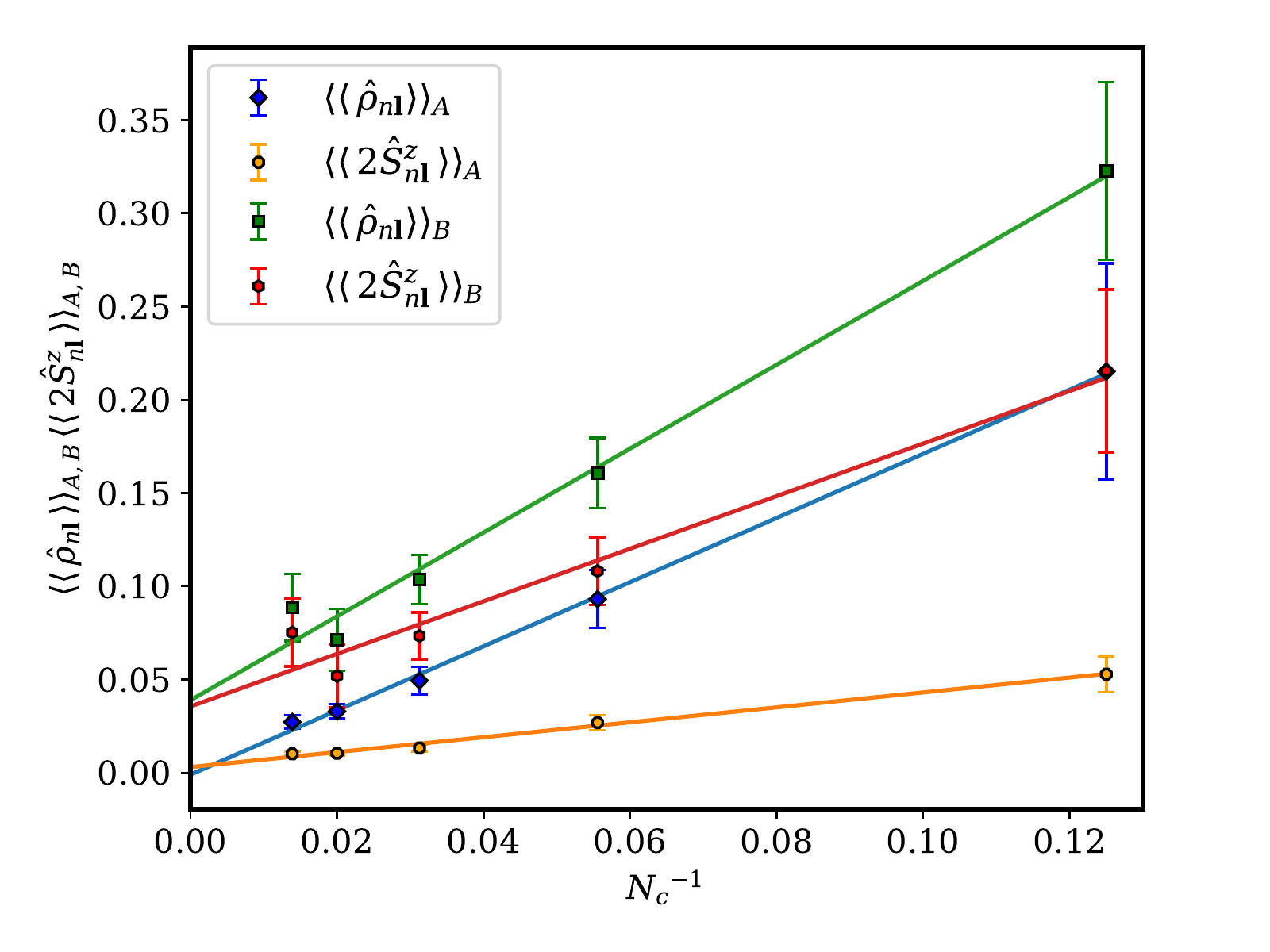}
    \caption{\label{fig:finite_densities} Finite size scaling of the long time average of the particle and spin densities. The long time average of the  particle and spin densities on sites $A$ and $B$ are shown as a function of the inverse number of unit cells $N_{\rm c}^{-1}$ in the finite cluster. The numerical data shown here are the same as in Fig.~\ref{fig:densities_2up_1down} and are obtained from the time evolution of the projected Hamiltonian with three particles ($N_\uparrow= 2$, $N_\downarrow = 1$).} 
\end{figure}

In Fig.~\ref{fig:finite_densities}, we can see the how the asymptotic value of the particle densities and spin densities scale with system size. By observing the trends of these quantities as they approach an infinite system size ($N_c^{-1} \rightarrow 0$), we can estimate the impact of the finite size effects. We can see that the particle and spin densities at site $A$, where the on-site pair was placed, approaches zero as the lattice size increases. At site $B$, where there is initially a single unpaired spin, the particle and spin densities approach a finite nonzero value as the system size increases. The long-time average of the observables reveal a continued partial breaking of ergodicity. The asymptotic distribution of the spin density is not uniform and peaks on the site in which the unpaired particle was initially located. This is in contrast with the distribution of the spin density at thermal equilibrium, which is uniform as a consequence of translational invariance. The finite-size scaling results indicate that the partial breaking of ergodicity, and memory of the particle and spin density at site $B$ will both remain in an infinite system, supporting the case for true many-body localization in an infinite system. It would be important to perform a finite size scaling analysis by also increasing the particle number, but this is currently beyond reach for exact diagonalization since even for four particles, the system sizes for which we can perform the time evolution is very restricted.

\section{Conclusion and discussion} \label{sec:conclusions}

In this work, we have provided strong evidence through analytic arguments and numerical results obtained with exact diagonalization that propagating bosonic two-body bound states can coexist with the localization of fermionic quasiparticles in lattice models with flat bands. The analytical results, namely exact eigenstates of the projected Hamiltonian of the form~\eqref{eq:exact_eigen_localized}, are valid for generic lattice models with flat bands under a few assumptions (spin rotational and time-reversal symmetries, uniform pairing condition) and are a simple consequence of the  $\mathrm{SU}(2)$ symmetry that emerges in the isolated flat band limit.~\cite{Tovmasyan2016}
We have also shown how these analytical results can be extended in the case of the dice lattice due to the compact nature of the Wannier functions.

In a previous work,~\cite{Tovmasyan2018} local integrals of motion have been explicitly constructed for a few one-dimensional lattice models with flat bands. As a consequence, fermionic quasiparticles are strictly localized in these models and spin transport is completely suppressed. Similar analytical results are not available for dimension two or higher, therefore we have focused our numerical investigations on the two-dimensional dice lattice. The presence of local integrals of motion in the one-dimensional models of Ref.~\onlinecite{Tovmasyan2018} implies that all the eigenstates are degenerate in the case of spin imbalance, with a degeneracy no less than the number of unit cells $N_{\rm c}$ of the finite cluster. In the case of the two-dimensional dice lattice, we have observed from the energy spectrum computed numerically, that the same degeneracy is \textit{approximately} present in the excited states. This is in contrast with the ground state multiplet, which is composed of states of the form~\eqref{eq:exact_eigen_localized} and is thus perfectly degenerate, see~\eqref{eq:main1}. With increasing system size the degeneracy of the excited states seems to be restored.
We interpret this finding as a signature of the presence of local integrals of motion in the projected Hamiltonian of the dice lattice,  with the slight breaking of the degeneracy in the excited states caused by finite size effects. Further studies are needed to better understand the nature of these finite size effects.

Another clear signature of many-body localization is provided by the time evolution of the particle density and the spin density starting from initial states comprising one on-site pair and one or two unpaired particles. The long time average of the observables reveals a partial breaking of ergodicity. For instance, the asymptotic distribution of the spin density is not uniform and is peaked on the site in which the unpaired particle (or particles) was initially located. This is in stark contrast with the distribution of the spin density at thermal equilibrium, which is expected to be uniform as a consequence of translational invariance. The persisting memory of the initial state in local observables is a characteristic feature of many-body localization and has been observed in experiments with ultracold gases.~\cite{Schreiber2015,Choi2016} 

Another interesting observation is that, in the case of the one-dimensional models of Ref.~\onlinecite{Tovmasyan2018}, spin transport is strictly forbidden as a consequence of the local conserved quantities, while this is not the case for the dice lattice, since, for initial states with $N_\uparrow + N_\downarrow \geq 3$, the spin density can become finite at later times even on a Wannier function in which it was initially zero. A possible explanation is that the localized states corresponding to the local conserved quantities have an extension comparable to the size of the system used in the exact diagonalization.

Besides providing interesting information from a theoretical standpoint, our numerical protocol suggests a possible experiment that could be performed using ultracold gases in optical lattices. This might be feasible in the near future since a practical scheme to implement the dice lattice with a finite magnetic flux has been suggested~\cite{Moller2018} and spin-resolved imaging at the single atom level is nowadays routinely performed. Ultracold gases have the potential to simulate the dice lattice Hamiltonian for much larger system sizes and particle numbers than can be done with exact diagonalization or other currently available numerical methods.

Our work can be extended in several possible directions. For instance, it would be interesting to consider other lattice models to confirm that the numerical results obtained in the case of the dice lattice are generic. Studying other models is probably more difficult numerically than the dice lattice because the Wannier functions are, in general, not compactly localized. This means that the projected Hamiltonian would contain many more terms than in the case of the dice lattice and would be less sparse. Thus, exact diagonalization becomes numerically heavier. In this respect, it is interesting to note that strong finite size effects are still present even for the largest system sizes that we are able to reach at present with exact diagonalization. Thus, improving the efficiency of numerical simulations is of paramount importance. The finite size scaling analysis presented in Sec.~\ref{sec:finite_size} provides some evidence that the signature of many-body localization that we observe persists in the limit of infinite size, however it would be important to perform the same analysis with a larger particle number. This is currently out of reach for the numerical methods at our disposal.

From our point of view, the key open question is how to rigorously prove the existence of local conserved quantities, or better yet, how to explicitly construct them. In that sense, it would be interesting to borrow methods from the field of many-body localization, in which local conserved quantities have been investigated extensively,~\cite{Imbrie2016,Serbyn2013,Huse2014,Rademaker2017,Imbrie2017}  and adapt them to our case. The existence of an extensive number of local conserved quantities is generally considered the defining property of many-body localization. However, even in the most extensively studied models, such as one-dimensional spin chains, there is no consensus on whether a many-body localization transition takes place, precisely because rigorously proving the existence of local integrals of motion in a many-body system is a highly nontrivial task. Indeed, in recent works it has been suggested~\cite{Suntajs2020,Suntajs2020a} that the critical disorder strength of the many-body localization transition increases with the system size, meaning no transition takes place in the thermodynamic limit. These claims were subsequently criticised~\cite{Panda2020,Sierant2020,Abanin2021} on the basis that it is difficult to extrapolate from the information obtained using small systems.
The delicate issue of the existence of many-body localization in the thermodynamic limit is still actively debated and no consensus has been reached yet~\cite{Sels2021,Sels2023}.

We believe that lattice models with flat bands provide an interesting platform to tackle the difficult problem of many-body localization and that of its fate in the thermodynamic limit. On one hand, similar to disordered systems, we observe pronounced finite size effects, however, localization is realized in a translationally invariant setting which allows us to derive exact analytical results. A particularly important exact result in the case of the dice lattice is that one term of the projected Hamiltonian possesses an extensive number of local integrals of motion, namely $[\ham_{\rm tri.}, \hat{\vb{S}}^2_{n\vb{l}}] = 0$.
Moreover, the numerical results suggest that these integrals of motion are not destroyed, but simply deformed by the other terms of the Hamiltonian $\ham_{\rm kag.}+\ham_{\rm tri.-kag.}$, and in fact play an important role in determining the system dynamics. Our hope is that the lessons learned in the future in the relatively controlled setting of translationally invariant lattice models with flat bands can be subsequently transferred to disordered systems. Our work is an initial step in this promising research direction.

We conclude by noting that the idea of localized superconductors, that is, superconductors with localized quasiparticle excitations due to strong enough disorder, has already been proposed long ago~\cite{Ma1985,Ma1986}. In this work, we provide evidence that localization of quasiparticles can occur also in two-dimensional lattice models with flat bands even in the absence of disorder. Strictly speaking, the localized quasiparticles are themselves a source disorder~\cite{Danieli2022} and an interesting question is how they affect the supercurrent flow. We believe that the coexistence of antithetic phenomena, such as superfluidity and localization, in the same system is a rather interesting occurrence that deserves further attention. Moreover, superconductors with suppressed quasiparticle transport have a number of interesting technological applications, see for instance, Ref.~\onlinecite{Pyykkonen2022}. 

\acknowledgments

This work has been supported by the Academy of Finland under Grants No. 330384 and No. 336369. We acknowledge the computational resources provided by the Aalto Science-IT project. We thank P{\"a}ivi T{\"o}rm{\"a} for helpful conversations and comments on our work. We thank Phillip Weinberg for the support provided for the QuSpin package (\url{http://quspin.github.io/QuSpin/}).

\appendix

\section{Two-body problem}
\label{app:two_body_problem}

An important task is to solve the two-body problem and find the dispersion of two-body bound states. This is a good first step to understand the many-body physics. Note that the two-body problem in the case of the dice lattice has been discussed also in Ref.~\cite{Vidal2001}.

Most of the two-particle states of the form
\begin{equation}
	\label{eq:basis1}
	\ket{n_1\vb{j}_1\sigma_1, n_2\vb{j}_2\sigma_2} = \hat{d}_{n_1\vb{j}_1\sigma_1}^\dagger\hat{d}^\dagger_{n_2\vb{j}_2\sigma_2}\ket{\emptyset}\,.
\end{equation} 
are trivially eigenstates with eigenvalue zero of the projected Hamiltonian~\eqref{eq:dice_proj_Ham}. This is the case if $\sigma_1 = \sigma_2$ or if the Wannier functions labeled by $n_1\vb{j}_1$ and $n_2\vb{j}_2$ have zero overlap, as discussed in Section~\ref{sec:uniform_pairing_dice_lat}. The only states that are not eigenstates of the projected Hamiltonian are the ones obtained by applying the on-site singlet and bond singlet creation operators to the vacuum
\begin{gather}
	\label{eq:basis_two_body}
	\ket{n\vb{l}} = \hat{B}^+_{n\vb{l}}\ket{\emptyset}\,,\\
	\label{eq:tb_bond_singlet}
	 \ket{n_1\vb{l}_1,n_2\vb{l}_2} =  \frac{1}{\sqrt{2}}\hat{B}^+_{\langle n_1\vb{l}_1,n_2\vb{l}_2\rangle}\ket{\emptyset}\,.
\end{gather}
The bond singlet states $\ket{n_1\vb{l}_1,n_2\vb{l}_2}$ have been normalized and are defined only for a pair $\langle n_1\vb{l}_1,n_2\vb{l}_2\rangle$ of nearest-neighbor Wannier functions. 
The two-body problem is solved if we can diagonalize the projected Hamiltonian in the subspace spanned by the above orthonormal set of two-particle states.

\begin{figure}
	\includegraphics[scale=0.92]{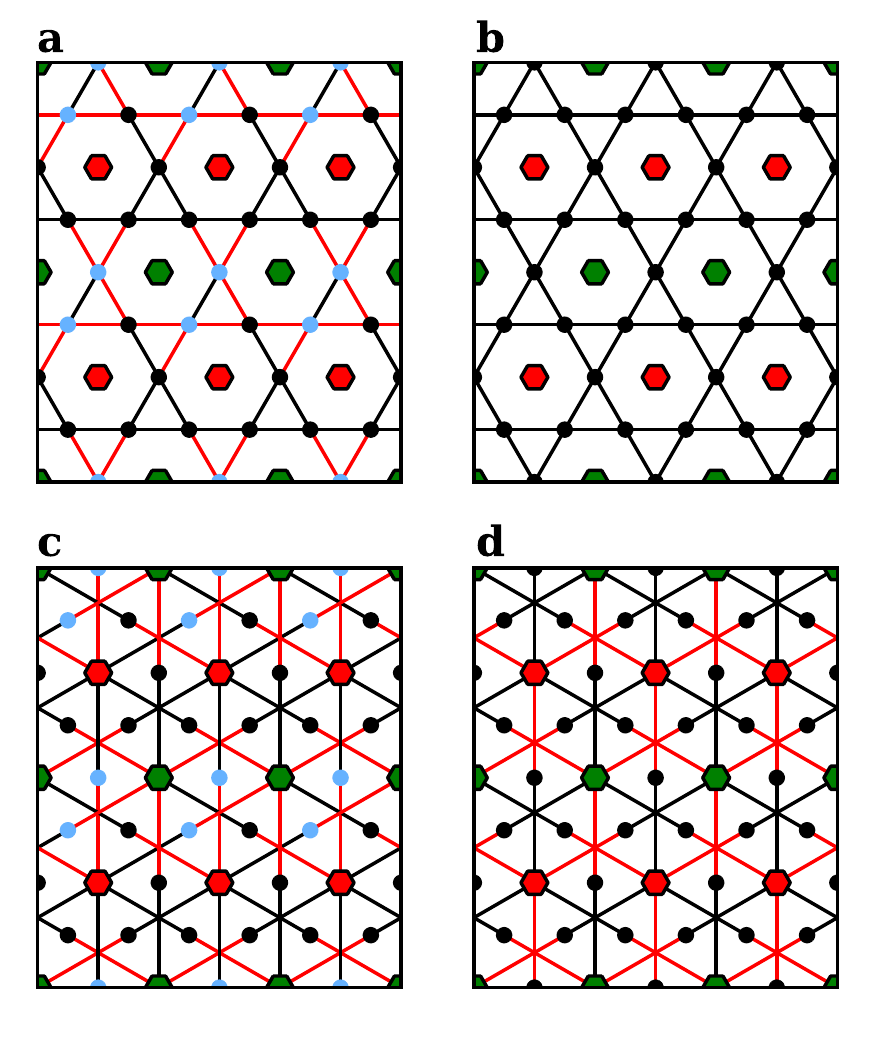}
	\caption{\label{fig:transformation} Illustration of the transformation~\eqref{eq:gauge_trans} that results in a two-body Hamiltonian with the same symmetry as the triangular lattice (the lattice formed by the green and red sites, ignoring the site colors). Panels \textbf{a} and \textbf{c} represent the Hamiltonians $\ham_{\rm kag.}$ \eqref{eq:H_kag} (bond singlet hopping) and $\ham_{\rm tri.-kag.}$ \eqref{eq:H_tri-kag} (bond singlet/on-site pair conversion), respectively.
	According to~\eqref{eq:gauge_trans}, the bond singlet states~\eqref{eq:tb_bond_singlet} corresponding to the blue sites on panel \textbf{a} and \textbf{c} on the left acquire a minus sign. Specifically, these are the sites in the kagome lattice labelled by $\langle (1, \vb{l}), (1,\vb{l}+\vb{e}_1)\rangle$ and $\langle (2, \vb{l}), (1,\vb{l}+\vb{e}_2)\rangle$. The black and red bonds in the upper panels (\textbf{a} and \textbf{b}) represent the terms of the form $\hat{B}_{\langle n_1\vb{l}_1,n_2\vb{l}_2\rangle}^+\hat{B}_{\langle n_1\vb{l}_1,n_3\vb{l}_3\rangle}^-$ in~\eqref{eq:H_kag}, with the color encoding the sign of the hopping as in Fig.~\ref{fig:full_lattice}. In the lower panels (\textbf{c} and \textbf{d}) the terms of the form $\hat{B}_{n_1\vb{l}_1}^+\hat{B}_{\langle n_2\vb{l}_2,n_3\vb{l}_3\rangle}^-$ in~\eqref{eq:H_tri-kag} are represented. After the transformation, all of the amplitudes of the terms of the form $\hat{B}_{\langle n_1\vb{l}_1,n_2\vb{l}_2\rangle}^+\hat{B}_{\langle n_1\vb{l}_1,n_3\vb{l}_3\rangle}^-$  have the same sign (panel \textbf{b}). In the case of the terms of the form $\hat{B}_{n_1\vb{l}_1}^+\hat{B}_{\langle n_2\vb{l}_2,n_3\vb{l}_3\rangle}^-$ the sign can vary, but the sign structure respects the symmetry of the triangular lattice after the transformation (panel \textbf{d}).}
\end{figure}

\begin{figure}
	\includegraphics[scale=0.95]{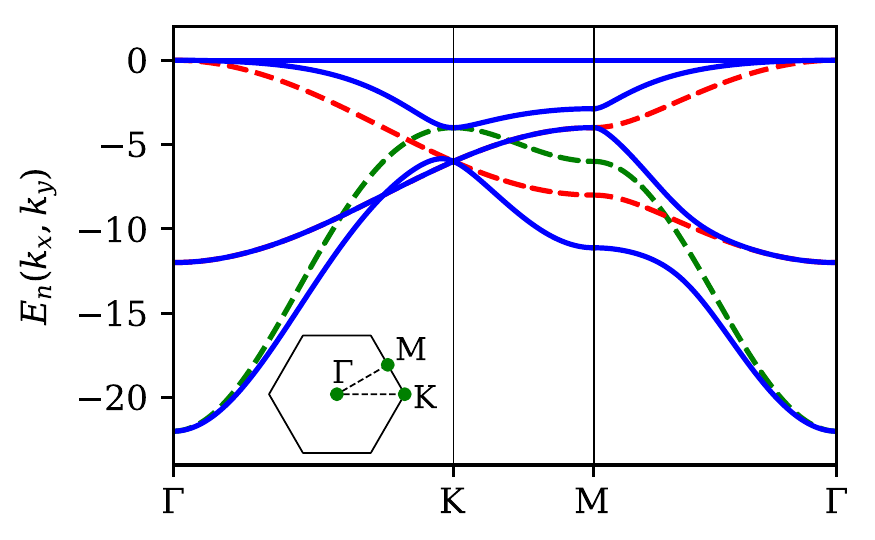}
	\caption{\label{fig:two-body_dispersion} The dispersions $E_{n}(\vb{k})$  of two-body bound states obtained from the diagonalization of~\eqref{eq:two-body_Ham} are shown as blue lines for $A = 10$. The Brillouin zone corresponding to the triangular lattice~\eqref{eq:fund_vec_v} and its high-symmetry points are shown in the inset. A notable feature is the presence of a flat band at zero energy, whose corresponding eigenvector is given in~\eqref{eq:flat_band_two-body}. The dashed red lines are dispersions obtained by diagonalizing the Hamiltonian that describes the hopping of the bond singlets between the sites of the kagome lattice, that is the lower $3\times 3$ diagonal block in~\eqref{eq:two-body_Ham}. The green dashed line is the dispersion of the on-site singlets, if these are decoupled from the bond singlets. This dispersion is given by the top diagonal element of the two-body Hamiltonian $[H^{(2)}(\vb{k})]_{1,1} = -A - 4\sum_{i = 1}^3 \cos (\vb{k}\cdot \vb{v}_i)$.} 
\end{figure}

To proceed, one needs to take advantage of translational symmetry. The two-body Hamiltonian naturally has the same symmetry of the original dice lattice from which it is derived. This translational symmetry consists in shifts of the unit cell index labeling the Wannier functions, that is $\vb{l} \to \vb{l} + \vb{e}_{i}$ with $\vb{e}_1 = (1,0)^T$ and $\vb{e}_2 = (0,1)^T$, and the fundamental vectors of the lattice are given by~\eqref{eq:fund_vec_rect}. We show that, by a suitable redefinition of the two-particle basis states in~\eqref{eq:tb_bond_singlet}, one obtains a two-body Hamiltonian with the same translational symmetry as the triangular lattice formed by the Wannier function centers, which is the Bravais lattice with fundamental vectors given by~\eqref{eq:fund_vec_v}. The redefinition of the two-particle basis states consists of a gauge transformation in which the bond singlet states~\eqref{eq:tb_bond_singlet} are multiplied by a phase factor
\begin{equation}
	\label{eq:gauge_trans}
	\ket{n_1\vb{l}_1,n_2\vb{l}_2} \to \chi(n_1\vb{l}_1,n_2\vb{l}_2) \ket{n_1\vb{l}_1,n_2\vb{l}_2}\,,
\end{equation}
where $\chi(n_1\vb{l}_1,n_2\vb{l}_2) = -1$ for $\langle n_1\vb{l}_1,n_2\vb{l}_2\rangle  = \langle (1, \vb{l}), (1,\vb{l}+\vb{e}_1)\rangle$ or $\langle (2, \vb{l}), (1,\vb{l}+\vb{e}_2)\rangle$ and $\chi(n_1\vb{l}_1,n_2\vb{l}_2) = +1$ otherwise.
The bond singlets that acquire a minus sign ($\chi_{\langle n_1\vb{l}_1,n_2\vb{l}_2\rangle} = -1$) are denoted as blue dots in panels \textbf{a} and \textbf{c} of Fig.~\ref{fig:transformation}. As illustrated in this figure, all the terms of the form $\hat{B}_{\langle n_1\vb{l}_1,n_2\vb{l}_2\rangle}^+\hat{B}_{\langle n_1\vb{l}_1,n_3\vb{l}_3\rangle}^-$ have the same sign after the gauge transformation. On the other hand, the signs of the terms of the form $\hat{B}_{n_1\vb{l}_1}^+\hat{B}_{\langle n_2\vb{l}_2,n_3\vb{l}_3\rangle}^-$ can still be positive or negative, however the sign structure has the symmetry of the triangular lattice after the redefinition~\eqref{eq:gauge_trans}, as shown in panel \textbf{d} of Fig.~\ref{fig:transformation}.

The triangular symmetry of the two-body Hamiltonian means that we no longer need to distinguish between the green and red sites in Fig.~\ref{fig:full_lattice}, which can now be labeled by using the unit cell index only, while the band index $n$ distinguishing inequivalent Wannier functions, is dropped. Note that the new unit cell index has a different meaning compared to the old one. Indeed, the state $\hat{B}^+_{\vb{l}}\ket{\emptyset}$ denotes an on-site singlet on the Wannier function centered at the position $\vb{r}_{\vb{l}}  = l_1\vb{v}_1 + l_2\vb{v}_2$, where the fundamental vectors $\vb{v}_i$ of the triangular lattice have been introduced in~\eqref{eq:fund_vec_v}. Similarly, the state $\hat{B}^+_{\langle\vb{l}_1,\vb{l}_2\rangle}\ket{\emptyset}$ denotes a singlet that lives on the two Wannier functions centered at positions $\vb{r}_{\vb{l}_1}$ and $\vb{r}_{\vb{l}_2}$ in the triangular lattice. For the purpose of solving the two-body problem, the phase factor $\chi$ introduced above is absorbed in the definition of the new bond singlet operators $\hat{B}^+_{\langle\vb{l}_1,\vb{l}_2\rangle}$. Then, we can construct plane wave linear combinations
\begin{gather}
	\ket{\vb{k};\vb{0}} = \frac{1}{\sqrt{N_{\rm c}}}\sum_{\vb{l}} e^{i\vb{k}\cdot\vb{r}_{\vb{l}}} \hat{B}^+_{\vb{l}}\ket{\emptyset}\,\\
	\ket{\vb{k};\vb{j}} = \frac{1}{\sqrt{2N_{\rm c}}}\sum_{\vb{l}} e^{i\vb{k}\cdot\vb{r}_{\vb{l}}} \hat{B}^+_{\langle \vb{l},(\vb{l}+\vb{j})\rangle}\ket{\emptyset}\,,\quad \vb{j}\neq \vb{0}\,.
\end{gather}
We choose the following orthonormal set of plane wave states to construct the two-body Hamiltonian
\begin{equation}
	\label{eq:basis_reduced}
	\begin{split}
	&\ket{\vb{k},1} = \ket{\vb{k}; \vb{0}}\,,\quad \ket{\vb{k},2} = \ket{\vb{k}; \vb{e}_1}\,,\\
	&\ket{\vb{k},3} = \ket{\vb{k}; \vb{e}_2}\,,\quad
	\ket{\vb{k},4} = \ket{\vb{k}; \vb{e}_2-\vb{e}_1}\,.
	\end{split}
\end{equation}
Then, the momentum space two-body Hamiltonian in this basis reads
\begin{widetext}
\begin{equation}
	\label{eq:two-body_Ham}
	\begin{split}
		&H^{(2)}(\vb{k}) = -\pmqty{
			A + 4\sum_{i = 1}^3 \cos (\vb{k}\cdot \vb{v}_i)& \sqrt{2}\qty(e^{i\vb{k}\cdot\vb{v}_3}-e^{-i\vb{k}\cdot\vb{v}_2})& \sqrt{2}\qty(e^{-i\vb{k}\cdot\vb{v}_3} -e^{-i\vb{k}\cdot\vb{v}_1}) & 
			\sqrt{2}\qty(e^{-i\vb{k}\cdot \vb{v}_2} - e^{i\vb{k}\cdot \vb{v}_1})\\
			\sqrt{2}\qty(e^{-i\vb{k}\cdot\vb{v}_3}-e^{i\vb{k}\cdot\vb{v}_2})  & 4 & 2+2e^{-i\vb{k}\cdot\vb{v}_3} & 2e^{i\vb{k}\cdot \vb{v}_1}+2e^{-i\vb{k}\cdot \vb{v}_3}\\
			\sqrt{2}\qty(e^{i\vb{k}\cdot\vb{v}_3} -e^{i\vb{k}\cdot\vb{v}_1})  & 2+2e^{i\vb{k}\cdot \vb{v}_3} & 4& 2+2e^{i\vb{k}\cdot \vb{v}_1}\\
			\sqrt{2}\qty(e^{i\vb{k}\cdot \vb{v}_2} - e^{-i\vb{k}\cdot \vb{v}_1})  & 2e^{-i\vb{k}\cdot \vb{v}_1}+2e^{i\vb{k}\cdot \vb{v}_3} & 2+2e^{-i\vb{k}\cdot \vb{v}_1}&	4
		}\,.
	\end{split}
\end{equation}
\end{widetext}
The dispersion of the two-body bound states obtained by the diagonalization of $H^{(2)}(\vb{k})$ is shown in Fig.~\ref{fig:two-body_dispersion}. A notable feature is the presence of a flat band at zero energy. The flat band corresponds to the following eigenvector of the two-body Hamiltonian
\begin{equation}
	\label{eq:flat_band_two-body}
	\ket{G_{\rm f.b.}(\vb{k})} = 
	\frac{1}{\sqrt{2\qty\big[3-\sum_{i = 1}^3\cos(\vb{k}\cdot \vb{v}_i)]}}\pmqty{0 \\1-e^{i\vb{k}\cdot \vb{v}_1}\\ -1 + e^{i\vb{k}\cdot \vb{v}_2} \\
		1-e^{i\vb{k}\cdot \vb{v}_3}}\,,
\end{equation}
which is a linear combination of the bond singlets only, that is, of the states $\ket{\vb{k}, l = 2,3,4}$ in~\eqref{eq:basis_reduced} but not of the on-site singlet states ($\ket{\vb{k},1}$). Therefore, the state~\eqref{eq:flat_band_two-body} is an eigenstate also of the Hamiltonian that describes the hopping of a bond singlet between nearest-neighbor sites of the kagome lattice, that is the lower $3\times 3$ diagonal block in~\eqref{eq:two-body_Ham}. The dispersion obtained by diagonalizing only this block is shown in Fig.~\ref{fig:two-body_dispersion} as well. 
The structure of the states of the kagome lattice flat band has been studied in detail.~\cite{Bergman2008} As in the case of the dice lattice, the flat band subspace is spanned by compact wave functions, but with the essential difference that they are not orthogonal since they overlap at most on a single site of the kagome lattice. The presence of the flat band of two-body states in~\eqref{eq:flat_band_two-body} is not unexpected since it is a consequence of the fact that the flat band two-body Hamiltonian has a rank which is not larger than the number of orbitals per unit cell $N_{\rm orb}$ in the original lattice. 

For $\vb{k} = \vb{0}$ the lower $3\times 3$ diagonal block in the two-body Hamiltonian, corresponding to the bond singlets states $\ket{\vb{k}, l = 2,3,4}$, completely decouples from the on-site singlet state $\ket{\vb{k},1}$. This is in agreement with the general exact result presented in Sec.~\ref{eq:ket_pair}, according to which the two-body state $\ket{N_{\rm p}=1} = \hat{b}^\dg \ket{\emptyset}$ is an exact eigenstate representing a Cooper pair with zero momentum. It is only in the dice lattice that we have the freedom of tuning the energy of this state by varying the parameter $A$. This is a consequence of the fact that the uniform pairing condition can be relaxed in the case of the dice lattice, see Sec.~\ref{sec:uniform_pairing_dice_lat}. For $\vb{k}\neq \vb{0}$, the two-body bound states are, in general, linear combinations of on-site and bond singlets, with the exception of the flat band state~\eqref{eq:flat_band_two-body}.

\input{dice_lattice.bbl}

\end{document}

%% file: dice_lattice.bbl
%

%% file: dice_lattice.bbl
\begin{thebibliography}{95}%
\makeatletter
\providecommand \@ifxundefined [1]{%
 \@ifx{#1\undefined}
}%
\providecommand \@ifnum [1]{%
 \ifnum #1\expandafter \@firstoftwo
 \else \expandafter \@secondoftwo
 \fi
}%
\providecommand \@ifx [1]{%
 \ifx #1\expandafter \@firstoftwo
 \else \expandafter \@secondoftwo
 \fi
}%
\providecommand \natexlab [1]{#1}%
\providecommand \enquote  [1]{``#1''}%
\providecommand \bibnamefont  [1]{#1}%
\providecommand \bibfnamefont [1]{#1}%
\providecommand \citenamefont [1]{#1}%
\providecommand \href@noop [0]{\@secondoftwo}%
\providecommand \href [0]{\begingroup \@sanitize@url \@href}%
\providecommand \@href[1]{\@@startlink{#1}\@@href}%
\providecommand \@@href[1]{\endgroup#1\@@endlink}%
\providecommand \@sanitize@url [0]{\catcode `\\12\catcode `\$12\catcode
  `\&12\catcode `\#12\catcode `\^12\catcode `\_12\catcode `\%12\relax}%
\providecommand \@@startlink[1]{}%
\providecommand \@@endlink[0]{}%
\providecommand \url  [0]{\begingroup\@sanitize@url \@url }%
\providecommand \@url [1]{\endgroup\@href {#1}{\urlprefix }}%
\providecommand \urlprefix  [0]{URL }%
\providecommand \Eprint [0]{\href }%
\providecommand \doibase [0]{https://doi.org/}%
\providecommand \selectlanguage [0]{\@gobble}%
\providecommand \bibinfo  [0]{\@secondoftwo}%
\providecommand \bibfield  [0]{\@secondoftwo}%
\providecommand \translation [1]{[#1]}%
\providecommand \BibitemOpen [0]{}%
\providecommand \bibitemStop [0]{}%
\providecommand \bibitemNoStop [0]{.\EOS\space}%
\providecommand \EOS [0]{\spacefactor3000\relax}%
\providecommand \BibitemShut  [1]{\csname bibitem#1\endcsname}%
\let\auto@bib@innerbib\@empty
\bibitem [{\citenamefont {Bednorz}\ and\ \citenamefont
  {M{\"u}ller}(1986)}]{Bednorz1986}%
  \BibitemOpen
  \bibfield  {author} {\bibinfo {author} {\bibfnamefont {J.~G.}\ \bibnamefont
  {Bednorz}}\ and\ \bibinfo {author} {\bibfnamefont {K.~A.}\ \bibnamefont
  {M{\"u}ller}},\ }\bibfield  {title} {\bibinfo {title} {Possible {{highTc}}
  superconductivity in the {{Ba}}-{{La}}-{{Cu}}-{{O}} system},\ }\href
  {https://doi.org/10.1007/BF01303701} {\bibfield  {journal} {\bibinfo
  {journal} {Zeitschrift f\"ur Physik B Condensed Matter}\ }\textbf {\bibinfo
  {volume} {64}},\ \bibinfo {pages} {189} (\bibinfo {year} {1986})}\BibitemShut
  {NoStop}%
\bibitem [{\citenamefont {{Hubbard, John}}(1963)}]{HubbardJohn1963}%
  \BibitemOpen
  \bibfield  {author} {\bibinfo {author} {\bibnamefont {{Hubbard, John}}},\
  }\bibfield  {title} {\bibinfo {title} {Electron correlations in narrow energy
  bands},\ }\href {https://doi.org/10.1098/rspa.1963.0204} {\bibfield
  {journal} {\bibinfo  {journal} {Proceedings of the Royal Society of London,
  Series A}\ }\bibinfo {series} {A},\ \textbf {\bibinfo {volume} {276}},\
  \bibinfo {pages} {238} (\bibinfo {year} {1963})}\BibitemShut {NoStop}%
\bibitem [{\citenamefont {LeBlanc}\ \emph {et~al.}(2015)\citenamefont
  {LeBlanc}, \citenamefont {Antipov}, \citenamefont {Becca}, \citenamefont
  {Bulik}, \citenamefont {Chan}, \citenamefont {Chung}, \citenamefont {Deng},
  \citenamefont {Ferrero}, \citenamefont {Henderson}, \citenamefont
  {{Jim{\'e}nez-Hoyos}}, \citenamefont {Kozik}, \citenamefont {Liu},
  \citenamefont {Millis}, \citenamefont {Prokof'ev}, \citenamefont {Qin},
  \citenamefont {Scuseria}, \citenamefont {Shi}, \citenamefont {Svistunov},
  \citenamefont {Tocchio}, \citenamefont {Tupitsyn}, \citenamefont {White},
  \citenamefont {Zhang}, \citenamefont {Zheng}, \citenamefont {Zhu},\ and\
  \citenamefont {Gull}}]{LeBlanc2015}%
  \BibitemOpen
  \bibfield  {author} {\bibinfo {author} {\bibfnamefont {J.~P.~F.}\
  \bibnamefont {LeBlanc}}, \bibinfo {author} {\bibfnamefont {A.~E.}\
  \bibnamefont {Antipov}}, \bibinfo {author} {\bibfnamefont {F.}~\bibnamefont
  {Becca}}, \bibinfo {author} {\bibfnamefont {I.~W.}\ \bibnamefont {Bulik}},
  \bibinfo {author} {\bibfnamefont {G.~K.-L.}\ \bibnamefont {Chan}}, \bibinfo
  {author} {\bibfnamefont {C.-M.}\ \bibnamefont {Chung}}, \bibinfo {author}
  {\bibfnamefont {Y.}~\bibnamefont {Deng}}, \bibinfo {author} {\bibfnamefont
  {M.}~\bibnamefont {Ferrero}}, \bibinfo {author} {\bibfnamefont {T.~M.}\
  \bibnamefont {Henderson}}, \bibinfo {author} {\bibfnamefont {C.~A.}\
  \bibnamefont {{Jim{\'e}nez-Hoyos}}}, \bibinfo {author} {\bibfnamefont
  {E.}~\bibnamefont {Kozik}}, \bibinfo {author} {\bibfnamefont {X.-W.}\
  \bibnamefont {Liu}}, \bibinfo {author} {\bibfnamefont {A.~J.}\ \bibnamefont
  {Millis}}, \bibinfo {author} {\bibfnamefont {N.~V.}\ \bibnamefont
  {Prokof'ev}}, \bibinfo {author} {\bibfnamefont {M.}~\bibnamefont {Qin}},
  \bibinfo {author} {\bibfnamefont {G.~E.}\ \bibnamefont {Scuseria}}, \bibinfo
  {author} {\bibfnamefont {H.}~\bibnamefont {Shi}}, \bibinfo {author}
  {\bibfnamefont {B.~V.}\ \bibnamefont {Svistunov}}, \bibinfo {author}
  {\bibfnamefont {L.~F.}\ \bibnamefont {Tocchio}}, \bibinfo {author}
  {\bibfnamefont {I.~S.}\ \bibnamefont {Tupitsyn}}, \bibinfo {author}
  {\bibfnamefont {S.~R.}\ \bibnamefont {White}}, \bibinfo {author}
  {\bibfnamefont {S.}~\bibnamefont {Zhang}}, \bibinfo {author} {\bibfnamefont
  {B.-X.}\ \bibnamefont {Zheng}}, \bibinfo {author} {\bibfnamefont
  {Z.}~\bibnamefont {Zhu}},\ and\ \bibinfo {author} {\bibfnamefont
  {E.}~\bibnamefont {Gull}},\ }\bibfield  {title} {\bibinfo {title} {Solutions
  of the {{Two-Dimensional Hubbard Model}}: {{Benchmarks}} and {{Results}} from
  a {{Wide Range}} of {{Numerical Algorithms}}},\ }\href
  {https://doi.org/10.1103/PhysRevX.5.041041} {\bibfield  {journal} {\bibinfo
  {journal} {Physical Review X}\ }\textbf {\bibinfo {volume} {5}},\ \bibinfo
  {pages} {041041} (\bibinfo {year} {2015})}\BibitemShut {NoStop}%
\bibitem [{\citenamefont {Arovas}\ \emph {et~al.}(2022)\citenamefont {Arovas},
  \citenamefont {Berg}, \citenamefont {Kivelson},\ and\ \citenamefont
  {Raghu}}]{Arovas2022}%
  \BibitemOpen
  \bibfield  {author} {\bibinfo {author} {\bibfnamefont {D.~P.}\ \bibnamefont
  {Arovas}}, \bibinfo {author} {\bibfnamefont {E.}~\bibnamefont {Berg}},
  \bibinfo {author} {\bibfnamefont {S.~A.}\ \bibnamefont {Kivelson}},\ and\
  \bibinfo {author} {\bibfnamefont {S.}~\bibnamefont {Raghu}},\ }\bibfield
  {title} {\bibinfo {title} {The {{Hubbard Model}}},\ }\href
  {https://doi.org/10.1146/annurev-conmatphys-031620-102024} {\bibfield
  {journal} {\bibinfo  {journal} {Annual Review of Condensed Matter Physics}\
  }\textbf {\bibinfo {volume} {13}},\ \bibinfo {pages} {239} (\bibinfo {year}
  {2022})}\BibitemShut {NoStop}%
\bibitem [{\citenamefont {Kivelson}\ \emph {et~al.}(1998)\citenamefont
  {Kivelson}, \citenamefont {Fradkin},\ and\ \citenamefont
  {Emery}}]{Kivelson1998}%
  \BibitemOpen
  \bibfield  {author} {\bibinfo {author} {\bibfnamefont {S.~A.}\ \bibnamefont
  {Kivelson}}, \bibinfo {author} {\bibfnamefont {E.}~\bibnamefont {Fradkin}},\
  and\ \bibinfo {author} {\bibfnamefont {V.~J.}\ \bibnamefont {Emery}},\
  }\bibfield  {title} {\bibinfo {title} {Electronic liquid-crystal phases of a
  doped {{Mott}} insulator},\ }\href {https://doi.org/10.1038/31177} {\bibfield
   {journal} {\bibinfo  {journal} {Nature}\ }\textbf {\bibinfo {volume}
  {393}},\ \bibinfo {pages} {550} (\bibinfo {year} {1998})}\BibitemShut
  {NoStop}%
\bibitem [{\citenamefont {Fradkin}\ \emph {et~al.}(2015)\citenamefont
  {Fradkin}, \citenamefont {Kivelson},\ and\ \citenamefont
  {Tranquada}}]{Fradkin2015}%
  \BibitemOpen
  \bibfield  {author} {\bibinfo {author} {\bibfnamefont {E.}~\bibnamefont
  {Fradkin}}, \bibinfo {author} {\bibfnamefont {S.~A.}\ \bibnamefont
  {Kivelson}},\ and\ \bibinfo {author} {\bibfnamefont {J.~M.}\ \bibnamefont
  {Tranquada}},\ }\bibfield  {title} {\bibinfo {title} {Colloquium: {{Theory}}
  of intertwined orders in high temperature superconductors},\ }\href
  {https://doi.org/10.1103/RevModPhys.87.457} {\bibfield  {journal} {\bibinfo
  {journal} {Reviews of Modern Physics}\ }\textbf {\bibinfo {volume} {87}},\
  \bibinfo {pages} {457} (\bibinfo {year} {2015})}\BibitemShut {NoStop}%
\bibitem [{\citenamefont {Vanhala}\ and\ \citenamefont
  {T{\"o}rm{\"a}}(2018)}]{Vanhala2018a}%
  \BibitemOpen
  \bibfield  {author} {\bibinfo {author} {\bibfnamefont {T.~I.}\ \bibnamefont
  {Vanhala}}\ and\ \bibinfo {author} {\bibfnamefont {P.}~\bibnamefont
  {T{\"o}rm{\"a}}},\ }\bibfield  {title} {\bibinfo {title} {Dynamical
  mean-field theory study of stripe order and \$d\$-wave superconductivity in
  the two-dimensional {{Hubbard}} model},\ }\href
  {https://doi.org/10.1103/PhysRevB.97.075112} {\bibfield  {journal} {\bibinfo
  {journal} {Physical Review B}\ }\textbf {\bibinfo {volume} {97}},\ \bibinfo
  {pages} {075112} (\bibinfo {year} {2018})}\BibitemShut {NoStop}%
\bibitem [{\citenamefont {Esslinger}(2010)}]{Esslinger2010}%
  \BibitemOpen
  \bibfield  {author} {\bibinfo {author} {\bibfnamefont {T.}~\bibnamefont
  {Esslinger}},\ }\bibfield  {title} {\bibinfo {title} {Fermi-{{Hubbard
  Physics}} with {{Atoms}} in an {{Optical Lattice}}},\ }\href
  {https://doi.org/10.1146/annurev-conmatphys-070909-104059} {\bibfield
  {journal} {\bibinfo  {journal} {Annual Review of Condensed Matter Physics}\
  }\textbf {\bibinfo {volume} {1}},\ \bibinfo {pages} {129} (\bibinfo {year}
  {2010})}\BibitemShut {NoStop}%
\bibitem [{\citenamefont {Mazurenko}\ \emph {et~al.}(2017)\citenamefont
  {Mazurenko}, \citenamefont {Chiu}, \citenamefont {Ji}, \citenamefont
  {Parsons}, \citenamefont {{Kan{\'a}sz-Nagy}}, \citenamefont {Schmidt},
  \citenamefont {Grusdt}, \citenamefont {Demler}, \citenamefont {Greif},\ and\
  \citenamefont {Greiner}}]{Mazurenko2017}%
  \BibitemOpen
  \bibfield  {author} {\bibinfo {author} {\bibfnamefont {A.}~\bibnamefont
  {Mazurenko}}, \bibinfo {author} {\bibfnamefont {C.~S.}\ \bibnamefont {Chiu}},
  \bibinfo {author} {\bibfnamefont {G.}~\bibnamefont {Ji}}, \bibinfo {author}
  {\bibfnamefont {M.~F.}\ \bibnamefont {Parsons}}, \bibinfo {author}
  {\bibfnamefont {M.}~\bibnamefont {{Kan{\'a}sz-Nagy}}}, \bibinfo {author}
  {\bibfnamefont {R.}~\bibnamefont {Schmidt}}, \bibinfo {author} {\bibfnamefont
  {F.}~\bibnamefont {Grusdt}}, \bibinfo {author} {\bibfnamefont
  {E.}~\bibnamefont {Demler}}, \bibinfo {author} {\bibfnamefont
  {D.}~\bibnamefont {Greif}},\ and\ \bibinfo {author} {\bibfnamefont
  {M.}~\bibnamefont {Greiner}},\ }\bibfield  {title} {\bibinfo {title} {A
  cold-atom {{Fermi}}\textendash{{Hubbard}} antiferromagnet},\ }\href
  {https://doi.org/10.1038/nature22362} {\bibfield  {journal} {\bibinfo
  {journal} {Nature}\ }\textbf {\bibinfo {volume} {545}},\ \bibinfo {pages}
  {462} (\bibinfo {year} {2017})}\BibitemShut {NoStop}%
\bibitem [{\citenamefont {Schollw{\"o}ck}(2011)}]{Schollwock2011}%
  \BibitemOpen
  \bibfield  {author} {\bibinfo {author} {\bibfnamefont {U.}~\bibnamefont
  {Schollw{\"o}ck}},\ }\bibfield  {title} {\bibinfo {title} {The density-matrix
  renormalization group in the age of matrix product states},\ }\href
  {https://doi.org/10.1016/j.aop.2010.09.012} {\bibfield  {journal} {\bibinfo
  {journal} {Annals of Physics}\ }\bibinfo {series} {January 2011 {{Special
  Issue}}},\ \textbf {\bibinfo {volume} {326}},\ \bibinfo {pages} {96}
  (\bibinfo {year} {2011})}\BibitemShut {NoStop}%
\bibitem [{\citenamefont {Stoudenmire}\ and\ \citenamefont
  {White}(2012)}]{Stoudenmire2012}%
  \BibitemOpen
  \bibfield  {author} {\bibinfo {author} {\bibfnamefont {E.}~\bibnamefont
  {Stoudenmire}}\ and\ \bibinfo {author} {\bibfnamefont {S.~R.}\ \bibnamefont
  {White}},\ }\bibfield  {title} {\bibinfo {title} {Studying {{Two-Dimensional
  Systems}} with the {{Density Matrix Renormalization Group}}},\ }\href
  {https://doi.org/10.1146/annurev-conmatphys-020911-125018} {\bibfield
  {journal} {\bibinfo  {journal} {Annual Review of Condensed Matter Physics}\
  }\textbf {\bibinfo {volume} {3}},\ \bibinfo {pages} {111} (\bibinfo {year}
  {2012})}\BibitemShut {NoStop}%
\bibitem [{\citenamefont {Qin}\ \emph {et~al.}(2020)\citenamefont {Qin},
  \citenamefont {Chung}, \citenamefont {Shi}, \citenamefont {Vitali},
  \citenamefont {Hubig}, \citenamefont {Schollw{\"o}ck}, \citenamefont
  {White},\ and\ \citenamefont {Zhang}}]{Qin2020}%
  \BibitemOpen
  \bibfield  {author} {\bibinfo {author} {\bibfnamefont {M.}~\bibnamefont
  {Qin}}, \bibinfo {author} {\bibfnamefont {C.-M.}\ \bibnamefont {Chung}},
  \bibinfo {author} {\bibfnamefont {H.}~\bibnamefont {Shi}}, \bibinfo {author}
  {\bibfnamefont {E.}~\bibnamefont {Vitali}}, \bibinfo {author} {\bibfnamefont
  {C.}~\bibnamefont {Hubig}}, \bibinfo {author} {\bibfnamefont
  {U.}~\bibnamefont {Schollw{\"o}ck}}, \bibinfo {author} {\bibfnamefont
  {S.~R.}\ \bibnamefont {White}},\ and\ \bibinfo {author} {\bibfnamefont
  {S.}~\bibnamefont {Zhang}},\ }\bibfield  {title} {\bibinfo {title} {Absence
  of {{Superconductivity}} in the {{Pure Two-Dimensional Hubbard Model}}},\
  }\href {https://doi.org/10.1103/PhysRevX.10.031016} {\bibfield  {journal}
  {\bibinfo  {journal} {Physical Review X}\ }\textbf {\bibinfo {volume} {10}},\
  \bibinfo {pages} {031016} (\bibinfo {year} {2020})}\BibitemShut {NoStop}%
\bibitem [{\citenamefont {Ponsioen}\ \emph {et~al.}(2019)\citenamefont
  {Ponsioen}, \citenamefont {Chung},\ and\ \citenamefont
  {Corboz}}]{Ponsioen2019}%
  \BibitemOpen
  \bibfield  {author} {\bibinfo {author} {\bibfnamefont {B.}~\bibnamefont
  {Ponsioen}}, \bibinfo {author} {\bibfnamefont {S.~S.}\ \bibnamefont
  {Chung}},\ and\ \bibinfo {author} {\bibfnamefont {P.}~\bibnamefont
  {Corboz}},\ }\bibfield  {title} {\bibinfo {title} {Period 4 stripe in the
  extended two-dimensional {{Hubbard}} model},\ }\href
  {https://doi.org/10.1103/PhysRevB.100.195141} {\bibfield  {journal} {\bibinfo
   {journal} {Physical Review B}\ }\textbf {\bibinfo {volume} {100}},\ \bibinfo
  {pages} {195141} (\bibinfo {year} {2019})}\BibitemShut {NoStop}%
\bibitem [{\citenamefont {Jiang}\ and\ \citenamefont
  {Devereaux}(2019)}]{Jiang2019}%
  \BibitemOpen
  \bibfield  {author} {\bibinfo {author} {\bibfnamefont {H.-C.}\ \bibnamefont
  {Jiang}}\ and\ \bibinfo {author} {\bibfnamefont {T.~P.}\ \bibnamefont
  {Devereaux}},\ }\bibfield  {title} {\bibinfo {title} {Superconductivity in
  the doped {{Hubbard}} model and its interplay with next-nearest hopping
  {\emph{t}} {${'}$}},\ }\href {https://doi.org/10.1126/science.aal5304}
  {\bibfield  {journal} {\bibinfo  {journal} {Science}\ }\textbf {\bibinfo
  {volume} {365}},\ \bibinfo {pages} {1424} (\bibinfo {year}
  {2019})}\BibitemShut {NoStop}%
\bibitem [{\citenamefont {Varma}\ \emph {et~al.}(1987)\citenamefont {Varma},
  \citenamefont {{Schmitt-Rink}},\ and\ \citenamefont {Abrahams}}]{Varma1987}%
  \BibitemOpen
  \bibfield  {author} {\bibinfo {author} {\bibfnamefont {C.}~\bibnamefont
  {Varma}}, \bibinfo {author} {\bibfnamefont {S.}~\bibnamefont
  {{Schmitt-Rink}}},\ and\ \bibinfo {author} {\bibfnamefont {E.}~\bibnamefont
  {Abrahams}},\ }\bibfield  {title} {\bibinfo {title} {Charge transfer
  excitations and superconductivity in ``ionic'' metals},\ }\href
  {https://doi.org/10.1016/0038-1098(87)90407-8} {\bibfield  {journal}
  {\bibinfo  {journal} {Solid State Communications}\ }\textbf {\bibinfo
  {volume} {62}},\ \bibinfo {pages} {681} (\bibinfo {year} {1987})}\BibitemShut
  {NoStop}%
\bibitem [{\citenamefont {Mattheiss}(1987)}]{Mattheiss1987}%
  \BibitemOpen
  \bibfield  {author} {\bibinfo {author} {\bibfnamefont {L.~F.}\ \bibnamefont
  {Mattheiss}},\ }\bibfield  {title} {\bibinfo {title} {Electronic band
  properties and superconductivity in {{La}} 2 - y {{X}} y {{CuO}} 4},\ }\href
  {https://doi.org/10.1103/PhysRevLett.58.1028} {\bibfield  {journal} {\bibinfo
   {journal} {Physical Review Letters}\ }\textbf {\bibinfo {volume} {58}},\
  \bibinfo {pages} {1028} (\bibinfo {year} {1987})}\BibitemShut {NoStop}%
\bibitem [{\citenamefont {Emery}(1987)}]{Emery1987}%
  \BibitemOpen
  \bibfield  {author} {\bibinfo {author} {\bibfnamefont {V.~J.}\ \bibnamefont
  {Emery}},\ }\bibfield  {title} {\bibinfo {title} {Theory of high- {{T}} c
  superconductivity in oxides},\ }\href
  {https://doi.org/10.1103/PhysRevLett.58.2794} {\bibfield  {journal} {\bibinfo
   {journal} {Physical Review Letters}\ }\textbf {\bibinfo {volume} {58}},\
  \bibinfo {pages} {2794} (\bibinfo {year} {1987})}\BibitemShut {NoStop}%
\bibitem [{\citenamefont {Kung}\ \emph {et~al.}(2016)\citenamefont {Kung},
  \citenamefont {Chen}, \citenamefont {Wang}, \citenamefont {Huang},
  \citenamefont {Nowadnick}, \citenamefont {Moritz}, \citenamefont {Scalettar},
  \citenamefont {Johnston},\ and\ \citenamefont {Devereaux}}]{Kung2016}%
  \BibitemOpen
  \bibfield  {author} {\bibinfo {author} {\bibfnamefont {Y.~F.}\ \bibnamefont
  {Kung}}, \bibinfo {author} {\bibfnamefont {C.-C.}\ \bibnamefont {Chen}},
  \bibinfo {author} {\bibfnamefont {Y.}~\bibnamefont {Wang}}, \bibinfo {author}
  {\bibfnamefont {E.~W.}\ \bibnamefont {Huang}}, \bibinfo {author}
  {\bibfnamefont {E.~A.}\ \bibnamefont {Nowadnick}}, \bibinfo {author}
  {\bibfnamefont {B.}~\bibnamefont {Moritz}}, \bibinfo {author} {\bibfnamefont
  {R.~T.}\ \bibnamefont {Scalettar}}, \bibinfo {author} {\bibfnamefont
  {S.}~\bibnamefont {Johnston}},\ and\ \bibinfo {author} {\bibfnamefont
  {T.~P.}\ \bibnamefont {Devereaux}},\ }\bibfield  {title} {\bibinfo {title}
  {Characterizing the three-orbital {{Hubbard}} model with determinant quantum
  {{Monte Carlo}}},\ }\href {https://doi.org/10.1103/PhysRevB.93.155166}
  {\bibfield  {journal} {\bibinfo  {journal} {Physical Review B}\ }\textbf
  {\bibinfo {volume} {93}},\ \bibinfo {pages} {155166} (\bibinfo {year}
  {2016})}\BibitemShut {NoStop}%
\bibitem [{\citenamefont {Val'kov}\ \emph {et~al.}(2016)\citenamefont
  {Val'kov}, \citenamefont {Dzebisashvili}, \citenamefont {Korovushkin},\ and\
  \citenamefont {Barabanov}}]{Valkov2016}%
  \BibitemOpen
  \bibfield  {author} {\bibinfo {author} {\bibfnamefont {V.~V.}\ \bibnamefont
  {Val'kov}}, \bibinfo {author} {\bibfnamefont {D.~M.}\ \bibnamefont
  {Dzebisashvili}}, \bibinfo {author} {\bibfnamefont {M.~M.}\ \bibnamefont
  {Korovushkin}},\ and\ \bibinfo {author} {\bibfnamefont {A.~F.}\ \bibnamefont
  {Barabanov}},\ }\bibfield  {title} {\bibinfo {title} {Stability of the
  superconducting \$\$\{d\_\{\{x\^2\} - \{y\^2\}\}\}\$\$ phase in high-{{T}} c
  superconductors with respect to the intersite coulomb repulsion of holes at
  oxygen},\ }\href {https://doi.org/10.1134/S0021364016060114} {\bibfield
  {journal} {\bibinfo  {journal} {JETP Letters}\ }\textbf {\bibinfo {volume}
  {103}},\ \bibinfo {pages} {385} (\bibinfo {year} {2016})}\BibitemShut
  {NoStop}%
\bibitem [{\citenamefont {Adolphs}\ \emph {et~al.}(2016)\citenamefont
  {Adolphs}, \citenamefont {Moser}, \citenamefont {Sawatzky},\ and\
  \citenamefont {Berciu}}]{Adolphs2016}%
  \BibitemOpen
  \bibfield  {author} {\bibinfo {author} {\bibfnamefont {C.~P.~J.}\
  \bibnamefont {Adolphs}}, \bibinfo {author} {\bibfnamefont {S.}~\bibnamefont
  {Moser}}, \bibinfo {author} {\bibfnamefont {G.~A.}\ \bibnamefont
  {Sawatzky}},\ and\ \bibinfo {author} {\bibfnamefont {M.}~\bibnamefont
  {Berciu}},\ }\bibfield  {title} {\bibinfo {title} {Non-{{Zhang-Rice Singlet
  Character}} of the {{First Ionization State}} of {{T-CuO}}},\ }\href
  {https://doi.org/10.1103/PhysRevLett.116.087002} {\bibfield  {journal}
  {\bibinfo  {journal} {Physical Review Letters}\ }\textbf {\bibinfo {volume}
  {116}},\ \bibinfo {pages} {087002} (\bibinfo {year} {2016})}\BibitemShut
  {NoStop}%
\bibitem [{\citenamefont {Leykam}\ \emph {et~al.}(2018)\citenamefont {Leykam},
  \citenamefont {Andreanov},\ and\ \citenamefont {Flach}}]{Leykam2018}%
  \BibitemOpen
  \bibfield  {author} {\bibinfo {author} {\bibfnamefont {D.}~\bibnamefont
  {Leykam}}, \bibinfo {author} {\bibfnamefont {A.}~\bibnamefont {Andreanov}},\
  and\ \bibinfo {author} {\bibfnamefont {S.}~\bibnamefont {Flach}},\ }\bibfield
   {title} {\bibinfo {title} {Artificial flat band systems: From lattice models
  to experiments},\ }\href {https://doi.org/10.1080/23746149.2018.1473052}
  {\bibfield  {journal} {\bibinfo  {journal} {Advances in Physics: X}\ }\textbf
  {\bibinfo {volume} {3}},\ \bibinfo {pages} {1473052} (\bibinfo {year}
  {2018})}\BibitemShut {NoStop}%
\bibitem [{\citenamefont {Huber}\ and\ \citenamefont
  {Altman}(2010)}]{Huber2010}%
  \BibitemOpen
  \bibfield  {author} {\bibinfo {author} {\bibfnamefont {S.~D.}\ \bibnamefont
  {Huber}}\ and\ \bibinfo {author} {\bibfnamefont {E.}~\bibnamefont {Altman}},\
  }\bibfield  {title} {\bibinfo {title} {Bose condensation in flat bands},\
  }\href {https://doi.org/10.1103/PhysRevB.82.184502} {\bibfield  {journal}
  {\bibinfo  {journal} {Physical Review B}\ }\textbf {\bibinfo {volume} {82}},\
  \bibinfo {pages} {184502} (\bibinfo {year} {2010})}\BibitemShut {NoStop}%
\bibitem [{\citenamefont {Tovmasyan}\ \emph {et~al.}(2016)\citenamefont
  {Tovmasyan}, \citenamefont {Peotta}, \citenamefont {T{\"o}rm{\"a}},\ and\
  \citenamefont {Huber}}]{Tovmasyan2016}%
  \BibitemOpen
  \bibfield  {author} {\bibinfo {author} {\bibfnamefont {M.}~\bibnamefont
  {Tovmasyan}}, \bibinfo {author} {\bibfnamefont {S.}~\bibnamefont {Peotta}},
  \bibinfo {author} {\bibfnamefont {P.}~\bibnamefont {T{\"o}rm{\"a}}},\ and\
  \bibinfo {author} {\bibfnamefont {S.~D.}\ \bibnamefont {Huber}},\ }\bibfield
  {title} {\bibinfo {title} {Effective theory and emergent {{SU}} ( 2 )
  symmetry in the flat bands of attractive {{Hubbard}} models},\ }\href
  {https://doi.org/10.1103/PhysRevB.94.245149} {\bibfield  {journal} {\bibinfo
  {journal} {Physical Review B}\ }\textbf {\bibinfo {volume} {94}},\ \bibinfo
  {pages} {245149} (\bibinfo {year} {2016})}\BibitemShut {NoStop}%
\bibitem [{\citenamefont {Tovmasyan}\ \emph {et~al.}(2018)\citenamefont
  {Tovmasyan}, \citenamefont {Peotta}, \citenamefont {Liang}, \citenamefont
  {T{\"o}rm{\"a}},\ and\ \citenamefont {Huber}}]{Tovmasyan2018}%
  \BibitemOpen
  \bibfield  {author} {\bibinfo {author} {\bibfnamefont {M.}~\bibnamefont
  {Tovmasyan}}, \bibinfo {author} {\bibfnamefont {S.}~\bibnamefont {Peotta}},
  \bibinfo {author} {\bibfnamefont {L.}~\bibnamefont {Liang}}, \bibinfo
  {author} {\bibfnamefont {P.}~\bibnamefont {T{\"o}rm{\"a}}},\ and\ \bibinfo
  {author} {\bibfnamefont {S.~D.}\ \bibnamefont {Huber}},\ }\bibfield  {title}
  {\bibinfo {title} {Preformed pairs in flat {{Bloch}} bands},\ }\href
  {https://doi.org/10.1103/PhysRevB.98.134513} {\bibfield  {journal} {\bibinfo
  {journal} {Physical Review B}\ }\textbf {\bibinfo {volume} {98}},\ \bibinfo
  {pages} {134513} (\bibinfo {year} {2018})}\BibitemShut {NoStop}%
\bibitem [{\citenamefont {{Herzog-Arbeitman}}\ \emph
  {et~al.}(2022)\citenamefont {{Herzog-Arbeitman}}, \citenamefont {Chew},
  \citenamefont {Huhtinen}, \citenamefont {T{\"o}rm{\"a}},\ and\ \citenamefont
  {Bernevig}}]{Herzog-Arbeitman2022}%
  \BibitemOpen
  \bibfield  {author} {\bibinfo {author} {\bibfnamefont {J.}~\bibnamefont
  {{Herzog-Arbeitman}}}, \bibinfo {author} {\bibfnamefont {A.}~\bibnamefont
  {Chew}}, \bibinfo {author} {\bibfnamefont {K.-E.}\ \bibnamefont {Huhtinen}},
  \bibinfo {author} {\bibfnamefont {P.}~\bibnamefont {T{\"o}rm{\"a}}},\ and\
  \bibinfo {author} {\bibfnamefont {B.~A.}\ \bibnamefont {Bernevig}},\ }\href
  {https://doi.org/10.48550/arXiv.2209.00007} {\bibinfo {title} {Many-{{Body
  Superconductivity}} in {{Topological Flat Bands}}}} (\bibinfo {year}
  {2022}),\ \Eprint {https://arxiv.org/abs/arXiv:2209.00007}
  {arxiv:arXiv:2209.00007} \BibitemShut {NoStop}%
\bibitem [{\citenamefont {Heikkil{\"a}}\ \emph {et~al.}(2011)\citenamefont
  {Heikkil{\"a}}, \citenamefont {Kopnin},\ and\ \citenamefont
  {Volovik}}]{Heikkila2011}%
  \BibitemOpen
  \bibfield  {author} {\bibinfo {author} {\bibfnamefont {T.~T.}\ \bibnamefont
  {Heikkil{\"a}}}, \bibinfo {author} {\bibfnamefont {N.~B.}\ \bibnamefont
  {Kopnin}},\ and\ \bibinfo {author} {\bibfnamefont {G.~E.}\ \bibnamefont
  {Volovik}},\ }\bibfield  {title} {\bibinfo {title} {Flat bands in topological
  media},\ }\href {https://doi.org/10.1134/S0021364011150045} {\bibfield
  {journal} {\bibinfo  {journal} {JETP Letters}\ }\textbf {\bibinfo {volume}
  {94}},\ \bibinfo {pages} {233} (\bibinfo {year} {2011})}\BibitemShut
  {NoStop}%
\bibitem [{\citenamefont {Kopnin}\ \emph {et~al.}(2011)\citenamefont {Kopnin},
  \citenamefont {Heikkil{\"a}},\ and\ \citenamefont {Volovik}}]{Kopnin2011}%
  \BibitemOpen
  \bibfield  {author} {\bibinfo {author} {\bibfnamefont {N.~B.}\ \bibnamefont
  {Kopnin}}, \bibinfo {author} {\bibfnamefont {T.~T.}\ \bibnamefont
  {Heikkil{\"a}}},\ and\ \bibinfo {author} {\bibfnamefont {G.~E.}\ \bibnamefont
  {Volovik}},\ }\bibfield  {title} {\bibinfo {title} {High-temperature surface
  superconductivity in topological flat-band systems},\ }\href
  {https://doi.org/10.1103/PhysRevB.83.220503} {\bibfield  {journal} {\bibinfo
  {journal} {Physical Review B}\ }\textbf {\bibinfo {volume} {83}},\ \bibinfo
  {pages} {220503} (\bibinfo {year} {2011})}\BibitemShut {NoStop}%
\bibitem [{\citenamefont {Taie}\ \emph {et~al.}(2015)\citenamefont {Taie},
  \citenamefont {Ozawa}, \citenamefont {Ichinose}, \citenamefont {Nishio},
  \citenamefont {Nakajima},\ and\ \citenamefont {Takahashi}}]{Taie2015}%
  \BibitemOpen
  \bibfield  {author} {\bibinfo {author} {\bibfnamefont {S.}~\bibnamefont
  {Taie}}, \bibinfo {author} {\bibfnamefont {H.}~\bibnamefont {Ozawa}},
  \bibinfo {author} {\bibfnamefont {T.}~\bibnamefont {Ichinose}}, \bibinfo
  {author} {\bibfnamefont {T.}~\bibnamefont {Nishio}}, \bibinfo {author}
  {\bibfnamefont {S.}~\bibnamefont {Nakajima}},\ and\ \bibinfo {author}
  {\bibfnamefont {Y.}~\bibnamefont {Takahashi}},\ }\bibfield  {title} {\bibinfo
  {title} {Coherent driving and freezing of bosonic matter wave in an optical
  {{Lieb}} lattice},\ }\href {https://doi.org/10.1126/sciadv.1500854}
  {\bibfield  {journal} {\bibinfo  {journal} {Science Advances}\ }\textbf
  {\bibinfo {volume} {1}},\ \bibinfo {pages} {e1500854} (\bibinfo {year}
  {2015})}\BibitemShut {NoStop}%
\bibitem [{\citenamefont {Ozawa}\ \emph {et~al.}(2017)\citenamefont {Ozawa},
  \citenamefont {Taie}, \citenamefont {Ichinose},\ and\ \citenamefont
  {Takahashi}}]{Ozawa2017}%
  \BibitemOpen
  \bibfield  {author} {\bibinfo {author} {\bibfnamefont {H.}~\bibnamefont
  {Ozawa}}, \bibinfo {author} {\bibfnamefont {S.}~\bibnamefont {Taie}},
  \bibinfo {author} {\bibfnamefont {T.}~\bibnamefont {Ichinose}},\ and\
  \bibinfo {author} {\bibfnamefont {Y.}~\bibnamefont {Takahashi}},\ }\bibfield
  {title} {\bibinfo {title} {Interaction-{{Driven Shift}} and {{Distortion}} of
  a {{Flat Band}} in an {{Optical Lieb Lattice}}},\ }\href
  {https://doi.org/10.1103/PhysRevLett.118.175301} {\bibfield  {journal}
  {\bibinfo  {journal} {Physical Review Letters}\ }\textbf {\bibinfo {volume}
  {118}},\ \bibinfo {pages} {175301} (\bibinfo {year} {2017})}\BibitemShut
  {NoStop}%
\bibitem [{\citenamefont {Jo}\ \emph {et~al.}(2012)\citenamefont {Jo},
  \citenamefont {Guzman}, \citenamefont {Thomas}, \citenamefont {Hosur},
  \citenamefont {Vishwanath},\ and\ \citenamefont {{Stamper-Kurn}}}]{Jo2012}%
  \BibitemOpen
  \bibfield  {author} {\bibinfo {author} {\bibfnamefont {G.-B.}\ \bibnamefont
  {Jo}}, \bibinfo {author} {\bibfnamefont {J.}~\bibnamefont {Guzman}}, \bibinfo
  {author} {\bibfnamefont {C.~K.}\ \bibnamefont {Thomas}}, \bibinfo {author}
  {\bibfnamefont {P.}~\bibnamefont {Hosur}}, \bibinfo {author} {\bibfnamefont
  {A.}~\bibnamefont {Vishwanath}},\ and\ \bibinfo {author} {\bibfnamefont
  {D.~M.}\ \bibnamefont {{Stamper-Kurn}}},\ }\bibfield  {title} {\bibinfo
  {title} {Ultracold {{Atoms}} in a {{Tunable Optical Kagome Lattice}}},\
  }\href {https://doi.org/10.1103/PhysRevLett.108.045305} {\bibfield  {journal}
  {\bibinfo  {journal} {Physical Review Letters}\ }\textbf {\bibinfo {volume}
  {108}},\ \bibinfo {pages} {045305} (\bibinfo {year} {2012})}\BibitemShut
  {NoStop}%
\bibitem [{\citenamefont {Leung}\ \emph {et~al.}(2020)\citenamefont {Leung},
  \citenamefont {Schwarz}, \citenamefont {Chang}, \citenamefont {Brown},
  \citenamefont {Unnikrishnan},\ and\ \citenamefont
  {{Stamper-Kurn}}}]{Leung2020}%
  \BibitemOpen
  \bibfield  {author} {\bibinfo {author} {\bibfnamefont {T.-H.}\ \bibnamefont
  {Leung}}, \bibinfo {author} {\bibfnamefont {M.~N.}\ \bibnamefont {Schwarz}},
  \bibinfo {author} {\bibfnamefont {S.-W.}\ \bibnamefont {Chang}}, \bibinfo
  {author} {\bibfnamefont {C.~D.}\ \bibnamefont {Brown}}, \bibinfo {author}
  {\bibfnamefont {G.}~\bibnamefont {Unnikrishnan}},\ and\ \bibinfo {author}
  {\bibfnamefont {D.}~\bibnamefont {{Stamper-Kurn}}},\ }\bibfield  {title}
  {\bibinfo {title} {Interaction-{{Enhanced Group Velocity}} of {{Bosons}} in
  the {{Flat Band}} of an {{Optical Kagome Lattice}}},\ }\href
  {https://doi.org/10.1103/PhysRevLett.125.133001} {\bibfield  {journal}
  {\bibinfo  {journal} {Physical Review Letters}\ }\textbf {\bibinfo {volume}
  {125}},\ \bibinfo {pages} {133001} (\bibinfo {year} {2020})}\BibitemShut
  {NoStop}%
\bibitem [{\citenamefont {M{\"o}ller}\ and\ \citenamefont
  {Cooper}(2018)}]{Moller2018}%
  \BibitemOpen
  \bibfield  {author} {\bibinfo {author} {\bibfnamefont {G.}~\bibnamefont
  {M{\"o}ller}}\ and\ \bibinfo {author} {\bibfnamefont {N.~R.}\ \bibnamefont
  {Cooper}},\ }\bibfield  {title} {\bibinfo {title} {Synthetic gauge fields for
  lattices with multi-orbital unit cells: Routes towards a \$\textbackslash
  uppi\$-flux dice lattice with flat bands},\ }\href
  {https://doi.org/10.1088/1367-2630/aad134} {\bibfield  {journal} {\bibinfo
  {journal} {New Journal of Physics}\ }\textbf {\bibinfo {volume} {20}},\
  \bibinfo {pages} {073025} (\bibinfo {year} {2018})}\BibitemShut {NoStop}%
\bibitem [{\citenamefont {Cao}\ \emph {et~al.}(2018)\citenamefont {Cao},
  \citenamefont {Fatemi}, \citenamefont {Fang}, \citenamefont {Watanabe},
  \citenamefont {Taniguchi}, \citenamefont {Kaxiras},\ and\ \citenamefont
  {{Jarillo-Herrero}}}]{Cao2018}%
  \BibitemOpen
  \bibfield  {author} {\bibinfo {author} {\bibfnamefont {Y.}~\bibnamefont
  {Cao}}, \bibinfo {author} {\bibfnamefont {V.}~\bibnamefont {Fatemi}},
  \bibinfo {author} {\bibfnamefont {S.}~\bibnamefont {Fang}}, \bibinfo {author}
  {\bibfnamefont {K.}~\bibnamefont {Watanabe}}, \bibinfo {author}
  {\bibfnamefont {T.}~\bibnamefont {Taniguchi}}, \bibinfo {author}
  {\bibfnamefont {E.}~\bibnamefont {Kaxiras}},\ and\ \bibinfo {author}
  {\bibfnamefont {P.}~\bibnamefont {{Jarillo-Herrero}}},\ }\bibfield  {title}
  {\bibinfo {title} {Unconventional superconductivity in magic-angle graphene
  superlattices},\ }\href {https://doi.org/10.1038/nature26160} {\bibfield
  {journal} {\bibinfo  {journal} {Nature}\ }\textbf {\bibinfo {volume} {556}},\
  \bibinfo {pages} {43} (\bibinfo {year} {2018})}\BibitemShut {NoStop}%
\bibitem [{\citenamefont {T{\"o}rm{\"a}}\ \emph {et~al.}(2022)\citenamefont
  {T{\"o}rm{\"a}}, \citenamefont {Peotta},\ and\ \citenamefont
  {Bernevig}}]{Torma2022}%
  \BibitemOpen
  \bibfield  {author} {\bibinfo {author} {\bibfnamefont {P.}~\bibnamefont
  {T{\"o}rm{\"a}}}, \bibinfo {author} {\bibfnamefont {S.}~\bibnamefont
  {Peotta}},\ and\ \bibinfo {author} {\bibfnamefont {B.~A.}\ \bibnamefont
  {Bernevig}},\ }\bibfield  {title} {\bibinfo {title} {Superconductivity,
  superfluidity and quantum geometry in twisted multilayer systems},\ }\href
  {https://doi.org/10.1038/s42254-022-00466-y} {\bibfield  {journal} {\bibinfo
  {journal} {Nature Reviews Physics}\ }\textbf {\bibinfo {volume} {4}},\
  \bibinfo {pages} {528} (\bibinfo {year} {2022})}\BibitemShut {NoStop}%
\bibitem [{\citenamefont {Meng}\ \emph {et~al.}(2023)\citenamefont {Meng},
  \citenamefont {Wang}, \citenamefont {Han}, \citenamefont {Liu}, \citenamefont
  {Wen}, \citenamefont {Gao}, \citenamefont {Wang}, \citenamefont {Chin},\ and\
  \citenamefont {Zhang}}]{Meng2021}%
  \BibitemOpen
  \bibfield  {author} {\bibinfo {author} {\bibfnamefont {Z.}~\bibnamefont
  {Meng}}, \bibinfo {author} {\bibfnamefont {L.}~\bibnamefont {Wang}}, \bibinfo
  {author} {\bibfnamefont {W.}~\bibnamefont {Han}}, \bibinfo {author}
  {\bibfnamefont {F.}~\bibnamefont {Liu}}, \bibinfo {author} {\bibfnamefont
  {K.}~\bibnamefont {Wen}}, \bibinfo {author} {\bibfnamefont {C.}~\bibnamefont
  {Gao}}, \bibinfo {author} {\bibfnamefont {P.}~\bibnamefont {Wang}}, \bibinfo
  {author} {\bibfnamefont {C.}~\bibnamefont {Chin}},\ and\ \bibinfo {author}
  {\bibfnamefont {J.}~\bibnamefont {Zhang}},\ }\bibfield  {title} {\bibinfo
  {title} {Atomic bose--einstein condensate in twisted-bilayer optical
  lattices},\ }\href
  {https://doi.org/https://doi.org/10.1038/s41586-023-05695-4} {\bibfield
  {journal} {\bibinfo  {journal} {Nature}\ }\textbf {\bibinfo {volume} {615}},\
  \bibinfo {pages} {231} (\bibinfo {year} {2023})}\BibitemShut {NoStop}%
\bibitem [{\citenamefont {Julku}\ \emph {et~al.}(2016)\citenamefont {Julku},
  \citenamefont {Peotta}, \citenamefont {Vanhala}, \citenamefont {Kim},\ and\
  \citenamefont {T{\"o}rm{\"a}}}]{Julku2016}%
  \BibitemOpen
  \bibfield  {author} {\bibinfo {author} {\bibfnamefont {A.}~\bibnamefont
  {Julku}}, \bibinfo {author} {\bibfnamefont {S.}~\bibnamefont {Peotta}},
  \bibinfo {author} {\bibfnamefont {T.~I.}\ \bibnamefont {Vanhala}}, \bibinfo
  {author} {\bibfnamefont {D.-H.}\ \bibnamefont {Kim}},\ and\ \bibinfo {author}
  {\bibfnamefont {P.}~\bibnamefont {T{\"o}rm{\"a}}},\ }\bibfield  {title}
  {\bibinfo {title} {Geometric {{Origin}} of {{Superfluidity}} in the
  {{Lieb-Lattice Flat Band}}},\ }\href
  {https://doi.org/10.1103/PhysRevLett.117.045303} {\bibfield  {journal}
  {\bibinfo  {journal} {Physical Review Letters}\ }\textbf {\bibinfo {volume}
  {117}},\ \bibinfo {pages} {045303} (\bibinfo {year} {2016})}\BibitemShut
  {NoStop}%
\bibitem [{\citenamefont {Liang}\ \emph {et~al.}(2017)\citenamefont {Liang},
  \citenamefont {Vanhala}, \citenamefont {Peotta}, \citenamefont {Siro},
  \citenamefont {Harju},\ and\ \citenamefont {T{\"o}rm{\"a}}}]{Liang2017a}%
  \BibitemOpen
  \bibfield  {author} {\bibinfo {author} {\bibfnamefont {L.}~\bibnamefont
  {Liang}}, \bibinfo {author} {\bibfnamefont {T.~I.}\ \bibnamefont {Vanhala}},
  \bibinfo {author} {\bibfnamefont {S.}~\bibnamefont {Peotta}}, \bibinfo
  {author} {\bibfnamefont {T.}~\bibnamefont {Siro}}, \bibinfo {author}
  {\bibfnamefont {A.}~\bibnamefont {Harju}},\ and\ \bibinfo {author}
  {\bibfnamefont {P.}~\bibnamefont {T{\"o}rm{\"a}}},\ }\bibfield  {title}
  {\bibinfo {title} {Band geometry, {{Berry}} curvature, and superfluid
  weight},\ }\href {https://doi.org/10.1103/PhysRevB.95.024515} {\bibfield
  {journal} {\bibinfo  {journal} {Physical Review B}\ }\textbf {\bibinfo
  {volume} {95}},\ \bibinfo {pages} {024515} (\bibinfo {year}
  {2017})}\BibitemShut {NoStop}%
\bibitem [{\citenamefont {Hofmann}\ \emph {et~al.}(2020)\citenamefont
  {Hofmann}, \citenamefont {Berg},\ and\ \citenamefont
  {Chowdhury}}]{Hofmann2020}%
  \BibitemOpen
  \bibfield  {author} {\bibinfo {author} {\bibfnamefont {J.~S.}\ \bibnamefont
  {Hofmann}}, \bibinfo {author} {\bibfnamefont {E.}~\bibnamefont {Berg}},\ and\
  \bibinfo {author} {\bibfnamefont {D.}~\bibnamefont {Chowdhury}},\ }\bibfield
  {title} {\bibinfo {title} {Superconductivity, pseudogap, and phase separation
  in topological flat bands},\ }\href
  {https://doi.org/10.1103/PhysRevB.102.201112} {\bibfield  {journal} {\bibinfo
   {journal} {Physical Review B}\ }\textbf {\bibinfo {volume} {102}},\ \bibinfo
  {pages} {201112} (\bibinfo {year} {2020})}\BibitemShut {NoStop}%
\bibitem [{\citenamefont {Peri}\ \emph {et~al.}(2021)\citenamefont {Peri},
  \citenamefont {Song}, \citenamefont {Bernevig},\ and\ \citenamefont
  {Huber}}]{Peri2021}%
  \BibitemOpen
  \bibfield  {author} {\bibinfo {author} {\bibfnamefont {V.}~\bibnamefont
  {Peri}}, \bibinfo {author} {\bibfnamefont {Z.-D.}\ \bibnamefont {Song}},
  \bibinfo {author} {\bibfnamefont {B.~A.}\ \bibnamefont {Bernevig}},\ and\
  \bibinfo {author} {\bibfnamefont {S.~D.}\ \bibnamefont {Huber}},\ }\bibfield
  {title} {\bibinfo {title} {Fragile {{Topology}} and {{Flat-Band
  Superconductivity}} in the {{Strong-Coupling Regime}}},\ }\href
  {https://doi.org/10.1103/PhysRevLett.126.027002} {\bibfield  {journal}
  {\bibinfo  {journal} {Physical Review Letters}\ }\textbf {\bibinfo {volume}
  {126}},\ \bibinfo {pages} {027002} (\bibinfo {year} {2021})}\BibitemShut
  {NoStop}%
\bibitem [{\citenamefont {Vidal}\ \emph {et~al.}(2000)\citenamefont {Vidal},
  \citenamefont {Dou{\c c}ot}, \citenamefont {Mosseri},\ and\ \citenamefont
  {Butaud}}]{Vidal2000}%
  \BibitemOpen
  \bibfield  {author} {\bibinfo {author} {\bibfnamefont {J.}~\bibnamefont
  {Vidal}}, \bibinfo {author} {\bibfnamefont {B.}~\bibnamefont {Dou{\c c}ot}},
  \bibinfo {author} {\bibfnamefont {R.}~\bibnamefont {Mosseri}},\ and\ \bibinfo
  {author} {\bibfnamefont {P.}~\bibnamefont {Butaud}},\ }\bibfield  {title}
  {\bibinfo {title} {Interaction {{Induced Delocalization}} for {{Two
  Particles}} in a {{Periodic Potential}}},\ }\href
  {https://doi.org/10.1103/PhysRevLett.85.3906} {\bibfield  {journal} {\bibinfo
   {journal} {Physical Review Letters}\ }\textbf {\bibinfo {volume} {85}},\
  \bibinfo {pages} {3906} (\bibinfo {year} {2000})}\BibitemShut {NoStop}%
\bibitem [{\citenamefont {Dou{\c c}ot}\ and\ \citenamefont
  {Vidal}(2002)}]{Doucot2002}%
  \BibitemOpen
  \bibfield  {author} {\bibinfo {author} {\bibfnamefont {B.}~\bibnamefont
  {Dou{\c c}ot}}\ and\ \bibinfo {author} {\bibfnamefont {J.}~\bibnamefont
  {Vidal}},\ }\bibfield  {title} {\bibinfo {title} {Pairing of {{Cooper Pairs}}
  in a {{Fully Frustrated Josephson-Junction Chain}}},\ }\href
  {https://doi.org/10.1103/PhysRevLett.88.227005} {\bibfield  {journal}
  {\bibinfo  {journal} {Physical Review Letters}\ }\textbf {\bibinfo {volume}
  {88}},\ \bibinfo {pages} {227005} (\bibinfo {year} {2002})}\BibitemShut
  {NoStop}%
\bibitem [{\citenamefont {T{\"o}rm{\"a}}\ \emph {et~al.}(2018)\citenamefont
  {T{\"o}rm{\"a}}, \citenamefont {Liang},\ and\ \citenamefont
  {Peotta}}]{Torma2018}%
  \BibitemOpen
  \bibfield  {author} {\bibinfo {author} {\bibfnamefont {P.}~\bibnamefont
  {T{\"o}rm{\"a}}}, \bibinfo {author} {\bibfnamefont {L.}~\bibnamefont
  {Liang}},\ and\ \bibinfo {author} {\bibfnamefont {S.}~\bibnamefont
  {Peotta}},\ }\bibfield  {title} {\bibinfo {title} {Quantum metric and
  effective mass of a two-body bound state in a flat band},\ }\href
  {https://doi.org/10.1103/PhysRevB.98.220511} {\bibfield  {journal} {\bibinfo
  {journal} {Physical Review B}\ }\textbf {\bibinfo {volume} {98}},\ \bibinfo
  {pages} {220511} (\bibinfo {year} {2018})}\BibitemShut {NoStop}%
\bibitem [{\citenamefont {Gornyi}\ \emph {et~al.}(2005)\citenamefont {Gornyi},
  \citenamefont {Mirlin},\ and\ \citenamefont {Polyakov}}]{Gornyi2005}%
  \BibitemOpen
  \bibfield  {author} {\bibinfo {author} {\bibfnamefont {I.~V.}\ \bibnamefont
  {Gornyi}}, \bibinfo {author} {\bibfnamefont {A.~D.}\ \bibnamefont {Mirlin}},\
  and\ \bibinfo {author} {\bibfnamefont {D.~G.}\ \bibnamefont {Polyakov}},\
  }\bibfield  {title} {\bibinfo {title} {Interacting {{Electrons}} in
  {{Disordered Wires}}: {{Anderson Localization}} and {{Low- T Transport}}},\
  }\href {https://doi.org/10.1103/PhysRevLett.95.206603} {\bibfield  {journal}
  {\bibinfo  {journal} {Physical Review Letters}\ }\textbf {\bibinfo {volume}
  {95}},\ \bibinfo {pages} {206603} (\bibinfo {year} {2005})}\BibitemShut
  {NoStop}%
\bibitem [{\citenamefont {Basko}\ \emph {et~al.}(2006)\citenamefont {Basko},
  \citenamefont {Aleiner},\ and\ \citenamefont {Altshuler}}]{Basko2006}%
  \BibitemOpen
  \bibfield  {author} {\bibinfo {author} {\bibfnamefont {D.}~\bibnamefont
  {Basko}}, \bibinfo {author} {\bibfnamefont {I.}~\bibnamefont {Aleiner}},\
  and\ \bibinfo {author} {\bibfnamefont {B.}~\bibnamefont {Altshuler}},\
  }\bibfield  {title} {\bibinfo {title} {Metal\textendash insulator transition
  in a weakly interacting many-electron system with localized single-particle
  states},\ }\href {https://doi.org/10.1016/j.aop.2005.11.014} {\bibfield
  {journal} {\bibinfo  {journal} {Annals of Physics}\ }\textbf {\bibinfo
  {volume} {321}},\ \bibinfo {pages} {1126} (\bibinfo {year}
  {2006})}\BibitemShut {NoStop}%
\bibitem [{\citenamefont {Oganesyan}\ and\ \citenamefont
  {Huse}(2007)}]{Oganesyan2007}%
  \BibitemOpen
  \bibfield  {author} {\bibinfo {author} {\bibfnamefont {V.}~\bibnamefont
  {Oganesyan}}\ and\ \bibinfo {author} {\bibfnamefont {D.~A.}\ \bibnamefont
  {Huse}},\ }\bibfield  {title} {\bibinfo {title} {Localization of interacting
  fermions at high temperature},\ }\href
  {https://doi.org/10.1103/PhysRevB.75.155111} {\bibfield  {journal} {\bibinfo
  {journal} {Physical Review B}\ }\textbf {\bibinfo {volume} {75}},\ \bibinfo
  {pages} {155111} (\bibinfo {year} {2007})}\BibitemShut {NoStop}%
\bibitem [{\citenamefont {Pal}\ and\ \citenamefont {Huse}(2010)}]{Pal2010}%
  \BibitemOpen
  \bibfield  {author} {\bibinfo {author} {\bibfnamefont {A.}~\bibnamefont
  {Pal}}\ and\ \bibinfo {author} {\bibfnamefont {D.~A.}\ \bibnamefont {Huse}},\
  }\bibfield  {title} {\bibinfo {title} {Many-body localization phase
  transition},\ }\href {https://doi.org/10.1103/PhysRevB.82.174411} {\bibfield
  {journal} {\bibinfo  {journal} {Physical Review B}\ }\textbf {\bibinfo
  {volume} {82}},\ \bibinfo {pages} {174411} (\bibinfo {year}
  {2010})}\BibitemShut {NoStop}%
\bibitem [{\citenamefont {Imbrie}(2016)}]{Imbrie2016}%
  \BibitemOpen
  \bibfield  {author} {\bibinfo {author} {\bibfnamefont {J.~Z.}\ \bibnamefont
  {Imbrie}},\ }\bibfield  {title} {\bibinfo {title} {On {{Many-Body
  Localization}} for {{Quantum Spin Chains}}},\ }\href
  {https://doi.org/10.1007/s10955-016-1508-x} {\bibfield  {journal} {\bibinfo
  {journal} {Journal of Statistical Physics}\ }\textbf {\bibinfo {volume}
  {163}},\ \bibinfo {pages} {998} (\bibinfo {year} {2016})}\BibitemShut
  {NoStop}%
\bibitem [{\citenamefont {Nandkishore}\ and\ \citenamefont
  {Huse}(2015{\natexlab{a}})}]{Nandkishore2015}%
  \BibitemOpen
  \bibfield  {author} {\bibinfo {author} {\bibfnamefont {R.}~\bibnamefont
  {Nandkishore}}\ and\ \bibinfo {author} {\bibfnamefont {D.~A.}\ \bibnamefont
  {Huse}},\ }\bibfield  {title} {\bibinfo {title} {Many-{{Body Localization}}
  and {{Thermalization}} in {{Quantum Statistical Mechanics}}},\ }\href
  {https://doi.org/10.1146/annurev-conmatphys-031214-014726} {\bibfield
  {journal} {\bibinfo  {journal} {Annual Review of Condensed Matter Physics}\
  }\textbf {\bibinfo {volume} {6}},\ \bibinfo {pages} {15} (\bibinfo {year}
  {2015}{\natexlab{a}})}\BibitemShut {NoStop}%
\bibitem [{\citenamefont {Altman}\ and\ \citenamefont
  {Vosk}(2015)}]{Altman2015}%
  \BibitemOpen
  \bibfield  {author} {\bibinfo {author} {\bibfnamefont {E.}~\bibnamefont
  {Altman}}\ and\ \bibinfo {author} {\bibfnamefont {R.}~\bibnamefont {Vosk}},\
  }\bibfield  {title} {\bibinfo {title} {Universal {{Dynamics}} and
  {{Renormalization}} in {{Many-Body-Localized Systems}}},\ }\href
  {https://doi.org/10.1146/annurev-conmatphys-031214-014701} {\bibfield
  {journal} {\bibinfo  {journal} {Annual Review of Condensed Matter Physics}\
  }\textbf {\bibinfo {volume} {6}},\ \bibinfo {pages} {383} (\bibinfo {year}
  {2015})}\BibitemShut {NoStop}%
\bibitem [{\citenamefont {Alet}\ and\ \citenamefont
  {Laflorencie}(2018)}]{Alet2018}%
  \BibitemOpen
  \bibfield  {author} {\bibinfo {author} {\bibfnamefont {F.}~\bibnamefont
  {Alet}}\ and\ \bibinfo {author} {\bibfnamefont {N.}~\bibnamefont
  {Laflorencie}},\ }\bibfield  {title} {\bibinfo {title} {Many-body
  localization: {{An}} introduction and selected topics},\ }\href
  {https://doi.org/10.1016/j.crhy.2018.03.003} {\bibfield  {journal} {\bibinfo
  {journal} {Comptes Rendus Physique}\ }\bibinfo {series} {Quantum Simulation /
  {{Simulation}} Quantique},\ \textbf {\bibinfo {volume} {19}},\ \bibinfo
  {pages} {498} (\bibinfo {year} {2018})}\BibitemShut {NoStop}%
\bibitem [{\citenamefont {Kondov}\ \emph {et~al.}(2015)\citenamefont {Kondov},
  \citenamefont {McGehee}, \citenamefont {Xu},\ and\ \citenamefont
  {DeMarco}}]{Kondov2015}%
  \BibitemOpen
  \bibfield  {author} {\bibinfo {author} {\bibfnamefont {S.~S.}\ \bibnamefont
  {Kondov}}, \bibinfo {author} {\bibfnamefont {W.~R.}\ \bibnamefont {McGehee}},
  \bibinfo {author} {\bibfnamefont {W.}~\bibnamefont {Xu}},\ and\ \bibinfo
  {author} {\bibfnamefont {B.}~\bibnamefont {DeMarco}},\ }\bibfield  {title}
  {\bibinfo {title} {Disorder-{{Induced Localization}} in a {{Strongly
  Correlated Atomic Hubbard Gas}}},\ }\href
  {https://doi.org/10.1103/PhysRevLett.114.083002} {\bibfield  {journal}
  {\bibinfo  {journal} {Physical Review Letters}\ }\textbf {\bibinfo {volume}
  {114}},\ \bibinfo {pages} {083002} (\bibinfo {year} {2015})}\BibitemShut
  {NoStop}%
\bibitem [{\citenamefont {Schreiber}\ \emph {et~al.}(2015)\citenamefont
  {Schreiber}, \citenamefont {Hodgman}, \citenamefont {Bordia}, \citenamefont
  {L{\"u}schen}, \citenamefont {Fischer}, \citenamefont {Vosk}, \citenamefont
  {Altman}, \citenamefont {Schneider},\ and\ \citenamefont
  {Bloch}}]{Schreiber2015}%
  \BibitemOpen
  \bibfield  {author} {\bibinfo {author} {\bibfnamefont {M.}~\bibnamefont
  {Schreiber}}, \bibinfo {author} {\bibfnamefont {S.~S.}\ \bibnamefont
  {Hodgman}}, \bibinfo {author} {\bibfnamefont {P.}~\bibnamefont {Bordia}},
  \bibinfo {author} {\bibfnamefont {H.~P.}\ \bibnamefont {L{\"u}schen}},
  \bibinfo {author} {\bibfnamefont {M.~H.}\ \bibnamefont {Fischer}}, \bibinfo
  {author} {\bibfnamefont {R.}~\bibnamefont {Vosk}}, \bibinfo {author}
  {\bibfnamefont {E.}~\bibnamefont {Altman}}, \bibinfo {author} {\bibfnamefont
  {U.}~\bibnamefont {Schneider}},\ and\ \bibinfo {author} {\bibfnamefont
  {I.}~\bibnamefont {Bloch}},\ }\bibfield  {title} {\bibinfo {title}
  {Observation of many-body localization of interacting fermions in a
  quasirandom optical lattice},\ }\href
  {https://doi.org/10.1126/science.aaa7432} {\bibfield  {journal} {\bibinfo
  {journal} {Science}\ }\textbf {\bibinfo {volume} {349}},\ \bibinfo {pages}
  {842} (\bibinfo {year} {2015})}\BibitemShut {NoStop}%
\bibitem [{\citenamefont {Choi}\ \emph {et~al.}(2016)\citenamefont {Choi},
  \citenamefont {Hild}, \citenamefont {Zeiher}, \citenamefont {Schau{\ss}},
  \citenamefont {{Rubio-Abadal}}, \citenamefont {Yefsah}, \citenamefont
  {Khemani}, \citenamefont {Huse}, \citenamefont {Bloch},\ and\ \citenamefont
  {Gross}}]{Choi2016}%
  \BibitemOpen
  \bibfield  {author} {\bibinfo {author} {\bibfnamefont {J.-y.}\ \bibnamefont
  {Choi}}, \bibinfo {author} {\bibfnamefont {S.}~\bibnamefont {Hild}}, \bibinfo
  {author} {\bibfnamefont {J.}~\bibnamefont {Zeiher}}, \bibinfo {author}
  {\bibfnamefont {P.}~\bibnamefont {Schau{\ss}}}, \bibinfo {author}
  {\bibfnamefont {A.}~\bibnamefont {{Rubio-Abadal}}}, \bibinfo {author}
  {\bibfnamefont {T.}~\bibnamefont {Yefsah}}, \bibinfo {author} {\bibfnamefont
  {V.}~\bibnamefont {Khemani}}, \bibinfo {author} {\bibfnamefont {D.~A.}\
  \bibnamefont {Huse}}, \bibinfo {author} {\bibfnamefont {I.}~\bibnamefont
  {Bloch}},\ and\ \bibinfo {author} {\bibfnamefont {C.}~\bibnamefont {Gross}},\
  }\bibfield  {title} {\bibinfo {title} {Exploring the many-body localization
  transition in two dimensions},\ }\href
  {https://doi.org/10.1126/science.aaf8834} {\bibfield  {journal} {\bibinfo
  {journal} {Science}\ }\textbf {\bibinfo {volume} {352}},\ \bibinfo {pages}
  {1547} (\bibinfo {year} {2016})}\BibitemShut {NoStop}%
\bibitem [{\citenamefont {Serbyn}\ \emph {et~al.}(2013)\citenamefont {Serbyn},
  \citenamefont {Papi{\'c}},\ and\ \citenamefont {Abanin}}]{Serbyn2013}%
  \BibitemOpen
  \bibfield  {author} {\bibinfo {author} {\bibfnamefont {M.}~\bibnamefont
  {Serbyn}}, \bibinfo {author} {\bibfnamefont {Z.}~\bibnamefont {Papi{\'c}}},\
  and\ \bibinfo {author} {\bibfnamefont {D.~A.}\ \bibnamefont {Abanin}},\
  }\bibfield  {title} {\bibinfo {title} {Local {{Conservation Laws}} and the
  {{Structure}} of the {{Many-Body Localized States}}},\ }\href
  {https://doi.org/10.1103/PhysRevLett.111.127201} {\bibfield  {journal}
  {\bibinfo  {journal} {Physical Review Letters}\ }\textbf {\bibinfo {volume}
  {111}},\ \bibinfo {pages} {127201} (\bibinfo {year} {2013})}\BibitemShut
  {NoStop}%
\bibitem [{\citenamefont {Huse}\ \emph {et~al.}(2014)\citenamefont {Huse},
  \citenamefont {Nandkishore},\ and\ \citenamefont {Oganesyan}}]{Huse2014}%
  \BibitemOpen
  \bibfield  {author} {\bibinfo {author} {\bibfnamefont {D.~A.}\ \bibnamefont
  {Huse}}, \bibinfo {author} {\bibfnamefont {R.}~\bibnamefont {Nandkishore}},\
  and\ \bibinfo {author} {\bibfnamefont {V.}~\bibnamefont {Oganesyan}},\
  }\bibfield  {title} {\bibinfo {title} {Phenomenology of fully
  many-body-localized systems},\ }\href
  {https://doi.org/10.1103/PhysRevB.90.174202} {\bibfield  {journal} {\bibinfo
  {journal} {Physical Review B}\ }\textbf {\bibinfo {volume} {90}},\ \bibinfo
  {pages} {174202} (\bibinfo {year} {2014})}\BibitemShut {NoStop}%
\bibitem [{\citenamefont {Rademaker}\ \emph {et~al.}(2017)\citenamefont
  {Rademaker}, \citenamefont {Ortu{\~n}o},\ and\ \citenamefont
  {Somoza}}]{Rademaker2017}%
  \BibitemOpen
  \bibfield  {author} {\bibinfo {author} {\bibfnamefont {L.}~\bibnamefont
  {Rademaker}}, \bibinfo {author} {\bibfnamefont {M.}~\bibnamefont
  {Ortu{\~n}o}},\ and\ \bibinfo {author} {\bibfnamefont {A.~M.}\ \bibnamefont
  {Somoza}},\ }\bibfield  {title} {\bibinfo {title} {Many-body localization
  from the perspective of {{Integrals}} of {{Motion}}},\ }\href
  {https://doi.org/10.1002/andp.201600322} {\bibfield  {journal} {\bibinfo
  {journal} {Annalen der Physik}\ }\textbf {\bibinfo {volume} {529}},\ \bibinfo
  {pages} {1600322} (\bibinfo {year} {2017})}\BibitemShut {NoStop}%
\bibitem [{\citenamefont {Imbrie}\ \emph {et~al.}(2017)\citenamefont {Imbrie},
  \citenamefont {Ros},\ and\ \citenamefont {Scardicchio}}]{Imbrie2017}%
  \BibitemOpen
  \bibfield  {author} {\bibinfo {author} {\bibfnamefont {J.~Z.}\ \bibnamefont
  {Imbrie}}, \bibinfo {author} {\bibfnamefont {V.}~\bibnamefont {Ros}},\ and\
  \bibinfo {author} {\bibfnamefont {A.}~\bibnamefont {Scardicchio}},\
  }\bibfield  {title} {\bibinfo {title} {Local integrals of motion in many-body
  localized systems: {{Local}} integrals of motion in many-body localized
  systems},\ }\href {https://doi.org/10.1002/andp.201600278} {\bibfield
  {journal} {\bibinfo  {journal} {Annalen der Physik}\ }\textbf {\bibinfo
  {volume} {529}},\ \bibinfo {pages} {1600278} (\bibinfo {year}
  {2017})}\BibitemShut {NoStop}%
\bibitem [{\citenamefont {Schiulaz}\ \emph {et~al.}(2015)\citenamefont
  {Schiulaz}, \citenamefont {Silva},\ and\ \citenamefont
  {M{\"u}ller}}]{Schiulaz2015}%
  \BibitemOpen
  \bibfield  {author} {\bibinfo {author} {\bibfnamefont {M.}~\bibnamefont
  {Schiulaz}}, \bibinfo {author} {\bibfnamefont {A.}~\bibnamefont {Silva}},\
  and\ \bibinfo {author} {\bibfnamefont {M.}~\bibnamefont {M{\"u}ller}},\
  }\bibfield  {title} {\bibinfo {title} {Dynamics in many-body localized
  quantum systems without disorder},\ }\href
  {https://doi.org/10.1103/PhysRevB.91.184202} {\bibfield  {journal} {\bibinfo
  {journal} {Physical Review B}\ }\textbf {\bibinfo {volume} {91}},\ \bibinfo
  {pages} {184202} (\bibinfo {year} {2015})}\BibitemShut {NoStop}%
\bibitem [{\citenamefont {Smith}\ \emph {et~al.}(2017)\citenamefont {Smith},
  \citenamefont {Knolle}, \citenamefont {Kovrizhin},\ and\ \citenamefont
  {Moessner}}]{Smith2017}%
  \BibitemOpen
  \bibfield  {author} {\bibinfo {author} {\bibfnamefont {A.}~\bibnamefont
  {Smith}}, \bibinfo {author} {\bibfnamefont {J.}~\bibnamefont {Knolle}},
  \bibinfo {author} {\bibfnamefont {D.~L.}\ \bibnamefont {Kovrizhin}},\ and\
  \bibinfo {author} {\bibfnamefont {R.}~\bibnamefont {Moessner}},\ }\bibfield
  {title} {\bibinfo {title} {Disorder-{{Free Localization}}},\ }\href
  {https://doi.org/10.1103/PhysRevLett.118.266601} {\bibfield  {journal}
  {\bibinfo  {journal} {Physical Review Letters}\ }\textbf {\bibinfo {volume}
  {118}},\ \bibinfo {pages} {266601} (\bibinfo {year} {2017})}\BibitemShut
  {NoStop}%
\bibitem [{\citenamefont {Mondaini}\ and\ \citenamefont
  {Cai}(2017)}]{Mondaini2017}%
  \BibitemOpen
  \bibfield  {author} {\bibinfo {author} {\bibfnamefont {R.}~\bibnamefont
  {Mondaini}}\ and\ \bibinfo {author} {\bibfnamefont {Z.}~\bibnamefont {Cai}},\
  }\bibfield  {title} {\bibinfo {title} {Many-body self-localization in a
  translation-invariant {{Hamiltonian}}},\ }\href
  {https://doi.org/10.1103/PhysRevB.96.035153} {\bibfield  {journal} {\bibinfo
  {journal} {Physical Review B}\ }\textbf {\bibinfo {volume} {96}},\ \bibinfo
  {pages} {035153} (\bibinfo {year} {2017})}\BibitemShut {NoStop}%
\bibitem [{\citenamefont {Brenes}\ \emph {et~al.}(2018)\citenamefont {Brenes},
  \citenamefont {Dalmonte}, \citenamefont {Heyl},\ and\ \citenamefont
  {Scardicchio}}]{Brenes2018}%
  \BibitemOpen
  \bibfield  {author} {\bibinfo {author} {\bibfnamefont {M.}~\bibnamefont
  {Brenes}}, \bibinfo {author} {\bibfnamefont {M.}~\bibnamefont {Dalmonte}},
  \bibinfo {author} {\bibfnamefont {M.}~\bibnamefont {Heyl}},\ and\ \bibinfo
  {author} {\bibfnamefont {A.}~\bibnamefont {Scardicchio}},\ }\bibfield
  {title} {\bibinfo {title} {Many-{{Body Localization Dynamics}} from {{Gauge
  Invariance}}},\ }\href {https://doi.org/10.1103/PhysRevLett.120.030601}
  {\bibfield  {journal} {\bibinfo  {journal} {Physical Review Letters}\
  }\textbf {\bibinfo {volume} {120}},\ \bibinfo {pages} {030601} (\bibinfo
  {year} {2018})}\BibitemShut {NoStop}%
\bibitem [{\citenamefont {Osborne}\ \emph {et~al.}(2023)\citenamefont
  {Osborne}, \citenamefont {McCulloch},\ and\ \citenamefont
  {Halimeh}}]{Osborne2023}%
  \BibitemOpen
  \bibfield  {author} {\bibinfo {author} {\bibfnamefont {J.}~\bibnamefont
  {Osborne}}, \bibinfo {author} {\bibfnamefont {I.~P.}\ \bibnamefont
  {McCulloch}},\ and\ \bibinfo {author} {\bibfnamefont {J.~C.}\ \bibnamefont
  {Halimeh}},\ }\href {https://doi.org/10.48550/arXiv.2301.07720} {\bibinfo
  {title} {Disorder-{{Free Localization}} in \$2+1\${{D Lattice Gauge
  Theories}} with {{Dynamical Matter}}}} (\bibinfo {year} {2023}),\ \Eprint
  {https://arxiv.org/abs/arXiv:2301.07720} {arxiv:arXiv:2301.07720}
  \BibitemShut {NoStop}%
\bibitem [{\citenamefont {Danieli}\ \emph {et~al.}(2020)\citenamefont
  {Danieli}, \citenamefont {Andreanov},\ and\ \citenamefont
  {Flach}}]{Danieli2020}%
  \BibitemOpen
  \bibfield  {author} {\bibinfo {author} {\bibfnamefont {C.}~\bibnamefont
  {Danieli}}, \bibinfo {author} {\bibfnamefont {A.}~\bibnamefont {Andreanov}},\
  and\ \bibinfo {author} {\bibfnamefont {S.}~\bibnamefont {Flach}},\ }\bibfield
   {title} {\bibinfo {title} {Many-body flatband localization},\ }\href
  {https://doi.org/10.1103/PhysRevB.102.041116} {\bibfield  {journal} {\bibinfo
   {journal} {Physical Review B}\ }\textbf {\bibinfo {volume} {102}},\ \bibinfo
  {pages} {041116} (\bibinfo {year} {2020})}\BibitemShut {NoStop}%
\bibitem [{\citenamefont {Danieli}\ \emph {et~al.}(2022)\citenamefont
  {Danieli}, \citenamefont {Andreanov},\ and\ \citenamefont
  {Flach}}]{Danieli2022}%
  \BibitemOpen
  \bibfield  {author} {\bibinfo {author} {\bibfnamefont {C.}~\bibnamefont
  {Danieli}}, \bibinfo {author} {\bibfnamefont {A.}~\bibnamefont {Andreanov}},\
  and\ \bibinfo {author} {\bibfnamefont {S.}~\bibnamefont {Flach}},\ }\bibfield
   {title} {\bibinfo {title} {Many-body localization transition from flat-band
  fine tuning},\ }\href {https://doi.org/10.1103/PhysRevB.105.L041113}
  {\bibfield  {journal} {\bibinfo  {journal} {Physical Review B}\ }\textbf
  {\bibinfo {volume} {105}},\ \bibinfo {pages} {L041113} (\bibinfo {year}
  {2022})}\BibitemShut {NoStop}%
\bibitem [{\citenamefont {Horiguchi}\ and\ \citenamefont
  {Chen}(1974)}]{Horiguchi1974}%
  \BibitemOpen
  \bibfield  {author} {\bibinfo {author} {\bibfnamefont {T.}~\bibnamefont
  {Horiguchi}}\ and\ \bibinfo {author} {\bibfnamefont {C.~C.}\ \bibnamefont
  {Chen}},\ }\bibfield  {title} {\bibinfo {title} {Lattice {{Green}}'s function
  for the diced lattice},\ }\href {https://doi.org/10.1063/1.1666703}
  {\bibfield  {journal} {\bibinfo  {journal} {Journal of Mathematical Physics}\
  }\textbf {\bibinfo {volume} {15}},\ \bibinfo {pages} {659} (\bibinfo {year}
  {1974})}\BibitemShut {NoStop}%
\bibitem [{\citenamefont {Sutherland}(1986)}]{Sutherland1986}%
  \BibitemOpen
  \bibfield  {author} {\bibinfo {author} {\bibfnamefont {B.}~\bibnamefont
  {Sutherland}},\ }\bibfield  {title} {\bibinfo {title} {Localization of
  electronic wave functions due to local topology},\ }\href
  {https://doi.org/10.1103/PhysRevB.34.5208} {\bibfield  {journal} {\bibinfo
  {journal} {Physical Review B}\ }\textbf {\bibinfo {volume} {34}},\ \bibinfo
  {pages} {5208} (\bibinfo {year} {1986})}\BibitemShut {NoStop}%
\bibitem [{\citenamefont {Vidal}\ \emph {et~al.}(1998)\citenamefont {Vidal},
  \citenamefont {Mosseri},\ and\ \citenamefont {Dou{\c c}ot}}]{Vidal1998}%
  \BibitemOpen
  \bibfield  {author} {\bibinfo {author} {\bibfnamefont {J.}~\bibnamefont
  {Vidal}}, \bibinfo {author} {\bibfnamefont {R.}~\bibnamefont {Mosseri}},\
  and\ \bibinfo {author} {\bibfnamefont {B.}~\bibnamefont {Dou{\c c}ot}},\
  }\bibfield  {title} {\bibinfo {title} {Aharonov-{{Bohm Cages}} in
  {{Two-Dimensional Structures}}},\ }\href
  {https://doi.org/10.1103/PhysRevLett.81.5888} {\bibfield  {journal} {\bibinfo
   {journal} {Physical Review Letters}\ }\textbf {\bibinfo {volume} {81}},\
  \bibinfo {pages} {5888} (\bibinfo {year} {1998})}\BibitemShut {NoStop}%
\bibitem [{\citenamefont {Vidal}\ \emph {et~al.}(2001)\citenamefont {Vidal},
  \citenamefont {Butaud}, \citenamefont {Dou{\c c}ot},\ and\ \citenamefont
  {Mosseri}}]{Vidal2001}%
  \BibitemOpen
  \bibfield  {author} {\bibinfo {author} {\bibfnamefont {J.}~\bibnamefont
  {Vidal}}, \bibinfo {author} {\bibfnamefont {P.}~\bibnamefont {Butaud}},
  \bibinfo {author} {\bibfnamefont {B.}~\bibnamefont {Dou{\c c}ot}},\ and\
  \bibinfo {author} {\bibfnamefont {R.}~\bibnamefont {Mosseri}},\ }\bibfield
  {title} {\bibinfo {title} {Disorder and interactions in {{Aharonov-Bohm}}
  cages},\ }\href {https://doi.org/10.1103/PhysRevB.64.155306} {\bibfield
  {journal} {\bibinfo  {journal} {Physical Review B}\ }\textbf {\bibinfo
  {volume} {64}},\ \bibinfo {pages} {155306} (\bibinfo {year}
  {2001})}\BibitemShut {NoStop}%
\bibitem [{\citenamefont {M{\"o}ller}\ and\ \citenamefont
  {Cooper}(2012)}]{Moller2012}%
  \BibitemOpen
  \bibfield  {author} {\bibinfo {author} {\bibfnamefont {G.}~\bibnamefont
  {M{\"o}ller}}\ and\ \bibinfo {author} {\bibfnamefont {N.~R.}\ \bibnamefont
  {Cooper}},\ }\bibfield  {title} {\bibinfo {title} {Correlated {{Phases}} of
  {{Bosons}} in the {{Flat Lowest Band}} of the {{Dice Lattice}}},\ }\href
  {https://doi.org/10.1103/PhysRevLett.108.045306} {\bibfield  {journal}
  {\bibinfo  {journal} {Physical Review Letters}\ }\textbf {\bibinfo {volume}
  {108}},\ \bibinfo {pages} {045306} (\bibinfo {year} {2012})}\BibitemShut
  {NoStop}%
\bibitem [{\citenamefont {Payrits}\ and\ \citenamefont
  {Barnett}(2014)}]{Payrits2014}%
  \BibitemOpen
  \bibfield  {author} {\bibinfo {author} {\bibfnamefont {M.}~\bibnamefont
  {Payrits}}\ and\ \bibinfo {author} {\bibfnamefont {R.}~\bibnamefont
  {Barnett}},\ }\bibfield  {title} {\bibinfo {title} {Order-by-disorder
  degeneracy lifting of interacting bosons on the dice lattice},\ }\href
  {https://doi.org/10.1103/PhysRevA.90.013608} {\bibfield  {journal} {\bibinfo
  {journal} {Physical Review A}\ }\textbf {\bibinfo {volume} {90}},\ \bibinfo
  {pages} {013608} (\bibinfo {year} {2014})}\BibitemShut {NoStop}%
\bibitem [{\citenamefont {Andrijauskas}\ \emph {et~al.}(2015)\citenamefont
  {Andrijauskas}, \citenamefont {Anisimovas}, \citenamefont {Ra{\v
  c}i{\=u}nas}, \citenamefont {Mekys}, \citenamefont {Kudria{\v s}ov},
  \citenamefont {Spielman},\ and\ \citenamefont
  {Juzeli{\=u}nas}}]{Andrijauskas2015}%
  \BibitemOpen
  \bibfield  {author} {\bibinfo {author} {\bibfnamefont {T.}~\bibnamefont
  {Andrijauskas}}, \bibinfo {author} {\bibfnamefont {E.}~\bibnamefont
  {Anisimovas}}, \bibinfo {author} {\bibfnamefont {M.}~\bibnamefont {Ra{\v
  c}i{\=u}nas}}, \bibinfo {author} {\bibfnamefont {A.}~\bibnamefont {Mekys}},
  \bibinfo {author} {\bibfnamefont {V.}~\bibnamefont {Kudria{\v s}ov}},
  \bibinfo {author} {\bibfnamefont {I.~B.}\ \bibnamefont {Spielman}},\ and\
  \bibinfo {author} {\bibfnamefont {G.}~\bibnamefont {Juzeli{\=u}nas}},\
  }\bibfield  {title} {\bibinfo {title} {Three-level {{Haldane-like}} model on
  a dice optical lattice},\ }\href {https://doi.org/10.1103/PhysRevA.92.033617}
  {\bibfield  {journal} {\bibinfo  {journal} {Physical Review A}\ }\textbf
  {\bibinfo {volume} {92}},\ \bibinfo {pages} {033617} (\bibinfo {year}
  {2015})}\BibitemShut {NoStop}%
\bibitem [{\citenamefont {Wu}\ and\ \citenamefont {Zhang}(2021)}]{Wu2021}%
  \BibitemOpen
  \bibfield  {author} {\bibinfo {author} {\bibfnamefont {Y.-R.}\ \bibnamefont
  {Wu}}\ and\ \bibinfo {author} {\bibfnamefont {Y.-C.}\ \bibnamefont {Zhang}},\
  }\bibfield  {title} {\bibinfo {title} {Superfluid states in {$\alpha$}
  \textendash{} {{T3}} lattice*},\ }\href
  {https://doi.org/10.1088/1674-1056/abea8a} {\bibfield  {journal} {\bibinfo
  {journal} {Chinese Physics B}\ }\textbf {\bibinfo {volume} {30}},\ \bibinfo
  {pages} {060306} (\bibinfo {year} {2021})}\BibitemShut {NoStop}%
\bibitem [{\citenamefont {Marzari}\ \emph {et~al.}(2012)\citenamefont
  {Marzari}, \citenamefont {Mostofi}, \citenamefont {Yates}, \citenamefont
  {Souza},\ and\ \citenamefont {Vanderbilt}}]{Marzari2012}%
  \BibitemOpen
  \bibfield  {author} {\bibinfo {author} {\bibfnamefont {N.}~\bibnamefont
  {Marzari}}, \bibinfo {author} {\bibfnamefont {A.~A.}\ \bibnamefont
  {Mostofi}}, \bibinfo {author} {\bibfnamefont {J.~R.}\ \bibnamefont {Yates}},
  \bibinfo {author} {\bibfnamefont {I.}~\bibnamefont {Souza}},\ and\ \bibinfo
  {author} {\bibfnamefont {D.}~\bibnamefont {Vanderbilt}},\ }\bibfield  {title}
  {\bibinfo {title} {Maximally localized {{Wannier}} functions: {{Theory}} and
  applications},\ }\href {https://doi.org/10.1103/RevModPhys.84.1419}
  {\bibfield  {journal} {\bibinfo  {journal} {Reviews of Modern Physics}\
  }\textbf {\bibinfo {volume} {84}},\ \bibinfo {pages} {1419} (\bibinfo {year}
  {2012})}\BibitemShut {NoStop}%
\bibitem [{\citenamefont {Abilio}\ \emph {et~al.}(1999)\citenamefont {Abilio},
  \citenamefont {Butaud}, \citenamefont {Fournier}, \citenamefont {Pannetier},
  \citenamefont {Vidal}, \citenamefont {Tedesco},\ and\ \citenamefont
  {Dalzotto}}]{Abilio1999}%
  \BibitemOpen
  \bibfield  {author} {\bibinfo {author} {\bibfnamefont {C.~C.}\ \bibnamefont
  {Abilio}}, \bibinfo {author} {\bibfnamefont {P.}~\bibnamefont {Butaud}},
  \bibinfo {author} {\bibfnamefont {{\relax Th}.}~\bibnamefont {Fournier}},
  \bibinfo {author} {\bibfnamefont {B.}~\bibnamefont {Pannetier}}, \bibinfo
  {author} {\bibfnamefont {J.}~\bibnamefont {Vidal}}, \bibinfo {author}
  {\bibfnamefont {S.}~\bibnamefont {Tedesco}},\ and\ \bibinfo {author}
  {\bibfnamefont {B.}~\bibnamefont {Dalzotto}},\ }\bibfield  {title} {\bibinfo
  {title} {Magnetic {{Field Induced Localization}} in a {{Two-Dimensional
  Superconducting Wire Network}}},\ }\href
  {https://doi.org/10.1103/PhysRevLett.83.5102} {\bibfield  {journal} {\bibinfo
   {journal} {Physical Review Letters}\ }\textbf {\bibinfo {volume} {83}},\
  \bibinfo {pages} {5102} (\bibinfo {year} {1999})}\BibitemShut {NoStop}%
\bibitem [{\citenamefont {Korshunov}(2001)}]{Korshunov2001}%
  \BibitemOpen
  \bibfield  {author} {\bibinfo {author} {\bibfnamefont {S.~E.}\ \bibnamefont
  {Korshunov}},\ }\bibfield  {title} {\bibinfo {title} {Vortex ordering in
  fully frustrated superconducting systems with a dice lattice},\ }\href
  {https://doi.org/10.1103/PhysRevB.63.134503} {\bibfield  {journal} {\bibinfo
  {journal} {Physical Review B}\ }\textbf {\bibinfo {volume} {63}},\ \bibinfo
  {pages} {134503} (\bibinfo {year} {2001})}\BibitemShut {NoStop}%
\bibitem [{\citenamefont {Cataudella}\ and\ \citenamefont
  {Fazio}(2003)}]{Cataudella2003}%
  \BibitemOpen
  \bibfield  {author} {\bibinfo {author} {\bibfnamefont {V.}~\bibnamefont
  {Cataudella}}\ and\ \bibinfo {author} {\bibfnamefont {R.}~\bibnamefont
  {Fazio}},\ }\bibfield  {title} {\bibinfo {title} {Glassy dynamics of
  {{Josephson}} arrays on a dice lattice},\ }\href
  {https://doi.org/10.1209/epl/i2003-00180-y} {\bibfield  {journal} {\bibinfo
  {journal} {Europhysics Letters}\ }\textbf {\bibinfo {volume} {61}},\ \bibinfo
  {pages} {341} (\bibinfo {year} {2003})}\BibitemShut {NoStop}%
\bibitem [{\citenamefont {Korshunov}\ and\ \citenamefont {Dou{\c
  c}ot}(2004)}]{Korshunov2004}%
  \BibitemOpen
  \bibfield  {author} {\bibinfo {author} {\bibfnamefont {S.~E.}\ \bibnamefont
  {Korshunov}}\ and\ \bibinfo {author} {\bibfnamefont {B.}~\bibnamefont {Dou{\c
  c}ot}},\ }\bibfield  {title} {\bibinfo {title} {Structure of the
  superconducting state in a fully frustrated wire network with dice lattice
  geometry},\ }\href {https://doi.org/10.1103/PhysRevB.70.134507} {\bibfield
  {journal} {\bibinfo  {journal} {Physical Review B}\ }\textbf {\bibinfo
  {volume} {70}},\ \bibinfo {pages} {134507} (\bibinfo {year}
  {2004})}\BibitemShut {NoStop}%
\bibitem [{\citenamefont {Korshunov}(2005)}]{Korshunov2005}%
  \BibitemOpen
  \bibfield  {author} {\bibinfo {author} {\bibfnamefont {S.~E.}\ \bibnamefont
  {Korshunov}},\ }\bibfield  {title} {\bibinfo {title} {Fluctuation-induced
  vortex pattern and its disordering in the fully frustrated \${{XY}}\$ model
  on a dice lattice},\ }\href {https://doi.org/10.1103/PhysRevB.71.174501}
  {\bibfield  {journal} {\bibinfo  {journal} {Physical Review B}\ }\textbf
  {\bibinfo {volume} {71}},\ \bibinfo {pages} {174501} (\bibinfo {year}
  {2005})}\BibitemShut {NoStop}%
\bibitem [{\citenamefont {Weinberg}\ and\ \citenamefont
  {Bukov}(2017)}]{Weinberg2017}%
  \BibitemOpen
  \bibfield  {author} {\bibinfo {author} {\bibfnamefont {P.}~\bibnamefont
  {Weinberg}}\ and\ \bibinfo {author} {\bibfnamefont {M.}~\bibnamefont
  {Bukov}},\ }\bibfield  {title} {\bibinfo {title} {{{QuSpin}}: A {{Python}}
  package for dynamics and exact diagonalisation of quantum many body systems
  part {{I}}: Spin chains},\ }\href
  {https://doi.org/10.21468/SciPostPhys.2.1.003} {\bibfield  {journal}
  {\bibinfo  {journal} {SciPost Physics}\ }\textbf {\bibinfo {volume} {2}},\
  \bibinfo {pages} {003} (\bibinfo {year} {2017})}\BibitemShut {NoStop}%
\bibitem [{\citenamefont {Weinberg}\ and\ \citenamefont
  {Bukov}(2019)}]{Weinberg2019}%
  \BibitemOpen
  \bibfield  {author} {\bibinfo {author} {\bibfnamefont {P.}~\bibnamefont
  {Weinberg}}\ and\ \bibinfo {author} {\bibfnamefont {M.}~\bibnamefont
  {Bukov}},\ }\bibfield  {title} {\bibinfo {title} {{{QuSpin}}: A {{Python}}
  package for dynamics and exact diagonalisation of quantum many body systems.
  {{Part II}}: Bosons, fermions and higher spins},\ }\href
  {https://doi.org/10.21468/SciPostPhys.7.2.020} {\bibfield  {journal}
  {\bibinfo  {journal} {SciPost Physics}\ }\textbf {\bibinfo {volume} {7}},\
  \bibinfo {pages} {020} (\bibinfo {year} {2019})}\BibitemShut {NoStop}%
\bibitem [{\citenamefont {Moore}(2015)}]{moore2015ergodic}%
  \BibitemOpen
  \bibfield  {author} {\bibinfo {author} {\bibfnamefont {C.~C.}\ \bibnamefont
  {Moore}},\ }\bibfield  {title} {\bibinfo {title} {Ergodic theorem, ergodic
  theory, and statistical mechanics},\ }\href
  {https://doi.org/https://doi.org/10.1073/pnas.1421798112} {\bibfield
  {journal} {\bibinfo  {journal} {Proceedings of the National Academy of
  Sciences}\ }\textbf {\bibinfo {volume} {112}},\ \bibinfo {pages} {1907}
  (\bibinfo {year} {2015})}\BibitemShut {NoStop}%
\bibitem [{\citenamefont {Abanin}\ \emph {et~al.}(2019)\citenamefont {Abanin},
  \citenamefont {Altman}, \citenamefont {Bloch},\ and\ \citenamefont
  {Serbyn}}]{abanin2019colloquium}%
  \BibitemOpen
  \bibfield  {author} {\bibinfo {author} {\bibfnamefont {D.~A.}\ \bibnamefont
  {Abanin}}, \bibinfo {author} {\bibfnamefont {E.}~\bibnamefont {Altman}},
  \bibinfo {author} {\bibfnamefont {I.}~\bibnamefont {Bloch}},\ and\ \bibinfo
  {author} {\bibfnamefont {M.}~\bibnamefont {Serbyn}},\ }\bibfield  {title}
  {\bibinfo {title} {Colloquium: Many-body localization, thermalization, and
  entanglement},\ }\href {https://doi.org/10.1103/RevModPhys.91.021001}
  {\bibfield  {journal} {\bibinfo  {journal} {Reviews of Modern Physics}\
  }\textbf {\bibinfo {volume} {91}},\ \bibinfo {pages} {021001} (\bibinfo
  {year} {2019})}\BibitemShut {NoStop}%
\bibitem [{\citenamefont {Altman}(2018)}]{altman2018many}%
  \BibitemOpen
  \bibfield  {author} {\bibinfo {author} {\bibfnamefont {E.}~\bibnamefont
  {Altman}},\ }\bibfield  {title} {\bibinfo {title} {Many-body localization and
  quantum thermalization},\ }\href
  {https://doi.org/https://doi.org/10.1038/s41567-018-0305-7} {\bibfield
  {journal} {\bibinfo  {journal} {Nature Physics}\ }\textbf {\bibinfo {volume}
  {14}},\ \bibinfo {pages} {979} (\bibinfo {year} {2018})}\BibitemShut
  {NoStop}%
\bibitem [{\citenamefont {Nandkishore}\ and\ \citenamefont
  {Huse}(2015{\natexlab{b}})}]{nandkishore2015many}%
  \BibitemOpen
  \bibfield  {author} {\bibinfo {author} {\bibfnamefont {R.}~\bibnamefont
  {Nandkishore}}\ and\ \bibinfo {author} {\bibfnamefont {D.~A.}\ \bibnamefont
  {Huse}},\ }\bibfield  {title} {\bibinfo {title} {Many-body localization and
  thermalization in quantum statistical mechanics},\ }\href
  {https://doi.org/10.1146/annurev-conmatphys-031214-014726} {\bibfield
  {journal} {\bibinfo  {journal} {Annu. Rev. Condens. Matter Phys.}\ }\textbf
  {\bibinfo {volume} {6}},\ \bibinfo {pages} {15} (\bibinfo {year}
  {2015}{\natexlab{b}})},\ \Eprint
  {https://arxiv.org/abs/https://doi.org/10.1146/annurev-conmatphys-031214-014726}
  {https://doi.org/10.1146/annurev-conmatphys-031214-014726} \BibitemShut
  {NoStop}%
\bibitem [{\citenamefont {Sels}\ and\ \citenamefont
  {Polkovnikov}(2023)}]{Sels2023}%
  \BibitemOpen
  \bibfield  {author} {\bibinfo {author} {\bibfnamefont {D.}~\bibnamefont
  {Sels}}\ and\ \bibinfo {author} {\bibfnamefont {A.}~\bibnamefont
  {Polkovnikov}},\ }\bibfield  {title} {\bibinfo {title} {Whither many-body
  localization?},\ }\bibfield  {journal} {\bibinfo  {journal} {Journal Club for
  Condensed Matter Physics}\ }\href
  {https://doi.org/10.36471/JCCM_January_2023_01}
  {10.36471/JCCM\_January\_2023\_01} (\bibinfo {year} {2023})\BibitemShut
  {NoStop}%
\bibitem [{\citenamefont {{\v S}untajs}\ \emph
  {et~al.}(2020{\natexlab{a}})\citenamefont {{\v S}untajs}, \citenamefont
  {Bon{\v c}a}, \citenamefont {Prosen},\ and\ \citenamefont
  {Vidmar}}]{Suntajs2020}%
  \BibitemOpen
  \bibfield  {author} {\bibinfo {author} {\bibfnamefont {J.}~\bibnamefont {{\v
  S}untajs}}, \bibinfo {author} {\bibfnamefont {J.}~\bibnamefont {Bon{\v c}a}},
  \bibinfo {author} {\bibfnamefont {T.}~\bibnamefont {Prosen}},\ and\ \bibinfo
  {author} {\bibfnamefont {L.}~\bibnamefont {Vidmar}},\ }\bibfield  {title}
  {\bibinfo {title} {Quantum chaos challenges many-body localization},\ }\href
  {https://doi.org/10.1103/PhysRevE.102.062144} {\bibfield  {journal} {\bibinfo
   {journal} {Physical Review E}\ }\textbf {\bibinfo {volume} {102}},\ \bibinfo
  {pages} {062144} (\bibinfo {year} {2020}{\natexlab{a}})}\BibitemShut
  {NoStop}%
\bibitem [{\citenamefont {{\v S}untajs}\ \emph
  {et~al.}(2020{\natexlab{b}})\citenamefont {{\v S}untajs}, \citenamefont
  {Bon{\v c}a}, \citenamefont {Prosen},\ and\ \citenamefont
  {Vidmar}}]{Suntajs2020a}%
  \BibitemOpen
  \bibfield  {author} {\bibinfo {author} {\bibfnamefont {J.}~\bibnamefont {{\v
  S}untajs}}, \bibinfo {author} {\bibfnamefont {J.}~\bibnamefont {Bon{\v c}a}},
  \bibinfo {author} {\bibfnamefont {T.}~\bibnamefont {Prosen}},\ and\ \bibinfo
  {author} {\bibfnamefont {L.}~\bibnamefont {Vidmar}},\ }\bibfield  {title}
  {\bibinfo {title} {Ergodicity breaking transition in finite disordered spin
  chains},\ }\href {https://doi.org/10.1103/PhysRevB.102.064207} {\bibfield
  {journal} {\bibinfo  {journal} {Physical Review B}\ }\textbf {\bibinfo
  {volume} {102}},\ \bibinfo {pages} {064207} (\bibinfo {year}
  {2020}{\natexlab{b}})}\BibitemShut {NoStop}%
\bibitem [{\citenamefont {Panda}\ \emph {et~al.}(2020)\citenamefont {Panda},
  \citenamefont {Scardicchio}, \citenamefont {Schulz}, \citenamefont {Taylor},\
  and\ \citenamefont {{\v Z}nidari{\v c}}}]{Panda2020}%
  \BibitemOpen
  \bibfield  {author} {\bibinfo {author} {\bibfnamefont {R.~K.}\ \bibnamefont
  {Panda}}, \bibinfo {author} {\bibfnamefont {A.}~\bibnamefont {Scardicchio}},
  \bibinfo {author} {\bibfnamefont {M.}~\bibnamefont {Schulz}}, \bibinfo
  {author} {\bibfnamefont {S.~R.}\ \bibnamefont {Taylor}},\ and\ \bibinfo
  {author} {\bibfnamefont {M.}~\bibnamefont {{\v Z}nidari{\v c}}},\ }\bibfield
  {title} {\bibinfo {title} {Can we study the many-body localisation
  transition?},\ }\href {https://doi.org/10.1209/0295-5075/128/67003}
  {\bibfield  {journal} {\bibinfo  {journal} {Europhysics Letters}\ }\textbf
  {\bibinfo {volume} {128}},\ \bibinfo {pages} {67003} (\bibinfo {year}
  {2020})}\BibitemShut {NoStop}%
\bibitem [{\citenamefont {Sierant}\ \emph {et~al.}(2020)\citenamefont
  {Sierant}, \citenamefont {Delande},\ and\ \citenamefont
  {Zakrzewski}}]{Sierant2020}%
  \BibitemOpen
  \bibfield  {author} {\bibinfo {author} {\bibfnamefont {P.}~\bibnamefont
  {Sierant}}, \bibinfo {author} {\bibfnamefont {D.}~\bibnamefont {Delande}},\
  and\ \bibinfo {author} {\bibfnamefont {J.}~\bibnamefont {Zakrzewski}},\
  }\bibfield  {title} {\bibinfo {title} {Thouless {{Time Analysis}} of
  {{Anderson}} and {{Many-Body Localization Transitions}}},\ }\href
  {https://doi.org/10.1103/PhysRevLett.124.186601} {\bibfield  {journal}
  {\bibinfo  {journal} {Physical Review Letters}\ }\textbf {\bibinfo {volume}
  {124}},\ \bibinfo {pages} {186601} (\bibinfo {year} {2020})}\BibitemShut
  {NoStop}%
\bibitem [{\citenamefont {Abanin}\ \emph {et~al.}(2021)\citenamefont {Abanin},
  \citenamefont {Bardarson}, \citenamefont {De~Tomasi}, \citenamefont
  {Gopalakrishnan}, \citenamefont {Khemani}, \citenamefont {Parameswaran},
  \citenamefont {Pollmann}, \citenamefont {Potter}, \citenamefont {Serbyn},\
  and\ \citenamefont {Vasseur}}]{Abanin2021}%
  \BibitemOpen
  \bibfield  {author} {\bibinfo {author} {\bibfnamefont {D.~A.}\ \bibnamefont
  {Abanin}}, \bibinfo {author} {\bibfnamefont {J.~H.}\ \bibnamefont
  {Bardarson}}, \bibinfo {author} {\bibfnamefont {G.}~\bibnamefont
  {De~Tomasi}}, \bibinfo {author} {\bibfnamefont {S.}~\bibnamefont
  {Gopalakrishnan}}, \bibinfo {author} {\bibfnamefont {V.}~\bibnamefont
  {Khemani}}, \bibinfo {author} {\bibfnamefont {S.~A.}\ \bibnamefont
  {Parameswaran}}, \bibinfo {author} {\bibfnamefont {F.}~\bibnamefont
  {Pollmann}}, \bibinfo {author} {\bibfnamefont {A.~C.}\ \bibnamefont
  {Potter}}, \bibinfo {author} {\bibfnamefont {M.}~\bibnamefont {Serbyn}},\
  and\ \bibinfo {author} {\bibfnamefont {R.}~\bibnamefont {Vasseur}},\
  }\bibfield  {title} {\bibinfo {title} {Distinguishing localization from
  chaos: {{Challenges}} in finite-size systems},\ }\href
  {https://doi.org/10.1016/j.aop.2021.168415} {\bibfield  {journal} {\bibinfo
  {journal} {Annals of Physics}\ }\textbf {\bibinfo {volume} {427}},\ \bibinfo
  {pages} {168415} (\bibinfo {year} {2021})}\BibitemShut {NoStop}%
\bibitem [{\citenamefont {Sels}\ and\ \citenamefont
  {Polkovnikov}(2021)}]{Sels2021}%
  \BibitemOpen
  \bibfield  {author} {\bibinfo {author} {\bibfnamefont {D.}~\bibnamefont
  {Sels}}\ and\ \bibinfo {author} {\bibfnamefont {A.}~\bibnamefont
  {Polkovnikov}},\ }\bibfield  {title} {\bibinfo {title} {Dynamical obstruction
  to localization in a disordered spin chain},\ }\href
  {https://doi.org/10.1103/PhysRevE.104.054105} {\bibfield  {journal} {\bibinfo
   {journal} {Physical Review E}\ }\textbf {\bibinfo {volume} {104}},\ \bibinfo
  {pages} {054105} (\bibinfo {year} {2021})}\BibitemShut {NoStop}%
\bibitem [{\citenamefont {Ma}\ and\ \citenamefont {Lee}(1985)}]{Ma1985}%
  \BibitemOpen
  \bibfield  {author} {\bibinfo {author} {\bibfnamefont {M.}~\bibnamefont
  {Ma}}\ and\ \bibinfo {author} {\bibfnamefont {P.~A.}\ \bibnamefont {Lee}},\
  }\bibfield  {title} {\bibinfo {title} {Localized superconductors},\ }\href
  {https://doi.org/10.1103/PhysRevB.32.5658} {\bibfield  {journal} {\bibinfo
  {journal} {Physical Review B}\ }\textbf {\bibinfo {volume} {32}},\ \bibinfo
  {pages} {5658} (\bibinfo {year} {1985})}\BibitemShut {NoStop}%
\bibitem [{\citenamefont {Ma}\ \emph {et~al.}(1986)\citenamefont {Ma},
  \citenamefont {Halperin},\ and\ \citenamefont {Lee}}]{Ma1986}%
  \BibitemOpen
  \bibfield  {author} {\bibinfo {author} {\bibfnamefont {M.}~\bibnamefont
  {Ma}}, \bibinfo {author} {\bibfnamefont {B.~I.}\ \bibnamefont {Halperin}},\
  and\ \bibinfo {author} {\bibfnamefont {P.~A.}\ \bibnamefont {Lee}},\
  }\bibfield  {title} {\bibinfo {title} {Strongly disordered superfluids:
  {{Quantum}} fluctuations and critical behavior},\ }\href
  {https://doi.org/10.1103/PhysRevB.34.3136} {\bibfield  {journal} {\bibinfo
  {journal} {Physical Review B}\ }\textbf {\bibinfo {volume} {34}},\ \bibinfo
  {pages} {3136} (\bibinfo {year} {1986})}\BibitemShut {NoStop}%
\bibitem [{\citenamefont {Pyykk\"onen}\ \emph {et~al.}(2023)\citenamefont
  {Pyykk\"onen}, \citenamefont {Peotta},\ and\ \citenamefont
  {T\"orm\"a}}]{Pyykkonen2022}%
  \BibitemOpen
  \bibfield  {author} {\bibinfo {author} {\bibfnamefont {V.~A.~J.}\
  \bibnamefont {Pyykk\"onen}}, \bibinfo {author} {\bibfnamefont
  {S.}~\bibnamefont {Peotta}},\ and\ \bibinfo {author} {\bibfnamefont
  {P.}~\bibnamefont {T\"orm\"a}},\ }\bibfield  {title} {\bibinfo {title}
  {Suppression of nonequilibrium quasiparticle transport in flat-band
  superconductors},\ }\href {https://doi.org/10.1103/PhysRevLett.130.216003}
  {\bibfield  {journal} {\bibinfo  {journal} {Phys. Rev. Lett.}\ }\textbf
  {\bibinfo {volume} {130}},\ \bibinfo {pages} {216003} (\bibinfo {year}
  {2023})}\BibitemShut {NoStop}%
\bibitem [{\citenamefont {Bergman}\ \emph {et~al.}(2008)\citenamefont
  {Bergman}, \citenamefont {Wu},\ and\ \citenamefont {Balents}}]{Bergman2008}%
  \BibitemOpen
  \bibfield  {author} {\bibinfo {author} {\bibfnamefont {D.~L.}\ \bibnamefont
  {Bergman}}, \bibinfo {author} {\bibfnamefont {C.}~\bibnamefont {Wu}},\ and\
  \bibinfo {author} {\bibfnamefont {L.}~\bibnamefont {Balents}},\ }\bibfield
  {title} {\bibinfo {title} {Band touching from real-space topology in
  frustrated hopping models},\ }\href
  {https://doi.org/10.1103/PhysRevB.78.125104} {\bibfield  {journal} {\bibinfo
  {journal} {Physical Review B}\ }\textbf {\bibinfo {volume} {78}},\ \bibinfo
  {pages} {125104} (\bibinfo {year} {2008})}\BibitemShut {NoStop}%
\end{thebibliography}
